%% file: Wpaper-model.tex
\documentclass[11pt]{article}
\usepackage{makeidx}
\usepackage{amsmath}
\usepackage{amssymb}
\usepackage{amsfonts}
\usepackage{times}
\usepackage{color}
\usepackage{url}
\usepackage{graphicx}
\usepackage{subfigure}
\usepackage{setspace}
\usepackage{fullpage}

\begin{document}
\newtheorem{theorem}{Theorem}[section]
\newtheorem{corollary}{Corollary}[theorem]
\newtheorem{proposition}[theorem]{Proposition}
\newtheorem{lemma}[theorem]{Lemma}
\newtheorem{definition}{Definition}
\newenvironment{proof}{{\bf Proof:}}{\hfill{\rule{2mm}{2mm}}}
\newenvironment{appendproof}[1]{{\bf Proof of Lemma~\ref{#1}:}}{\hfill{\rule{2mm}{2mm}}}
\newenvironment{appendthmproof}[1]{{\bf Proof of Theorem~\ref{#1}:}}{\hfill{\rule{2mm}{2mm}}}
\newenvironment{appendcorproof}[1]{{\bf Proof of Corollary~\ref{#1}:}}{\hfill{\rule{2mm}{2mm}}}


\newcommand{\GNP}{Erd\"os-R\'enyi random graph}
\newcommand{\hide}[1]{}
\newcommand{\semihide}[1]{{\tiny #1}}
\newcommand{\note}[1]{\textsf{{\textcolor{red}{[[#1]]}}}}
\newcommand{\xhdr}[1]{\paragraph{\bf {#1}.}}

\newcommand{\ie}{\textit{i.e.},}
\newcommand{\eg}{\textit{e.g.},}

\newcommand{\Kron}{Kronecker graph}
\newcommand{\modelx}{MAG model}
\newcommand{\modeldesc}{$M(n, l, \mu, \Theta)$}
\newcommand{\modelpower}{$M(n, l, \vec{\mu}, \vec{\Theta})$}
\newcommand{\modelgraph}{MAG graph}

\newcommand{\lwokron}{\mu \alpha + (1- \mu) \beta}
\newcommand{\swokron}{\mu \beta + (1- \mu) \gamma}
\newcommand{\lkron}{\left( \lwokron \right)}
\newcommand{\skron}{\left( \swokron \right)}
\newcommand{\bkron}[1]{\lkron^{{#1}}\skron^{l - {#1}}}
\newcommand{\blkron}[1]{\lkron^{{#1}l}\skron^{(1-{#1})l}}
\newcommand{\bxkron}[1]{\lkron^{{#1}}\skron^{1 - {#1}}}
\newcommand{\EX}[1]{\mathbb{E}\left[{#1}\right]}
\newcommand{\EP}[1]{\mathbb{E}\left[ P \left[ {#1} \right]\right]}
\newcommand{\ECP}[2]{\mathbb{E}\left[P \left[ {#1} \right] | {#2} \right]}
\newcommand{\xset}{\char`\\}
\newcommand{\LGN}{n \rightarrow \infty}

\newcommand{\criconn}[1]{\lkron^{\frac{{#1}}{\log n}} \skron^{\frac{l-{#1}}{\log n}}}
\newcommand{\crisimconn}{\skron^{\rho}}
\newcommand{\cricpxconn}[1]{\left[\bxkron{#1}\right]^{\rho}}
\newcommand{\crilcc}{\left[\bxkron{\mu}\right]^{\rho}}

\newcommand{\RATIO}[1]{\left( \frac{\mu}{{#1}} \right)^{{#1}} \left( \frac{1-\mu}{1-{#1}} \right)^{1-{#1}}}
\newcommand{\RATIC}[1]{\left( \frac{\mu}{{#1}} \right)^{{#1}} \left( \frac{1-\mu}{1-({#1})} \right)^{1-({#1})}}
\newcommand{\HALF}{\frac{1}{2}}

\newcommand{\EN}[1]{{l \choose {#1}} \mu^{#1} (1 - \mu)^{l - {#1}}}
\newcommand{\EL}[1]{{l \choose {#1}l} \mu^{{#1}l} (1 - \mu)^{(1 - {#1})l}}
\newcommand{\ACHL}[1]{\frac{\sqrt{2\pi l}{(\frac{l}{e})^{l}}}{ \sqrt{2\pi {#1}l}(\frac{{#1}l}{e})^{{#1}l} \sqrt{2\pi (1-{#1})} \left( \frac{(1-{#1})l}{e}\right)^{(1-{#1})l} }}
\newcommand{\ACHSIML}[1]{\sqrt{2\pi l{#1}\left(1 - {#1}\right)}}
\newcommand{\WHP}{with high probability}

\title{Multiplicative Attribute Graph Model of Real-World Networks\thanks{A short version of this paper appeared in \textit{Proceedings of the Seventh Workshop on Algorithms and Models for the Web Graph (WAW'10)}~\cite{mh10mag}.}}

\author{Myunghwan Kim \ \ \ \ \ \ \ \ \ \ \ \ \ \ \ \ Jure Leskovec\\Stanford University}
\date{}
\maketitle

\begin{abstract}
\input{W000abstract}

\end{abstract}

\section{Introduction}
\label{sec:intro}
\input{W010intro}

\section{Formulating of the Multiplicative Attribute Graph (MAG) model}
\label{sec:mag}
\input{W030proposed}

\input{W040proofs}
\input{W050simulation}

\section{Conclusion}
\label{sec:conclusion}
\input{W060conclusion}

\bigskip
\section*{Acknowledgments}
We thank to Daniel McFarland for discussion and comments.
Myunghwan Kim was supported by the Kwanjeong Educational Foundation fellowship.
The research was supported in part by NSF grants
CNS-1010921,  
IIS-1016909,    
LLNL grant DE-AC52-07NA27344, 
Albert Yu \& Mary Bechmann Foundation, IBM, Lightspeed, Microsoft and Yahoo.

\clearpage
\bibliography{Wpaper-model}
\bibliographystyle{abbrv}

\appendix
\label{sec:appendix}
\input{W070appendix}

\end{document}

%% file: W000abstract.tex
Large scale real-world network data such as social and information networks are ubiquitous.
The study of such social and information networks seeks to find patterns and explain their emergence through tractable models.
In most networks, and especially in social networks, nodes have a rich set of attributes (\eg~age, gender) associated with them.
%

Here we present a model that we refer to as the Multiplicative
Attribute Graphs (MAG), which naturally captures the interactions between the network structure and the node attributes.
We consider a model where each node has a vector of categorical latent attributes associated with it.
The probability of an edge between a pair of nodes then depends on the product of individual attribute-attribute affinities.
The model yields itself to mathematical analysis and we derive thresholds for the connectivity and the emergence of the giant connected component, and show that the model gives rise to networks with a constant diameter.
We analyze the degree distribution to show that MAG model can produce networks with either log-normal or power-law degree distributions depending on certain conditions.

%% file: W010intro.tex
With the emergence of the Web, large online social computing applications have
become ubiquitous, which in turn gave rise to a wide range of massive real-world
social and information network data such as social networks, computer networks,
Internet networks, communication networks, e-mail interactions, Web graphs, and
so on. The unifying theme of studying real-world networks
is to find patterns of connectivity and explain them through
models.
The main objective is to answer questions such as ``What do real graphs look
like?'', ``How do they evolve over time? ``How can we synthesize realistic looking
graphs?'', ``How can we find models that explain the observed patterns?'', and ``What are
algorithmic consequences of the observations and models?''.

Research on empirical observations about the structure of networks and the
models giving rise to such structures go hand in hand.
The empirical analysis of large real-world networks aims to discover common
structural properties or patterns, such as heavy-tailed degree
distributions~\cite{faloutsos99powerlaw,broder00bowtie}, local clustering of
edges~\cite{watts98smallworld,jure09community}, small
diameter~\cite{barabasi99diameter,jure05dpl},
navigability~\cite{milgram67smallworld,kleinberg00navigation}, emergence of
community structure~\cite{jure08ncp}, and so on.
%
%

In parallel, there have been efforts to develop the network formation
mechanisms that naturally generate networks with the observed structural features.
In these network formation mechanisms, there have been two relatively
dichotomous modeling approaches.
Broadly speaking, the theoretical computer science and physics community have
mainly focused on relatively simple ``mechanistic'' but analytically tractable
network models where connectivity patterns observed in the real-world naturally
emerge from the model. The prime example in this line of work is the
Preferential Attachment model with its many
variants~\cite{barabasi99emergence,aiello00random,bollobas03survey,borgs07fitness,cooper03model},
which specifies a simple but very natural edge creation mechanism
that in the limit leads to networks with power-law degree distributions. Other
models of similar flavor include the Copying
Model~\cite{kumar00stochastic}, the Small-world
model~\cite{watts98smallworld,kleinberg00navigation}, Geometric Random
Graphs~\cite{flaxman04geometric}, the Forest Fire model~\cite{jure05dpl}, the
Random surfer model~\cite{blum06surfer}, and models of bipartite affiliation
networks~\cite{lattanzi09affiliaton}.
On the other hand, in statistics, machine learning and traditional social
network analysis, a different approach to modeling network data has emerged.
There the effort is in the development of statistically sound models that
consider the structure of the network as well as the features (\eg~age, gender)
of nodes and edges in the network. 
%
Examples of such models include the Exponential Random
Graphs~\cite{wasserman96pstar}, the Stochastic Block
Model~\cite{airoldi07blockmodel}, and the Latent Space
Model~\cite{hoff02latent}.

\xhdr{``Mechanistic'' and ``Statistical'' models}
Generally, there has been some gap between the above two lines of research. The
``mechanistic'' models are analytically tractable in a sense that one can
mathematically analyze properties of the networks that arise from the models.
These models emphasize the natural emergence of networks that have certain
structural properties found in real-world networks. However, such
models are usually not statistically interesting in a sense that they do not
nicely lend themselves to model parameter estimation and are generally too
simplistic to model heterogeneities between
individual nodes.

On the contrary, ``statistical'' models are generally analytically intractable 
and the network properties do not naturally emerge from the model in general.
However, these models are usually accompanied by statistical procedures for
model parameter estimation and very useful for testing various hypotheses about
the interaction of connectivity patterns and the properties of nodes and edges.

Although models of network structure and formation are seldom {\em both}
analytically tractable and statistically interesting,
an example of a model satisfying both features is the Kronecker graphs
model~\cite{jure05kronecker,weichsel62kronecker}, which is based on the
recursive tensor product of small graph adjacency matrices. The Kronecker graphs
model is analytically tractable in a sense that one can analyze global structural
properties of networks that emerge from the
model~\cite{mahdian07kronecker,jure10kronecker,bodine10kronecker}. In addition,
this model is statistically meaningful because there exists an efficient
parameter estimation technique based on maximum
likelihood~\cite{jure07kronfit,mh11kronem}. It has been empirically shown that with only four
parameters Kronecker graphs quite accurately model the global structural
properties of real-world networks such as degree distributions, edge clustering,
diameter and spectral properties of the graph adjacency matrices.

\xhdr{Modeling networks with rich node attribute information}
Network models investigate edge creation mechanisms, but generally a rich set
of attributes is associated with each node. This is especially true in social
networks, where not only people's connections but also their characteristics,
like age, gender, work place, habits, etc., have been collected. Similarly,
various types of profile information is provided by users in online social
networks. In this sense, both node characteristics and the network structure
need to be considered simultaneously.

The attempt to model the interaction between the network structure and node
attributes raises a wide range of questions. For instance, how do we account
for the heterogeneity in the population of the nodes or how do we combine node
features in an interesting way to obtain probabilities of individual links?
While the earlier work on a general class of latent space models formulated
such questions, most resulting models were either analytically tractable but
statistically uninteresting
or statistically very powerful but do not lend themselves to mathematical
analysis.

To bridge this gap, we propose a class of stochastic network models that we refer to as
Multiplicative Attribute Graphs (MAG). The model naturally captures the
interactions between the network structure and the node attributes in a clean
and tractable manner. We consider a model where each node has a vector of
categorical attributes associated with it. Individual attributes of nodes are
then combined in order to model the emergence of links. The model allows for
rich interaction between node features in a sense that one can simultaneously
model features that reflect homophily (\ie~love of the same) as well as
heterophily (\ie~love of the different).
For example, if people share certain features like hobby, they are more likely
to be friends. However, for some other features like gender, people may be more
likely to form a relationship with someone with the opposite characteristic.
The proposed MAG model is designed to capture both homophily and 
heterophily that naturally occur in social networks.
%

We proceed by formulating the model and show that it is both 
analytically tractable and statistically interesting. In the following sections, we
present our mathematical results. Section~\ref{sec:expedge} examines the number of edges and
shows that our model naturally obeys the Densification Power
Law~\cite{jure05dpl}. Section~\ref{sec:conn} examines the connectivity of 
\modelx, which includes the conditions not only when the network
contains a giant connected component but also when it becomes connected.
Section~\ref{sec:diam} shows that the diameter of the MAG model remains small even though the number
of nodes is large. 
Section~\ref{sec:degree} shows that networks emerging from the MAG model have a log-normal degree distribution.
Furthermore, Section~\ref{sec:power} describes a more general version of the model that can also capture
the power-law degree distribution. We view this as particularly interesting
in the light of a long-standing debate about how to distinguish
the power-law distribution from the log-normal distribution in empirical
data~\cite{mitzenmacher04history,mitzenmacher06powerlaw} and what implications
this would make for real-world networks. Also, our results imply that the \modelx~model is
flexible in a sense that networks with very different properties emerge
depending on the parameter configuration.
Finally, Section~\ref{sec:simul} verifies the properties of the \modelx~by simulation experiments.
The results of the simulations examine how the synthetic network
changes depending the parameters as well as how similar the network looks to
real-world networks.

%% file: W030proposed.tex
In this section, we begin with the introduction of the Multiplicative Attribute Graph (MAG) model.
Then, we formulate the general version of \modelx~and present the simplified version that we analyze throughout this paper.
Finally, we investigate the connection to some related works.


\subsection{General considerations}
\begin{figure}[t]
    \centering
    \includegraphics[width=0.5\textwidth]{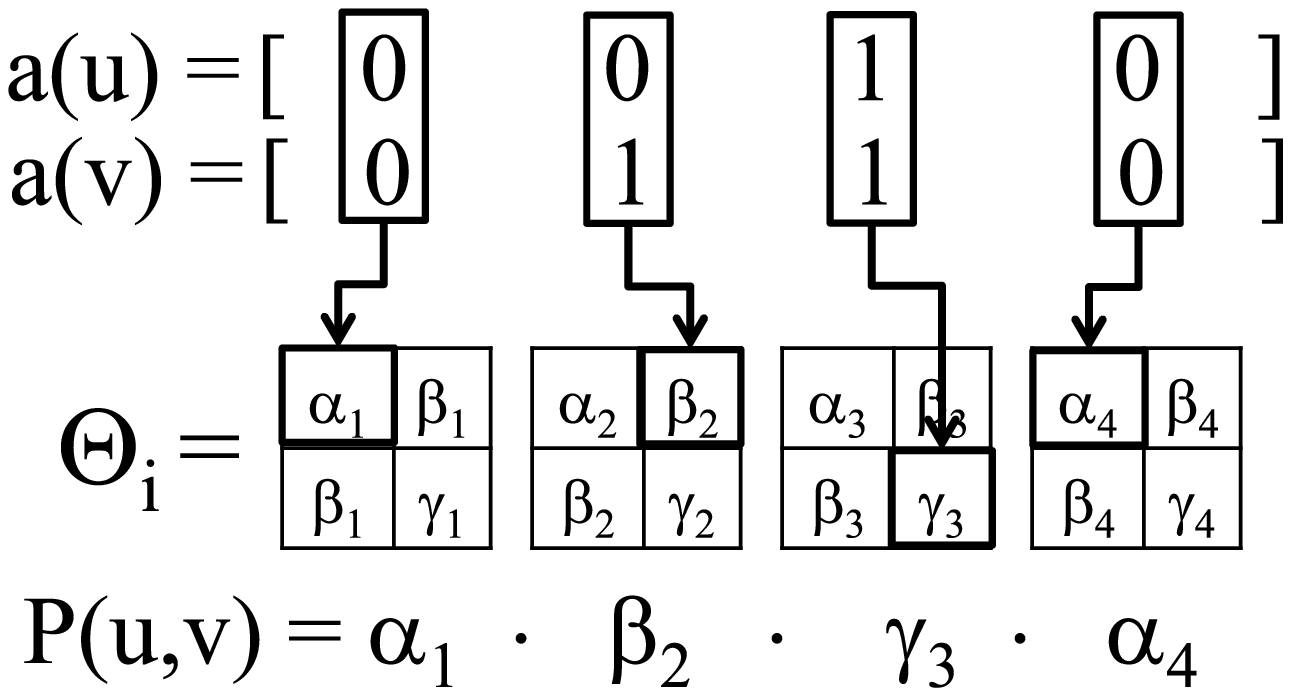}
    \caption{Schematic representation of the Multiplicative Attribute Graphs (MAG) model. Given a pair of nodes $u$ and $v$ with
    the corresponding binary attribute vectors $a(u)$ and $a(v)$, the probability of edge $P[u,v]$
    is the product over the entries of attribute-attribute affinity matrices $\Theta_i$
    where values of $a_i(u)$ and $a_i(v)$ ``select'' the appropriate entries (row/column) of $\Theta_i$.
	Note that this visualized model represents the undirected graph by make each $\Theta_i$ symmetric. However, the \modelx~in general represents the directed graph.}
    \label{fig:model}
\end{figure}

We consider a setting where each node $u$ has a vector $a(u)$ of $l$
categorical (\eg~binary) attributes associated with it.
For simple examples, one can think of such attribute vectors as a sequence of answers to $l$ yes/no
questions such as ``Are you female?'', ``Do you like ice cream?'', and so on.

The other essential ingredient of our model is to specify a mechanism that
generates the probability of an edge between two nodes based on their attribute
vectors. As mentioned before, we aim to be able to account for
the homophily of certain features as well as the heterophily of the
others by our model.
For this mechanism, we associate each attribute $i$ (\ie~$i$-th question)
with an attribute-attribute affinity matrix $\Theta_i$. 
Each entry of matrix $\Theta_i$ represents the affinity depending on the values of the $i$-th attribute 
between a pair of nodes.
More precisely, $\Theta_{i}[z_{1}, z_{2}]$ indicates the affinity between a pair of nodes,
each of which respectively takes value $z_{1}$ and $z_{2}$ for its $i$-th attribute.
For the binary attribute example in Figure~\ref{fig:model}, each $\Theta_i$ is a $2 \times 2$ matrix.
To obtain the affinity corresponding to the $i$-th attribute between node $u$ and $v$, 
the values of $i$-th attribute of both nodes select an appropriate cell of $\Theta_{i}$.
\hide{
Thus, if the attribute reflects homophily, the
corresponding matrix $\Theta_i$ would have large values on the diagonal
(\ie~the edge probability is high when the nodes' answers match),
whereas if the attribute represents heterophily the off-diagonal values of
$\Theta_i$ would be high (\ie~the edge probability is high when nodes
gave different answers to the same question). The top of Figure~\ref{fig:model}
illustrates the concept of node attributes acting as selectors of entries of
matrices $\Theta_i$.
}

\begin{figure}[t]
\centering
\begin{tabular}{cccc}
  \includegraphics[width=0.20\textwidth]{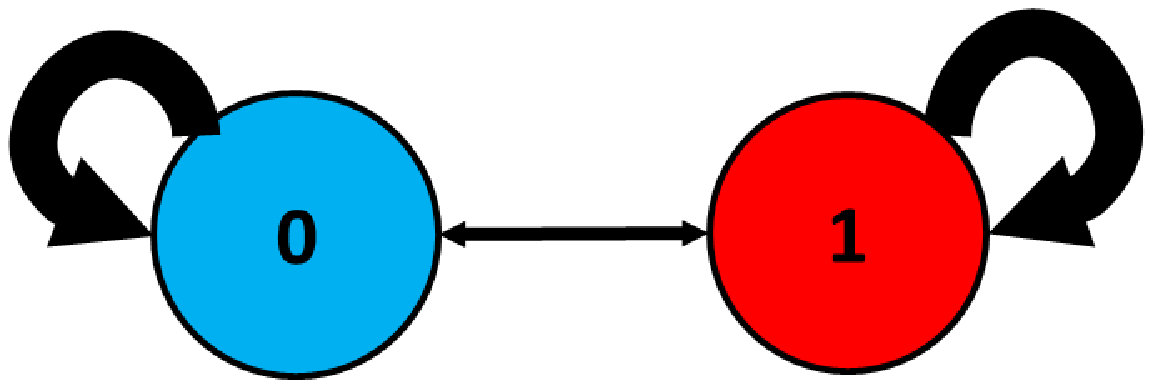} &
  \includegraphics[width=0.18\textwidth]{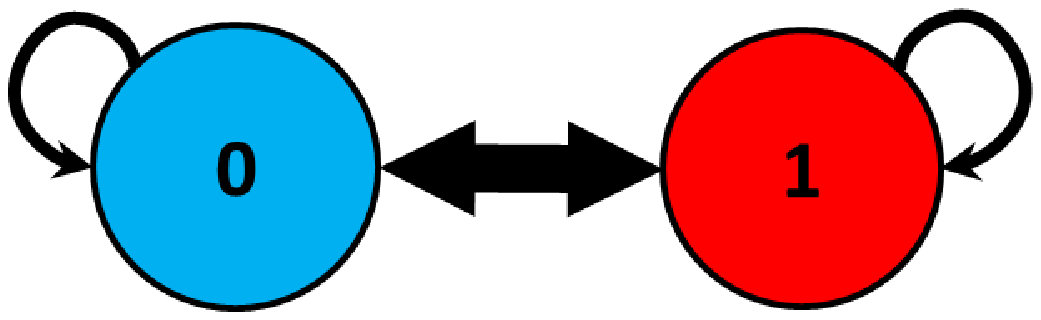} &
  \includegraphics[width=0.18\textwidth]{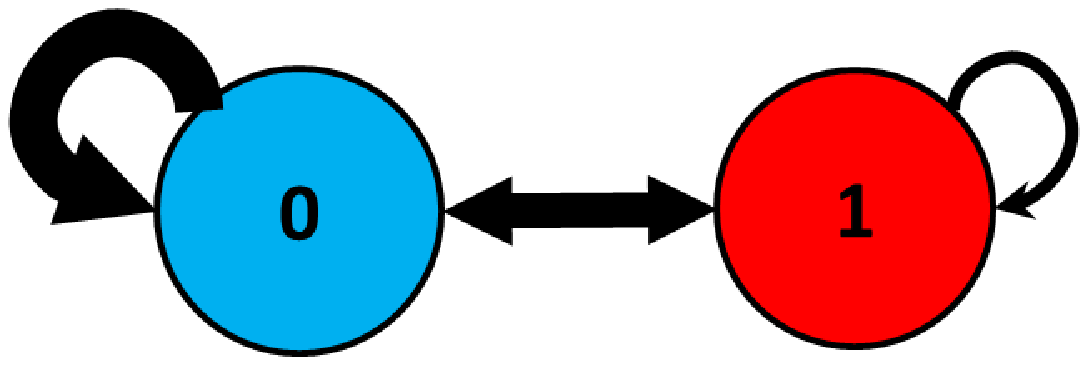} &
  \includegraphics[width=0.18\textwidth]{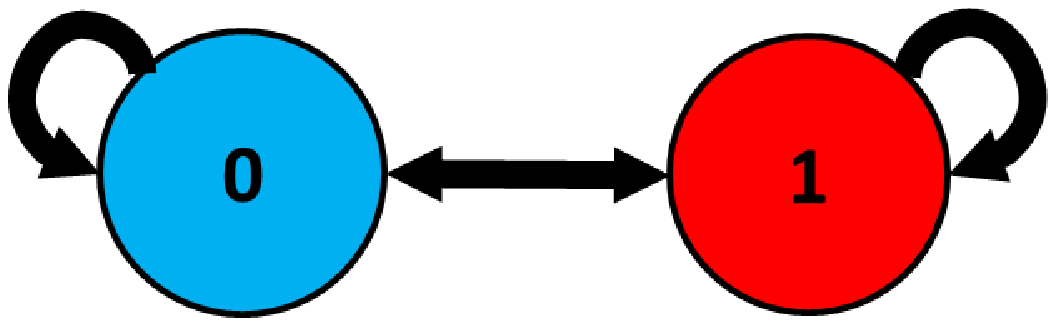} \\
  \includegraphics[width=0.12\textwidth]{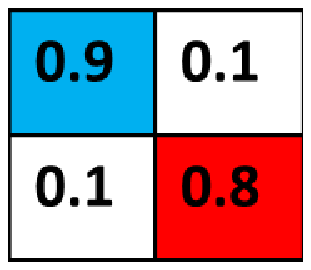} &
  \includegraphics[width=0.12\textwidth]{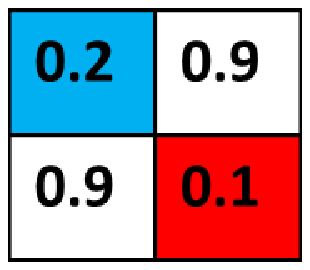} &
  \includegraphics[width=0.12\textwidth]{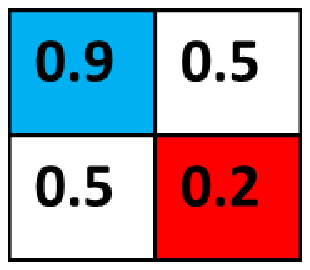} &
  \includegraphics[width=0.12\textwidth]{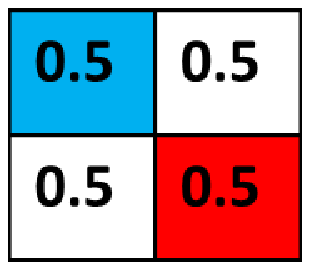} \\
  \small (a) Homophily & \small (b) Heterophily & \small (c) Core-Periphery & \small (d) Random\\
\end{tabular}
  \caption{Structures in which a node attribute can affect link affinity.
  The widths of arrows
  correspond to the affinities towards link formation.}
  \label{fig:structure}
\end{figure}

By these affinity matrices, we can capture the various types of structure in real-world social networks.
For example,
Figure~\ref{fig:structure} shows four possible linking affinities of a binary
attribute.
Top figure of each case visualizes the general structure of networks.
Each circle represents the group which shares the attribute value
and the width of each arrow indicates the affinity of the link formation in the given direction
(\eg~the arrow $0 \rightarrow 1$ indicates the affinity of link formation between a node with "0"-value of a given attribute and a node with "1"-value of that attribute.).
Then, under each figure, we represent the structure in the form of the affinity matrix.

To investigate one by one, 
Figure~\ref{fig:structure}(a) shows the homophily (love of the same)
attribute affinity and the corresponding affinity matrix $\Theta$. Notice large
values on the diagonal entries of $\Theta$, which means that link probability
is high when nodes share the same attribute value. Top of the figure
demonstrates that there will be many links between nodes that have the value of
the attribute set to ``0'' and many links between nodes that have the value
``1'', but there will be few links between nodes where one has value ``0'' and
the other ``1''.
Similarly, Figure~\ref{fig:structure}(b) shows the heterophily (love of the
different) affinity, where nodes that do not share the value of the attribute
are more likely to link, which gives rise to near-bipartite networks. Also,
Figure~\ref{fig:structure}(c) shows the core-periphery affinity, where links are
most likely to form between ``0'' nodes (\ie~members of the core) and least
likely to form between ``1'' nodes (\ie~members of the periphery). Notice that
links between the core and the periphery are more likely than the links between
the nodes of the periphery.
Additionally, Figure~\ref{fig:structure}(d) illustrates the uniformly random structure
that the \GNP~model generates.
By assiging the same value into every entry in each affinity matrix, we can build the \modelx~equivalent to the \GNP model.

From these examples, we notice that the \modelx~nicely provides the flexibility
in the network structure via the affinity matrices.
Although we presented the binary and undirected examples,
the \modelx~basically allows more complicated structure with larger cardinalities (\eg~$3 \times 3$ or $4 \times 4$)
as well as asymmetric structure through asymmetric affinity matrices.

\hide{
However, real-world networks may include various types of structures and thus
different attributes may interact in different ways. For example,
Figure~\ref{fig:structure} shows three possible linking affinities of a binary
attribute. Figure~\ref{fig:structure}(a) shows a homophily (love of the same)
attribute affinity and the corresponding affinity matrix $\Theta$. Notice large
values on the diagonal entries of $\Theta$, which means that link probability
is high when nodes share the same attribute value. Top of the figure
demonstrates that there will be many links between nodes that have the value of
the attribute set to ``0'' and many links between nodes that have the value
``1'', but there will be few links between nodes where one has value ``0'' and
the other ``1''.
Similarly, Figure~\ref{fig:structure}(b) shows a heterophily (love of the
different) affinity, where nodes that do not share the value of the attribute
are more likely to link, which gives rise to near-bipartite networks. Last,
Figure~\ref{fig:structure}(c) shows a core-periphery affinity, where links are
most likely to form between ``0'' nodes (\ie~members of the core) and least
likely to form between ``1'' nodes (\ie~members of the periphery). Notice that
links between the core and the periphery are more likely than the links between
the nodes of the periphery.
}

\subsection{The Multiplicative Attributes Graph (MAG) model}
Now we formulate a general version of the \modelx. To start with, let each
node $u$ have a vector of $l$ categorical attributes and let each attribute
have cardinality $d_i$ for $i = 1, 2, \cdots l$. We also have $l$ matrices,
$\Theta_{i} \in d_i \times d_i$ for $i = 1, 2, \cdots l$.
\hide{each of which is $\Theta_{i} \in d_i \times d_i$. }Each entry of $\Theta_i$ is 
the affinity of a real value between 0 and 1~\footnote{Note that there is no
condition for $\Theta_i$ to be stochastic, we only require each entry of
$\Theta_i$ to be on interval $(0,1)$.}.
Then, the probability of an edge $(u, v)$, $P[u,v]$, is defined as the
multiplication of affinities corresponding to individual attributes, \ie
\begin{equation}
P[u,v]=\prod_{i=1}^{l} \Theta_{i}\left[a_{i}(u), a_{i}(v)\right]
\label{eq:edgeprob}
\end{equation}
where $a_{i}(u)$ denotes the value of $i$-th attribute of node $u$. Note that
edges appear independently with probability determined by node attributes and
matrices $\Theta_i$. Figure~\ref{fig:model} illustrates the model.

One can think of the \modelx~in the following sense. In order to construct a
social network, we ask each node $u$ a series of multiple-choice questions and
the attribute vector $a(u)$ stores the answers fo these questions.
The answers of nodes $u$ and $v$ on a question $i$ 
select an entry of matrix $\Theta_i$, \ie~$u$'s answer selects a row and
$v$'s answer selects a column. One can thus think of matrices $\Theta_i$'s as
the attribute-attribute affinity matrices. Assuming that the questions are
appropriately chosen so that answers are independent of each other, 
the product over the entries of matrices $\Theta_i$ can be regarded as
 the probability of the edge between $u$ and $v$.

The choice of multiplicatively combining entries of $\Theta_i$ is
very natural. In particular, the social network literature defines a concept of Blau-spaces~\cite{McPherson83,mcpherson91} where socio-demographic attributes act as dimensions.
Organizing force in Blau space is homophily as it has been argued that the flow of information between a pair of nodes decreases with the ``distance'' in the corresponding Blau space. In this way, small pockets of nodes appear and lead to the development of social niches for human activity and social organization.
In this respect, multiplication is a natural way to combine node attribute data (\ie~the dimensions of the Blau space) so that even a single attribute can have profound impact on the linking structure (\ie~it creates a narrow social niche community).

The proposed \modelx~model is analytically tractable in a sense that we can
formally analyze the properties of the model. Moreover, the \modelx~is also
statistically interesting as it can account for the heterogeneities in the node population
and can be used to study the interaction between properties of nodes and their linking behavior.
Moreover, one can pose many interesting statistical inference questions:
Given attribute vectors of all nodes and the network
structure, how can we estimate the values of matrices $\Theta_i$? How can we infer the
attributes of unobserved nodes? Or, given a network, how can we estimate both the
node attributes and the matrices $\Theta_i$? \hide{ For example, an expectation
maximization (EM) based framework could be used to estimate the node attributes
as well as the attribute-attribute affinity matrices. }
However, the focus of this paper is in mathematical analysis
and we leave the questions of \modelx~parameter estimation for the future work.



\subsection{Simplified version of the model}
Next we delineate a simplified version of the model that we will
mathematically analyze in the further sections of the paper. First, while the
general \modelx~applies to directed networks, we consider the undirected
version of the model by requiring each $\Theta_{i}$ to be symmetric. Second, we
assume binary node attributes and thus affinity matrices $\Theta_{i}$ have 2 rows
and 2 columns. Third, to further reduce the number of parameters, we also
assume that the affinity matrices for all attributes are the same,
\ie~$\Theta_{i}=\Theta$ for all $i$. These three conditions imply that $\Theta = \left[
\begin{array}{cc} \alpha & \beta \\ \beta & \gamma \end{array} \right]$,
\ie~$\Theta[0, 0] = \alpha, \Theta[0, 1] = \Theta[1, 0] = \beta$, and
$\Theta[1, 1] = \gamma$ for $0 < \alpha, \beta, \gamma < 1$.
Furthermore, all our results will hold for $\alpha > \beta > \gamma$. 
The assumption $\alpha > \beta > \gamma$ is natural 
since most large real-world networks have a common onion-like 
``core-periphery'' structure~\cite{jure08ncp,jure09community,jure10kronecker}.
Figure~\ref{fig:structure}(c) exhibits this structure.
More precisely, the network is composed from denser and denser layers of edges as one moves
towards the core of the network. Basically, $\alpha > \beta > \gamma$ means
that more edges are likely to appear between nodes which share value $0$ on more
attributes and these nodes form the core of the network. Since more edges
appear between pairs of nodes with attribute combination ``0--1'' than between
those with ``1--1'', there are more edges between the core and the periphery
nodes (edges ``0--1'') than between the nodes of the periphery themselves
(edges ``1--1'').


Last, we also assume a simple generative model of node attributes where
each binary attribute vector is generated by $l$ independently and identically
distributed coin flips with bias $\mu$.
%
%
That is, we use an \textit{i.i.d.} Bernoulli distribution parameterized by
$\mu$ to model attribute vectors where the probability that the $i$-th
attribute of a node $u$ takes value 0 is $P\left(a_{i}(u) = 0\right) = \mu$ for $i = 1,
\cdots, l$ and $0 < \mu < 1$.

Putting it all together, the \modelx~\modeldesc~is fully specified by six
parameters: $n$ is the number of nodes, $l$ is the number of attributes of each
node, $\mu$ is the probability that an attribute takes a value of $1$, and
$\Theta = [\alpha~\beta;\beta~\gamma]$ specifies the attribute-attribute
affinity matrix.

We now study the properties of the random graphs that result from the
\modeldesc~where every unordered pair of nodes $(u, v)$ is independently
connected with probability $P[u,v]$ defined in Equation~(\ref{eq:edgeprob}).
Since the probability of an edge exponentially decreases in $l$, the most
interesting case occurs when $l = \rho \log n$ for some constant
$\rho$.\footnote{Throughout the paper, $\log(\cdot)$ indicates
$\log_{2}(\cdot)$ unless explicitly specified as $\ln(\cdot)$.}
This result perfectly agrees that the effective number of dimensions
which can represent online social networks is the order of $\log n$~\cite{bonato10waw}.



\subsection{Connections to other models}
We note that our model belongs to a general class of latent space network
models, where nodes have some discrete or continuous valued
attributes and the probability of linking depends on the values of attribute of
the two nodes. For example, the Latent Space Model~\cite{hoff02latent} assumes
that nodes reside in $d$-dimensional Euclidean space and the probability of an
edge depends on the Euclidean distance between the locations of the nodes.
Similarly, in Random Dot Product Graphs~\cite{young07dotprod}, the linking
probability depends on the inner product between the vectors associated with
node positions.
Furthermore, recently introduced Multifractal Network
Generator~\cite{Palla2010} can also be viewed
as a special case of \modelx~where the node attribute distribution
and the affinity matrix are equal for every attribute.

The \modelx~generalizes the Kronecker graphs model~\cite{jure10kronecker} in a
subtle way. The Kronecker graphs model takes a small (usually $2\times 2$)
initiator matrix $K$ and tensor-powers it $l$ times to obtain a matrix $G$ of
size $2^l \times 2^l$, interpreted as the stochastic graph adjacency
matrix. One can think of a Kronecker graph model as a special case of the
\modelx.

\begin{proposition}
\label{propose:kron} A Kronecker graph $G$ on $2^l$ nodes with a $2\times 2$
initiator matrix $K$ is equivalent to the following \modelgraph~$M$: Let us
number the nodes of $M$ as $0,\cdots,2^l-1$. Let the binary attribute vector of
a node $u$ of $M$ be a binary representation of its node id, and let
$\Theta_i=K$. Then individual edge probabilities $(u,v)$ of nodes in $G$ match
those in $M$, \ie~ $P_G[u,v]=P_M[u,v]$.
\end{proposition}

The above observation is interesting for several reasons. First, all results obtained
for Kronecker graphs naturally apply to a subclass of {\modelgraph}s where the
node's attribute values are the binary representation of its id. This
means that in a Kronecker graph version of the \modelx~each node has a unique
combination of attribute values (\ie~each node has different node id) and all
attribute value combinations are occupied (\ie~node ids range $0, \dots,
2^l-1$).

Second, building on this correspondence between Kronecker and {\modelgraph}s,
we also note that the estimates of the Kronecker initiator matrix $K$ nicely
transfer to matrix $\Theta$ of \modelx. For example, Kronecker initiator matrix
$K=[\alpha = 0.98, \beta = 0.58, \gamma = 0.05]$ accurately models the graph of
the internet connectivity, while the global network structure of the Epinions
online social network is captured by $K=[\alpha = 0.99, \beta = 0.53, \gamma =
0.13]$~\cite{jure07kronfit}. Thus, in the rest of the paper, we will consider
the above values of $K$ as the typical values that the matrix $\Theta$ would
normally take. In this respect, the assumption of $\alpha > \beta > \gamma$
naturally appears.

\hide{
Furthermore, the fact that most large real-world networks satisfy $\alpha >
\beta > \gamma$ tells us that such networks have an onion-like
``core-periphery'' structure~\cite{jure08ncp,jure10kronecker}. In other words,
the network is composed from denser and denser layers of edges as one moves
towards the core of the network. Basically, $\alpha > \beta > \gamma$ means
that more edges are likely to appear between nodes which share $1$'s on more
attributes and these nodes form the core of the network. Since more edges
appear between pairs of nodes with attribute combination ``1--0'' than between
those with ``0--0'', there are more edges between the core and the periphery
nodes (edges ``1--0'') than between the nodes of the periphery themselves
(edges ``0--0'').
}

In following sections, we analyze the properties of the \modelx.  We focus
mostly on the simplified version. Each section states the main theorem and
gives the overview of the proof.
We omit the full proofs in the main body of the paper and describe them in the
Appendix.

%% file: W040proofs.tex
\section{The Number of Edges}
\label{sec:expedge}
%


In this section, we derive the expression for the expected number of edges in \modelx.
Moreover, this formula can valdiate not only the assumption, $l = \rho \log n$,
but also a substantial social network property, namely the Densification Power Law.\hide{ with regard to graph density.}

\begin{theorem}
\label{thm:expedge}
For a \modelgraph~\modeldesc, the number of edges, $m$, satisfies
\[
  \EX{m} = \frac{n(n-1)}{2} \left(\mu^{2}\alpha + 2 \mu(1-\mu)\beta + (1-\mu)^{2} \gamma\right)^{l} + n \left(\mu \alpha + (1-\mu) \gamma\right)^{l} \, .
\]
\end{theorem}

The expression is divided into two diffrent terms.
The first term indicates the number of edges between distinct nodes, whereas the second term means the number of self-edges.
If we exclude self-edges, the number of edges would be therefore reduced to the first term.

Before the actual analysis, we define some useful notations that will be used throughout this paper. First, let $V$ be the set of nodes in the \modelgraph~\modeldesc. We refer to the \textit{weight} of a node $u$ as the number of $0$'s in its attribute vectors, and denote it as $|u|$
, \ie
$
|u| = \sum_{i = 1}^{l} \mathbf{1}\left\{ a_{i}(u) = 0 \right\}
$
where $\mathbf{1}\left\{\cdot\right\}$ is an indicator function.
We additionally define $W_j$ as a set which consists of all nodes with the same weight $j$, \ie~$W_{j} = \left\{u \in V : |u| = j\right\}$ for $j = 0, 1, \cdots, l$. Similarly, $S_j$ denotes the set of nodes with weight which is greater than or equal to $j$, \ie~$S_{j} = \left\{u \in V : |u| \geq j\right\}$. By definition, $S_{j} = \cup_{i = j}^{l} W_{i}$.

To complete the proof of Theorem~\ref{thm:expedge},
using the definition of the simplified \modelx,
we can derive the main lemmas as follows:
\begin{lemma}
\label{lemma:expprob}
For \textbf{distinct} $u, v \in V$,
$
\ECP{u, v}{u \in W_{i}} = \bkron{i}
$ \, .
\end{lemma}
%
\begin{lemma}
\label{lemma:expdeg}
For $u \in V$,
$
\EX{deg(u) | u \in W_{i}} = (n-1)\bkron{i} + 2\alpha^{i}\gamma^{l-i}
$ \, .
\end{lemma}

By using these lemmas, the outline of the proof for Theorem~\ref{thm:expedge} is as follows.
Since the number of edges is half of the degree sum, all we need to do is to sum $\EX{deg(u)}$ over the degree distribution.
However, because $\EX{deg(u)} = \EX{deg(v)}$ if the weights of $u$ and $v$ are the same,
we can add up $\EX{deg(u) | u \in W_{i}}$ over the \textit{weight} distribution, \ie~binomial distribution $Bin(l, \mu)$.

On the other hand, more significantly, Theorem~\ref{thm:expedge} can result in two substantial features of \modelx.
First, the assumption that $l = \rho \log n$ for a constant $\rho$
can be validated by the next two corollaries.
\begin{corollary}
\label{cor:landn}
$m \in o(n)$ \WHP\footnotemark~as $\LGN$,
if~ $\frac{l}{\log n} > -\frac{1}{\log \left(\mu^{2}\alpha + 2\mu(1-\mu)\beta + (1-\mu)^{2} \gamma\right)}$\,.
\footnotetext{It indicates the probability $1-o(1)$.}
\end{corollary}
\begin{corollary}
\label{cor:landn2}
$m \in \Theta(n^{2 - o(1)})$ \WHP~as $\LGN$,
if~ $l \in o(\log n)$.
\end{corollary}





Note that $\log \left(\mu^{2}\alpha + 2\mu(1-\mu)\beta + (1-\mu)^{2} \gamma\right) < 0$ because both $\mu$ and $\gamma$ are less than $1$.
Thus, in order for \modeldesc~to have a proper number of edges~(\eg~more than $n$), $l$ should be bounded by the order of $\log n$.
On the contrary, since most social networks are sparse, $l \in o(\log n)$ case can be also reasonably excluded.
In consequence, both Corollary~\ref{cor:landn} and Corollary~\ref{cor:landn2} provide the upper and lower bounds of $l$ for social networks.
These bounds eventually support the assumption of $l = \rho \log n$.

Although we do not technically define any process of \modelgraph~evolution,
we can interpret it in the folllowing way.
When a new node joins the network, its behavior is governed by the node attribute
distribution which is seemingly independent of the network structure. However, in a long
term, since the number of attributes grows slowly as the number of nodes
increases, the node attributes and the network structure are not independent.
This phenomenon is somewhat aligned with the real world.
When a new person enters the network, he or she
seems to act independently of other people, but people eventually constitue a structured
network in the large scale and their behaviors can be categorized into more
classes as the network evolves.

Second, under this assumption, the expected number of edges can be approximately restated as~\\
\[
	\frac{1}{2} n^{2 + \rho \log \left(\mu^{2}\alpha + 2 \mu(1-\mu)\beta + (1-\mu)^{2} \gamma\right)} \, .
\]
%
We find that this fact agrees with the Densification Power Law~\cite{jure05dpl},
one of the properties of social networks,
which indicates $m(t) \propto n(t)^{a}$ for $a > 1$.
For example, an instance of \modelx~with $\rho = 1, \mu = 0.5$ (Proposition~\ref{propose:kron}), would have the densification exponent $a = \log( |\Theta| )$ where $|\Theta|$ denotes the sum of all entries in $\Theta$.

The proofs are fully described in Appendix.


\section{Connectivity}
\label{sec:conn}

In the previous section, we observed that \modelx~obeys the Densification Power Law.
In this section, we mathematically investigate \modelx~for another general property of social networks,
the existence of a giant connected component.
Furthermore, we also examine the situation
where this giant component covers the entire network,
\ie~the network is connected.

%
We begin with the theorems
that \modelgraph~has a giant component
and further becomes connected.
%
%
\begin{theorem}
\label{thm:giant}
\emph{(Giant Component)}
Only one connected component of size $\Theta(n)$ exists in \modeldesc~\WHP~as $\LGN$ , if and only if
\begin{equation*}
\crilcc \geq \HALF \,.
\end{equation*}
\end{theorem}
\begin{theorem}
\label{thm:conn}
\emph{(Connectedness)}
Let the connectedness criterion function of \modeldesc~be
\[
F_{\mathrm{c}} (M) =
\left\{
	\begin{array}{l l}
	\crisimconn & \quad \text{when $(1 - \mu)^{\rho} \geq \HALF$} \\
	\cricpxconn{\nu} & \quad \text{otherwise}\\
	\end{array} \right. 
\]
where $\nu$ is a solution of 
$\left[\RATIO{\nu}\right]^{\rho} = \HALF$ in $(0, \mu)$.\\
Then, \modeldesc~is connected \WHP~as $\LGN$, if $F_{\mathrm{c}}(M) > \HALF$.
In contrast, \modeldesc~is disconnected \WHP~as $\LGN$,
if $F_{\mathrm{c}} (M) < \HALF$.
\end{theorem}

To show the above theorems, we first define the monotonicity property of \modelx.
\begin{theorem}
\label{thm:mono}
\emph{(Monotonicity)}
For $u, v \in V$,
$
P\left[u, v | |u| = i\right] \leq P\left[u, v | |u| = j\right]
$
if $i \leq j$.
\end{theorem}

Theorem~\ref{thm:mono} ultimately demonstrates that a node of larger weight is more likely to be connected with other nodes.
In other words, a node of large weight plays a "core" role in the network, whereas the node of small weight is regarded as "periphery".
This feature of the \modelx~has direct effects on the connectedness as well as on the existence of a giant component.
%
%

By the monotonicty property, the minimum degree is likely to be the degree of the minimum weight node.
Therefore, the disconnectedness could be proved by showing
that the expected degree of the minimum weight node is too small to be connected with any other node.
Conversely, if this lowest degree is large enough, say $\Omega(\log n)$,
then any subset of nodes would be connected with the other part of the graph.
Thus,
to show the connectedness,
the degree of the minimum weight node should be necessarily inspected,
using Lemma~\ref{lemma:expdeg}.
%

Note that the criterion in Theorem~\ref{thm:conn} is separated into two cases depending on $\mu$,
which tells whether or not the expected number of weight $0$ nodes, $\EX{|W_{0}|}$, is greater than $1$, because $|W_{j}|$ is a binomial random variable.
If this expectation is larger than $1$, then the minimum weight is likely to be close to $0$, \ie~$O(1)$.
Otherwise, if $\EX{|W_{0}|} < 1$,
the equation of $\nu$ describes the ratio of the minimum weight to $l$ as $\LGN$.
%
Therefore,
the condition for connectedness actually depends on the minimum weight node.
In fact, the proof of Theorem~\ref{thm:conn} is accomplished by computing the expected degree of this minimum weight node
and using some techniques introduced in~\cite{mahdian07kronecker}.

Similar explanation works for the existence of a giant component.
Instead of the minimum weight node,
Theorem~\ref{thm:giant} shows that the existence of $\Theta(n)$ component relies on the degree of the \textit{median} weight node.
We intuitively understand this in the following way.
We might throw away the lower half of nodes by degree.
If the degree of the median weight node is large enough,
then the half of the network is likely to be connected.
The connectedness of this half network implies the existence of $\Theta(n)$ component, the size of which is at least $\frac{n}{2}$.
In the proof, we actually examine the degrees of nodes
of three different weights: $\mu l$, $\mu l + l^{1/6}$, and $\mu l + l^{2/3}$.
The existence of $\Theta(n)$ component is determined by the degrees of these nodes.

However, the existence of $\Theta(n)$ component does not necessarily indicate
that it is a unique giant component, since there might be another
$\Theta(n)$ component.
Therefore, to prove Theorem~\ref{thm:giant} more strictly, the uniqueness of $\Theta(n)$ component has to follow the existence of it.
We can prove the uniqueness
by showing that if there are two connected subgraphs of size $\Theta(n)$ then they are connected each other with high probability.

The proofs of those three theorems are in Appendix.


\section{Diameter}
\label{sec:diam}

Another property of social networks is that the diameter of the network remains small although the number of nodes grows large.
We can show this property in \modelx~by applying the similar idea as in~\cite{mahdian07kronecker}.
\begin{theorem}
\label{thm:diam}
If~$\crisimconn > \HALF$, then \modeldesc~has a constant diameter \WHP~as $\LGN$.
\end{theorem}

This theorem does not specify the exact diameter,
but, under the given condition,
it guarantees the bounded diameter even though $\LGN$ by using the following lemmas:
\hide{
This theorem does not specify the exact value of diameter,
but, at least under the given condition,
it guarantees the bounded diameter even though $\LGN$.
The proof is based on the following theorem.
\begin{theorem}
\label{thm:randomdiam}
\cite{bollobas90diameter,klee81diameter}
For an \GNP~$G(n, p)$, if $(pn)^{d-1}/n \rightarrow 0$ and $(pn)^{d}/n \rightarrow \infty$ for a fixed integer $d$, then $G(n, p)$ has diameter $d$ with probability approaching 1 as $\LGN$.
\end{theorem}

Theorem~\ref{thm:randomdiam} describes only \GNP.
However, if we can assure that the edge probability between any pair of nodes in a \modelgraph~is greater than $p$, the diameter of that graph would be at most that of $G(n, p)$~\cite{mahdian07kronecker}.
Using this feature combined with Theorem~\ref{thm:randomdiam}, we can derive main lemmas as follow:
%
}
\begin{lemma}
\label{lemma:upperdiam}
If $\crisimconn > \HALF$,
for $\lambda = \frac{\mu \beta}{\swokron}$,
$S_{\lambda l}$ has a constant diameter \WHP~as $\LGN$.
\end{lemma}
\begin{lemma}
\label{lemma:directedge}
If $\crisimconn > \HALF$,
for $\lambda = \frac{\mu \beta}{\swokron}$,
all nodes in $V \xset S_{\lambda l}$ are
directly connected to $S_{\lambda l}$ \WHP~as $\LGN$.
\end{lemma}

By Lemma~\ref{lemma:directedge}, we can conclude that the diameter of the entire graph is limited to $(2 + $ diameter of $S_{\lambda l})$. Since by Lemma~\ref{lemma:upperdiam} the diameter of $S_{\lambda l}$ is constant with high probability under the given condition, the actual diameter is also constant.

The proofs are represented in Appendix.

\newcommand{\XYA}[1]{x^{#1}y^{l-{#1}}}
\newcommand{\XYL}[1]{x^{{#1}l}y^{(1-{#1})l}}
\newcommand{\XOY}{\left(\frac{x}{y}\right)}
\newcommand{\JS}{j^{*}}
\newcommand{\LNORM}{\mathrm{Log}-\mathcal{N}}

\section{Degree Distribution}
\label{sec:degree}

In this section, we analyze the degree distribution of the simplified \modelx~under some reasonable assumptions.\footnotemark
\footnotetext{We trivially exclude self-edges not only because computations become simple but also because other models usually do not include them.}
Depending on $\Theta$, \modelx~produces graphs of various degree distributions.
For instance, since the network becomes a sparse \GNP~if $\alpha \approx \beta \approx \gamma < 1$,
the degree distribution will approximately follow the binomial distribution.
For another extreme example, in case of $\alpha \approx 1$ and $\mu \approx 1$, the network will be close to a complete graph, which represents a degree distribution different from a sparse \GNP.
For this reason, we need to narrow down the conditions on $\mu$ and $\Theta$ as follows.
If $\mu$ is close to $0$ or $1$, then the graph becomes an \GNP~with edge probability $p = \alpha$ (when $\mu \approx 1$) or $\gamma$ (when $\mu \approx 0$).
Since the degree distribution of the \GNP~is binomial, we will exclude these extreme cases of $\mu$.
On the other hand, with regard to $\Theta$,
we assume that a reasonable configuration space for $\Theta$ would be where $\frac{\lwokron}{\swokron}$ is between $1.6$ and $3$.
For the previous \Kron~example, this ratio is actually about $2.44$.
Our approach for the condition on $\Theta$ can be also supported by real examples in~\cite{jure07kronfit}.
This condition is crucial for us, since in the analysis we use that $\left(\frac{\lwokron}{\swokron}\right)^{x}$ grows faster than the polynomial function of $x$. If $\frac{\lwokron}{\swokron}$ is close to 1, we cannot make use of this fact.
Assuming all these conditions on $\mu$ and $\Theta$, we result in the following theorem about the degree distribution.
\begin{theorem}
\label{thm:degdist}
In \modeldesc that follows above assumptions, 
if \[\crilcc > \HALF \, ,\]
then the tail of degree distribution, $p_{k}$, follows a log-normal, specifically,
\[
\mathcal{N}\left(\ln \left(n(\swokron)^{l}\right) + l \mu \ln R + \frac{l \mu (1-\mu) (\ln R)^{2}}{2}
, ~~l \mu (1-\mu) (\ln R)^{2} \right) \, ,
\]
for $R = \frac{\lwokron}{\swokron}$~as $\LGN$.
\end{theorem}

In other words, the degree distribution of \modelx~approximately follows a quadratic relationship on log-log scale.
This result is nice since some social networks follow the log-normal distribution.
For instance, the degree distribution of \textit{LiveJournal} network looks more parabolic than linear on log-log scale~\cite{lj05}.




In brief, as the expected degree is an exponential function of the node weight by Lemma~\ref{lemma:expdeg},
the degree distribution is mainly affected by the distribution of node weights.
Since the node weight follows a binomial distribution, 
it can be approximated to a normal distribution for sufficiently large $l$.
Because the logarithmic value of the expected degree is linear in the node weight
and this weight follows a binomial distribution,
the log value of degree approximately follows a normal distribution for large $l$.
This eventually indicates that the degree distribution roughly follows a log-normal.

Note that there exists a condition, $\crilcc > \HALF$, which is related to the existence of a giant component.
First, this condition is perfectly acceptable because real-world networks have a giant component.
Second, as we described in Section~\ref{sec:conn},
this condition ensures that the median degree is large enough.
Equivalently, it also indicates that the degrees of a half of the nodes are large enough.
If we refer to the tail of degree distribution as the degrees of nodes with degrees above the median degree,
then we can show Theorem~\ref{thm:degdist}.

\hide{
To analyze it more rigorously, we need the following theorem and its corollary.
\begin{theorem}
\label{thm:young}
\cite{young07dotprod}
$
P\left( deg(u) = k \right) = \int_{u \in V} {{n-1} \choose k} \left( \EP{u, v} \right)^{k} \left(1-\EP{u, v}\right)^{n-1-k} du
$ \, .
\end{theorem}
\begin{corollary}
\label{cor:degdist}
For $E_{j} = \lkron^{j}\skron^{l-j}$,\\
the probability of degree $k$ in \modeldesc~is~
$
p_{k} = \sum_{j = 0}^{l} \EN{j}  {{n-1} \choose k} E_{j}^{k} \left( 1 - E_{j} \right)^{n - 1 - k}
$ \, .
\end{corollary}

Corollary~\ref{cor:degdist} can be obtained by combining Lemma~\ref{lemma:expprob} with Theorem~\ref{thm:young},
but it seems difficult to find the exact closed form expression.
Therefore, we need some approximations such as Stirling approximation and normal approximation to prove Theorem~\ref{thm:degdist}.
In more detail, we seek the dominant term in Corollary~\ref{cor:degdist}
, and then show that this term governs $p_{k}$
and the log value of it is approximately a quadratic function of $\ln k$.
The quadratic function of $\ln k$ eventually represents the log-normality.
}

The full proofs for this analysis are described in Appendix.




\newcommand{\OP}[1]{p_{({#1})}}
\section{Extensions:~Power-Law Degree Distribution}
\label{sec:power}

So far we have handled the simplified version of \modelx~parameterized by only few variables.
Even with these few parameters, many well-known properties of social networks can be reproduced.
However, regarding to the degree distribution, even though the log-normal is one of the distributions that social networks commonly follow, a lot of social networks also follow the power-law degree distribution~\cite{faloutsos99powerlaw}.

In this section, we show that the \modelx~produces networks with the power-law degree distribution by releasing some constraints.
We do not attempt to analyze it in a rigorous manner, but give the intuition by suggesting an example of configuration.
%
We still hold the condition that every attribute is binary and independently sampled from Bernoulli distribution.
However, in contrast to the simplified version, 
we allow each attribute to have a different Bernoulli parameter as well as a different attribute-attribute affinity matrix associated wit it.
The formal definition of this model is as follows:
\[
P\left( a_{j}(u) = 0 \right) = \mu_{j}
, ~P[u, v] = \prod_{j = 1}^{l} \Theta_{j} \left[a_{j}(u), a_{j}(v) \right] \, .
\]
The number of parameters here is $4l$, which consist of $\mu_{j}$'s and $\Theta_{j}$'s for $j = 1, 2, \cdots, l$.
For convenience, we denote this power-law version of \modelx~as
\modelpower~where $\vec{\mu} = \{\mu_{1}, \cdots, \mu_{l} \}$
and $ \vec{\Theta} = \{\Theta_{1}, \cdots, \Theta_{l} \}$.
With these additional parameters, we are able to obtain the power law degree distribution as the following theorem describes.
\begin{theorem}
\label{thm:power}
For \modelpower,
if $\frac{\mu_j}{1-\mu_{j}} = \left(\frac{\mu_{j} \alpha_{j} + (1-\mu_{j})\beta_{j}}{\mu_{j} \beta_{j} + (1-\mu_{j})\gamma_{j}} \right)^{-\delta}$ for $\delta > 0$,
then \hide{there exists some configurations of $\vec{\mu}$ and $\vec{\Theta}$
that }the degree distribution satisfies $p_{k} \propto k^{-\delta-\HALF}$
as $\LGN$.
\end{theorem}
In order to investigate the degree distribution of this model, the following two lemmas are essential.
\begin{lemma}
\label{lemma:generalprob}
The probability that a node $u$ in \modelpower~has an attribute vector $a(u)$ is
\[
\prod_{i=1}^{l} (\mu_{i})^{\mathbf{1}\left\{a_{i}(u) = 0\right\}} (1 - \mu_{i})^{\mathbf{1}\left\{a_{i}(u) = 1\right\}} \, .
\]
\end{lemma}
\begin{lemma}
\label{lemma:generaldeg}
The expected degree of node $u$ in \modelpower~is
\[
\left(n-1\right) \prod_{i=1}^{l} \left(\mu_{i} \alpha_{i} + (1-\mu_{i})\beta_{i}\right)^{\mathbf{1}\left\{a_{i}(u) = 0\right\}}
\left(\mu_{i} \beta_{i} + \left(1-\mu_{i}\right) \gamma_{i} \right)^{\mathbf{1}\left\{a_{i}(u) = 1\right\}} \, .
\]
\end{lemma}
%
By Lemmas~\ref{lemma:generalprob} and~\ref{lemma:generaldeg},
if the condition in Theorem~\ref{thm:power} holds,
the probability that a node has the same attribute vector as node $u$
is proportional to $(-\delta)$-th power of the expected degree of $u$.
In addition, $(-\HALF)$-th power comes from the Stirling approximation for large $k$.
This roughly explains Theorem~\ref{thm:power}.

\hide{
To prove this theorem in detail, we can apply similar methods to Section~\ref{sec:degree}.
First, we compute the exact expression of $p_{k}$ based on Theorem~\ref{thm:young}.
Next, we approximate $p_{k}$ to the dominant term by using some algebra.
Finally, we obtain the expression of this dominant term,
which is roughly proportional to $k^{-\delta-\HALF}$.
}

The proof is given in Appendix and the result is also verified by simulation in Figure~\ref{fig:power}.

%% file: W050simulation.tex
\newcommand{\social}{Yahoo!-Flickr}
\section{Simulation}
\label{sec:simul}

In the previous sections, we performed theoretical analysis of the \modelx.
%
In this section, we use simulation experiments to further demonstrate the properties of networks that arise from the \modelx.
First, we generated synthetic \modelgraph s with varying parameter values
to explore how the network properties change as a function of those parameters.
We focus on the change of scalar network properties, like diameter and the fraction of nodes in the largest connected component of the graph,
as a function of the model parameter values.
%
Second, we also ran simulations with fixed parameter configurations to check other properties of \modelx~that we did not theoretically analyze.
In this way, we are able to qualitatively compare our model to a real-world network.

\subsection{MAG model parameter space}
%
%
Here we focus on the simplified version of the MAG model and examine how various network properties vary as a function of parameter settings. We fix all but one parameter and vary the remaining parameter.
We vary $\mu, \alpha, f$, and $n$ in \modeldesc, where $\alpha$ is the first entry of the affinity matrix $\Theta = [\alpha~\beta;\beta~\gamma]$ and $f$ indicates a scalar factor of $\Theta$,
\ie~$\Theta = f \cdot \Theta_0$ for a constant $\Theta_0 = [\alpha_{0}~\beta_{0};\beta_{0}~\gamma_{0}]$.

\begin{figure}[ttp]
\centering
\begin{tabular}{cccc}
\includegraphics[angle=90,width=0.04\textwidth]{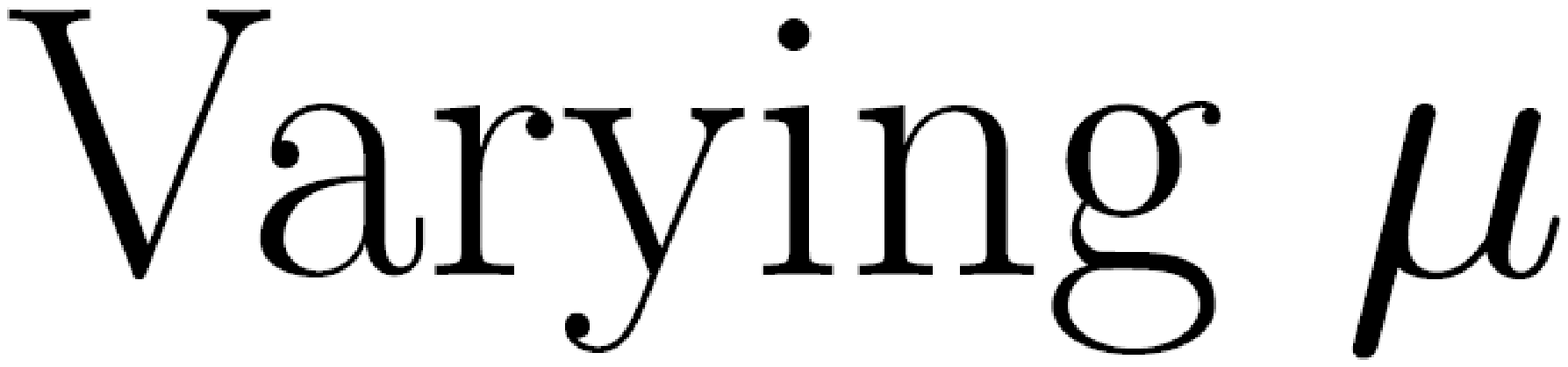}&
\includegraphics[width=0.28\textwidth]{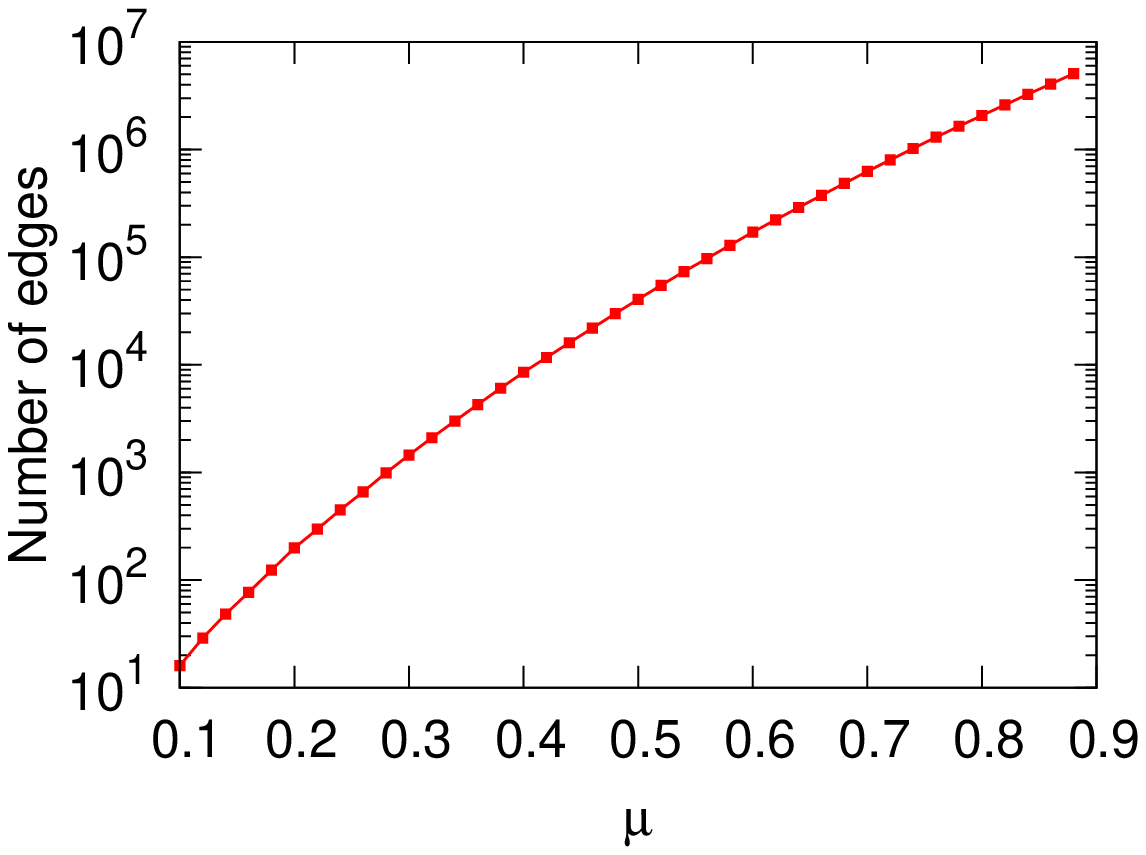} &
\includegraphics[width=0.28\textwidth]{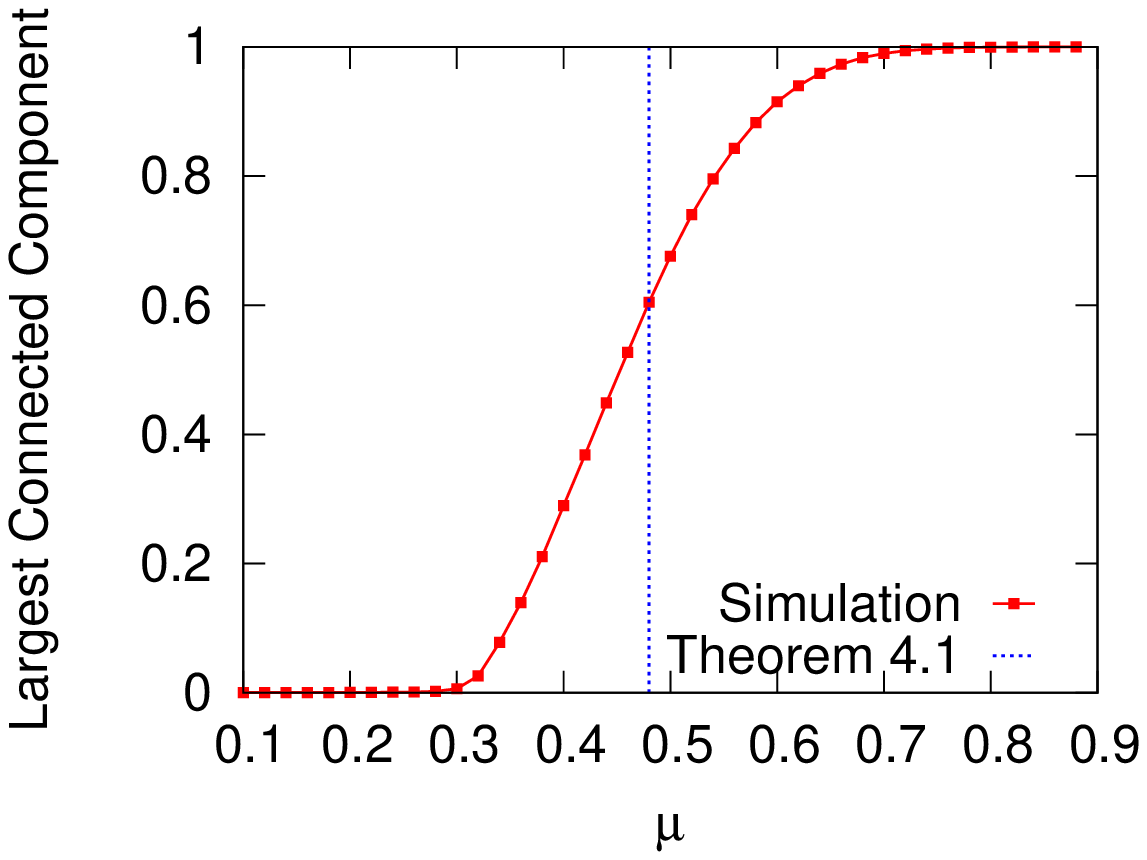} &
\includegraphics[width=0.28\textwidth]{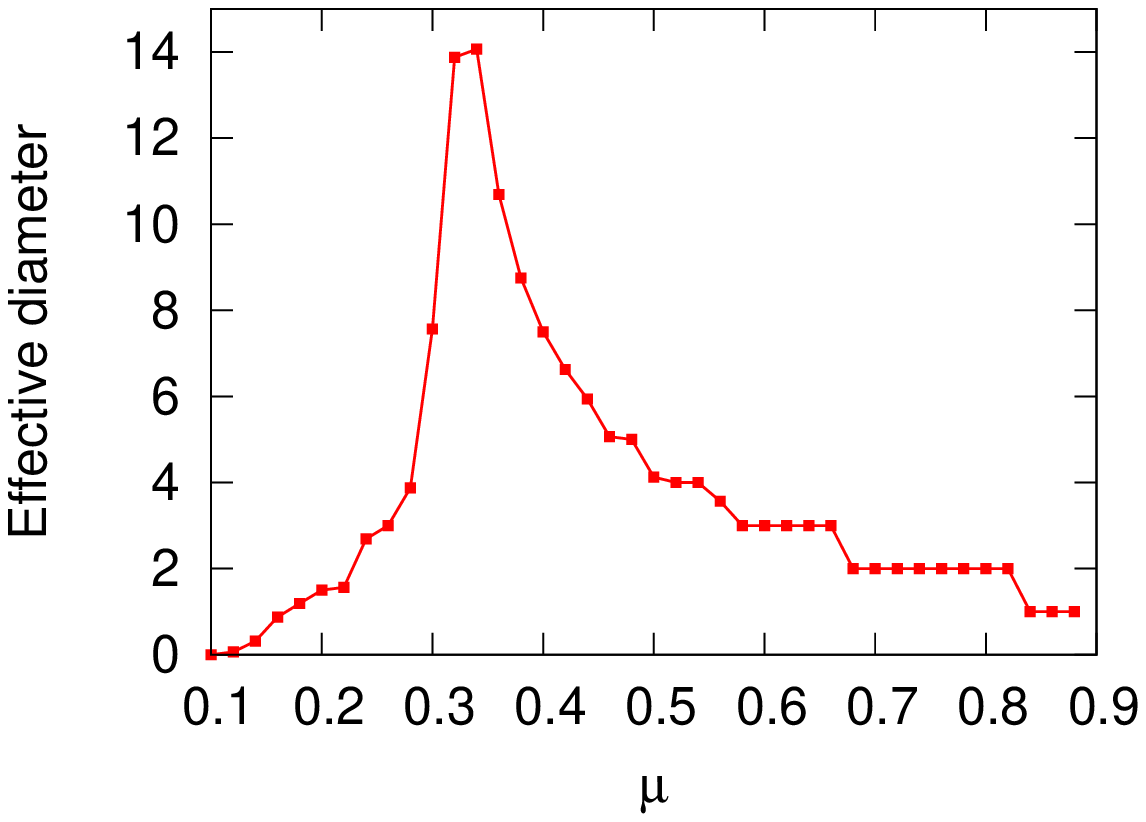} \\
\includegraphics[angle=90,width=0.04\textwidth]{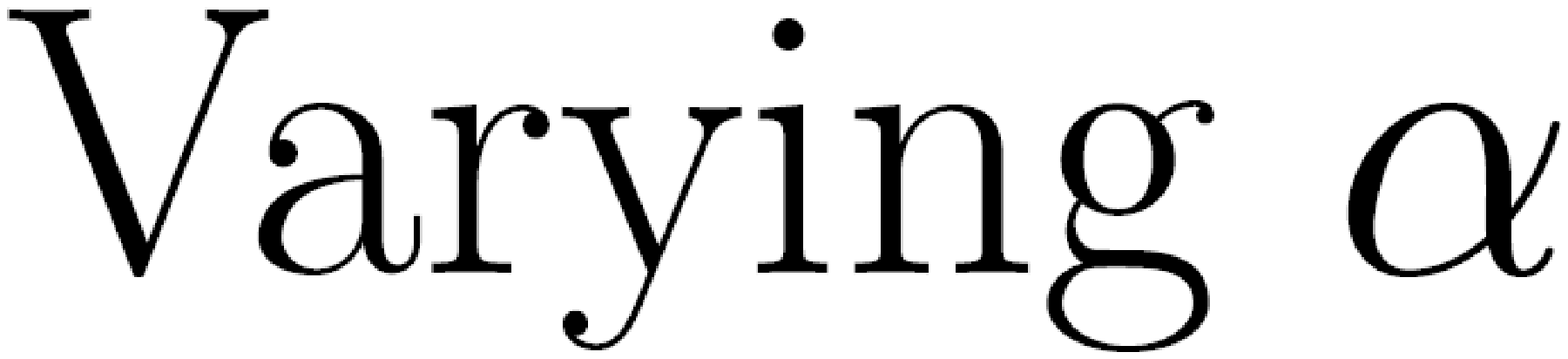}&
\includegraphics[width=0.28\textwidth]{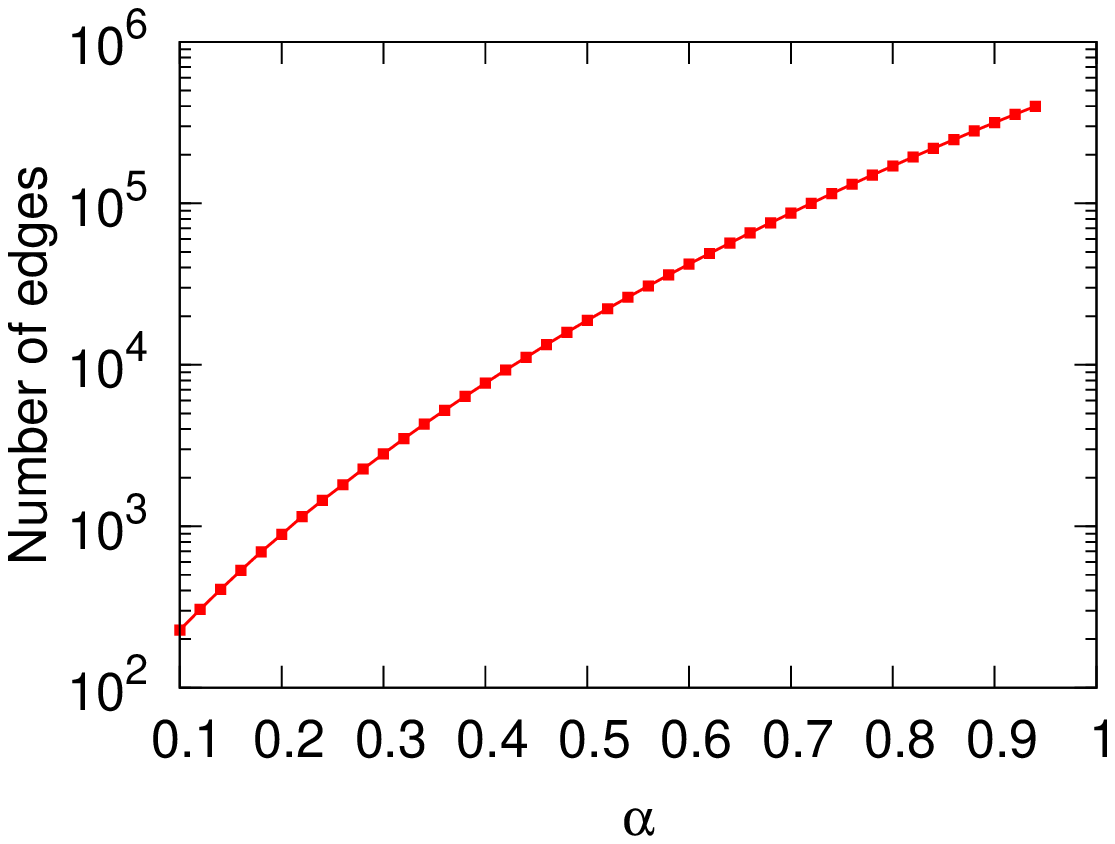} &
\includegraphics[width=0.28\textwidth]{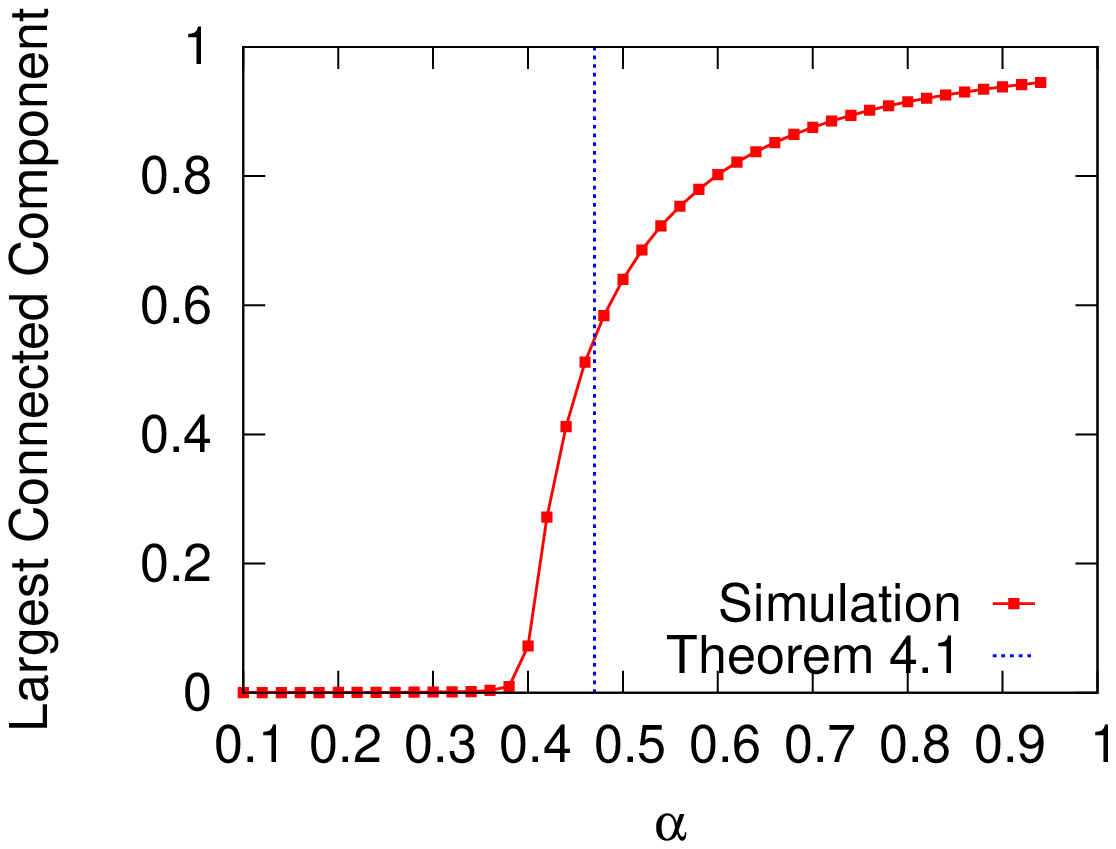} &
\includegraphics[width=0.28\textwidth]{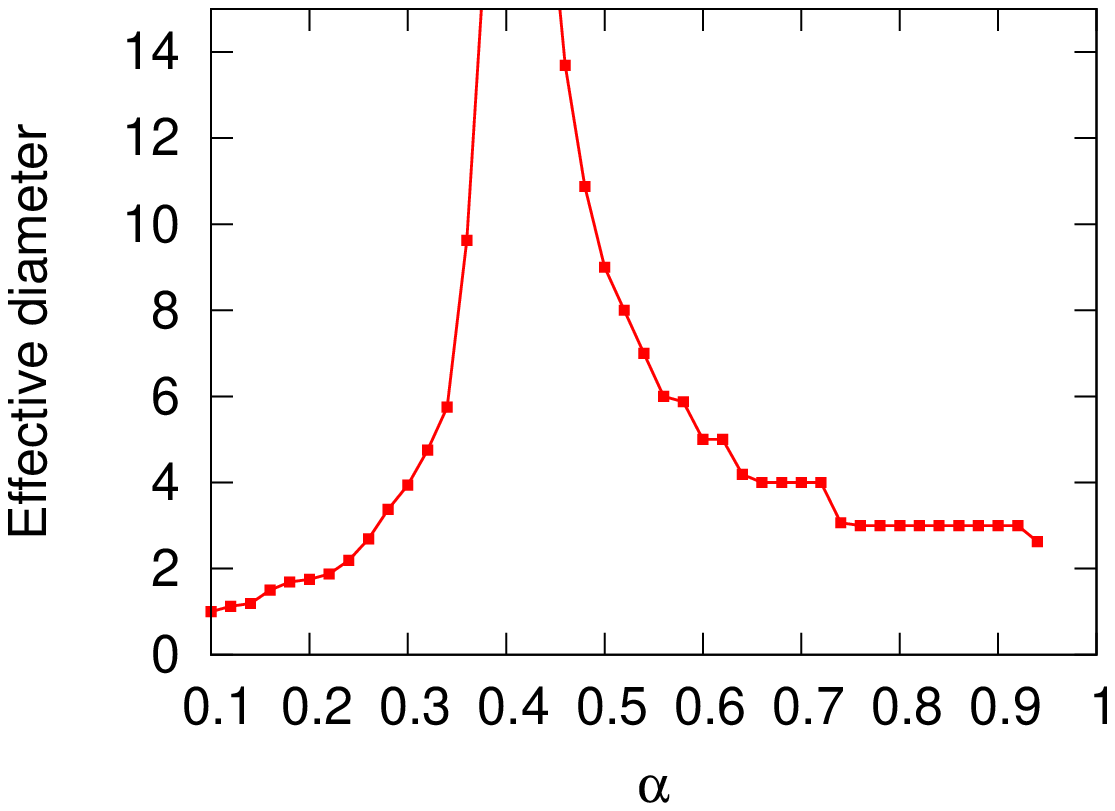} \\
\includegraphics[angle=90,width=0.04\textwidth]{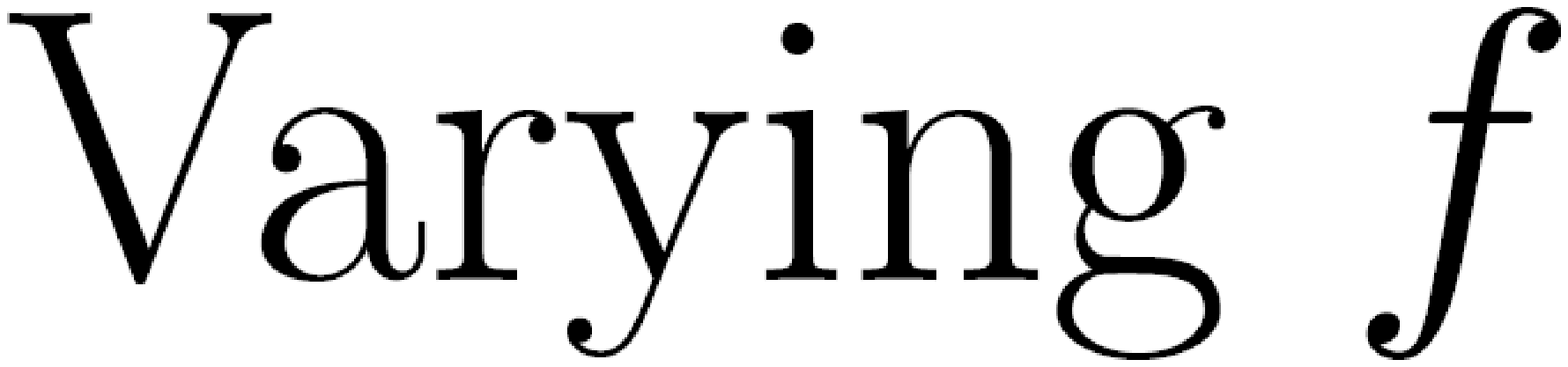}&
\includegraphics[width=0.28\textwidth]{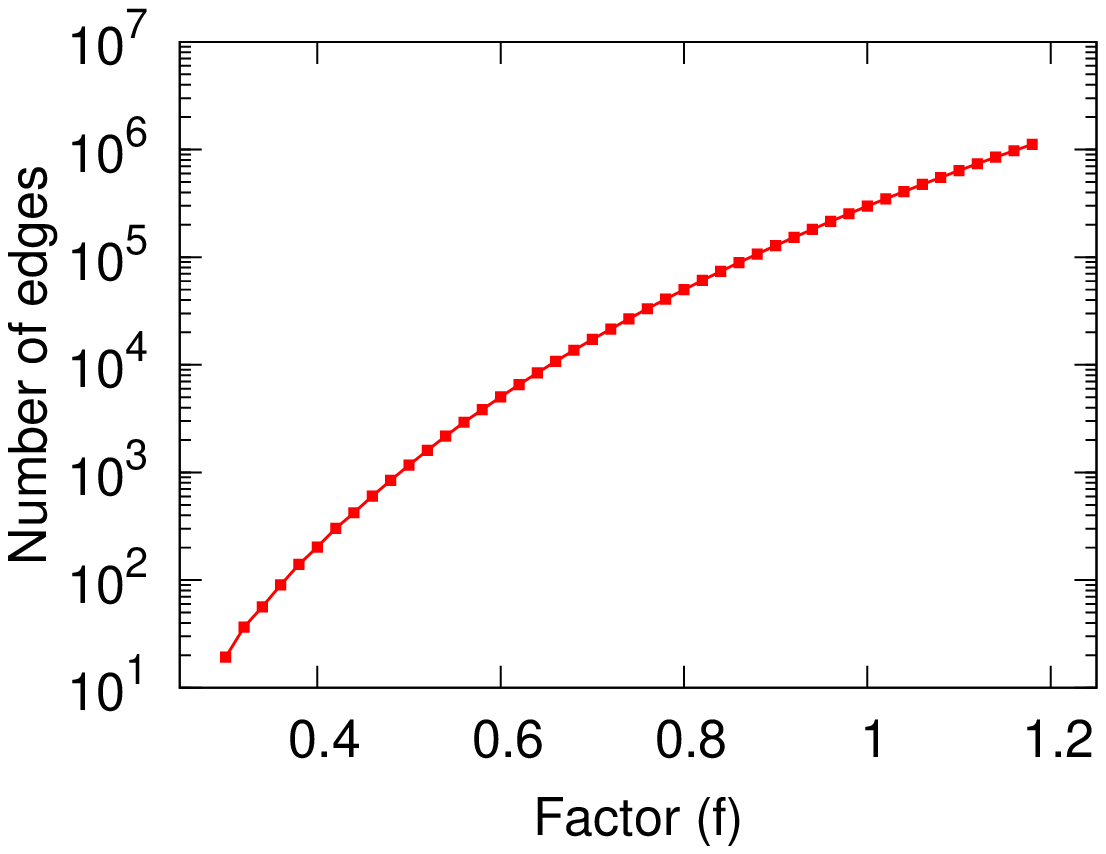} & 
\includegraphics[width=0.28\textwidth]{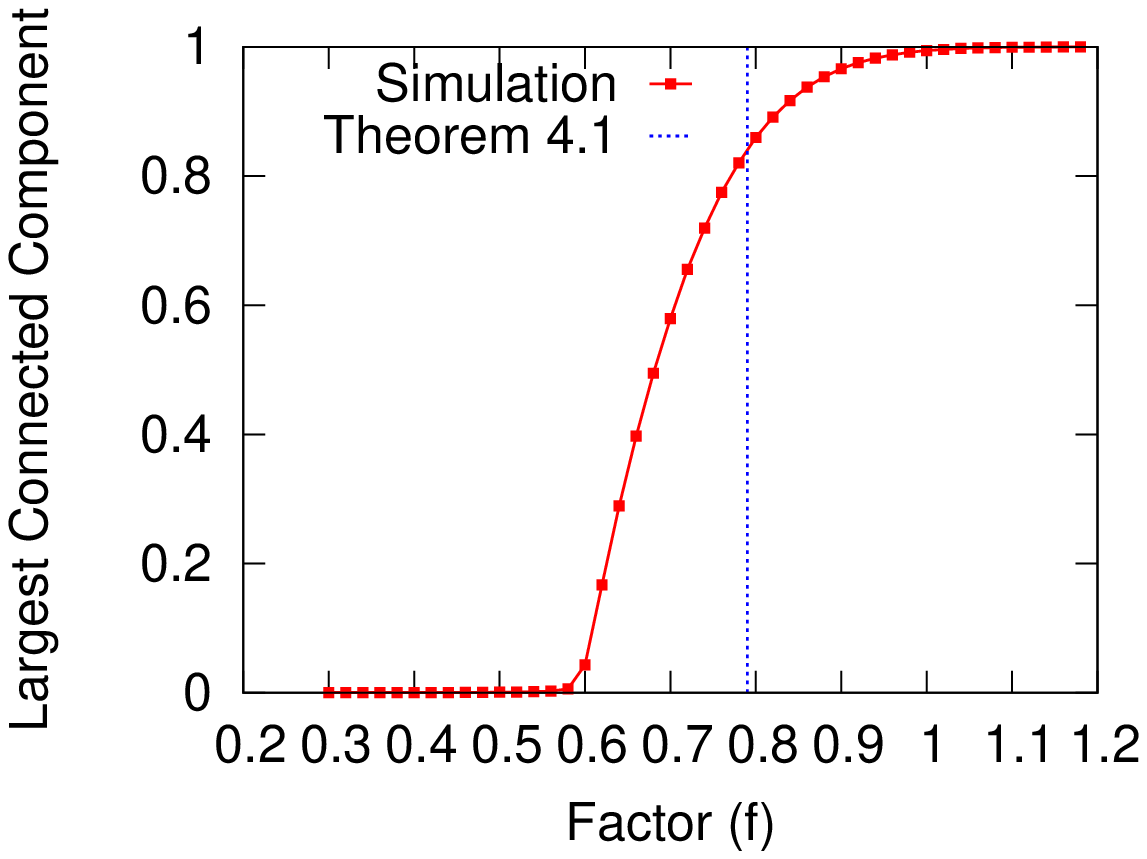} &
\includegraphics[width=0.28\textwidth]{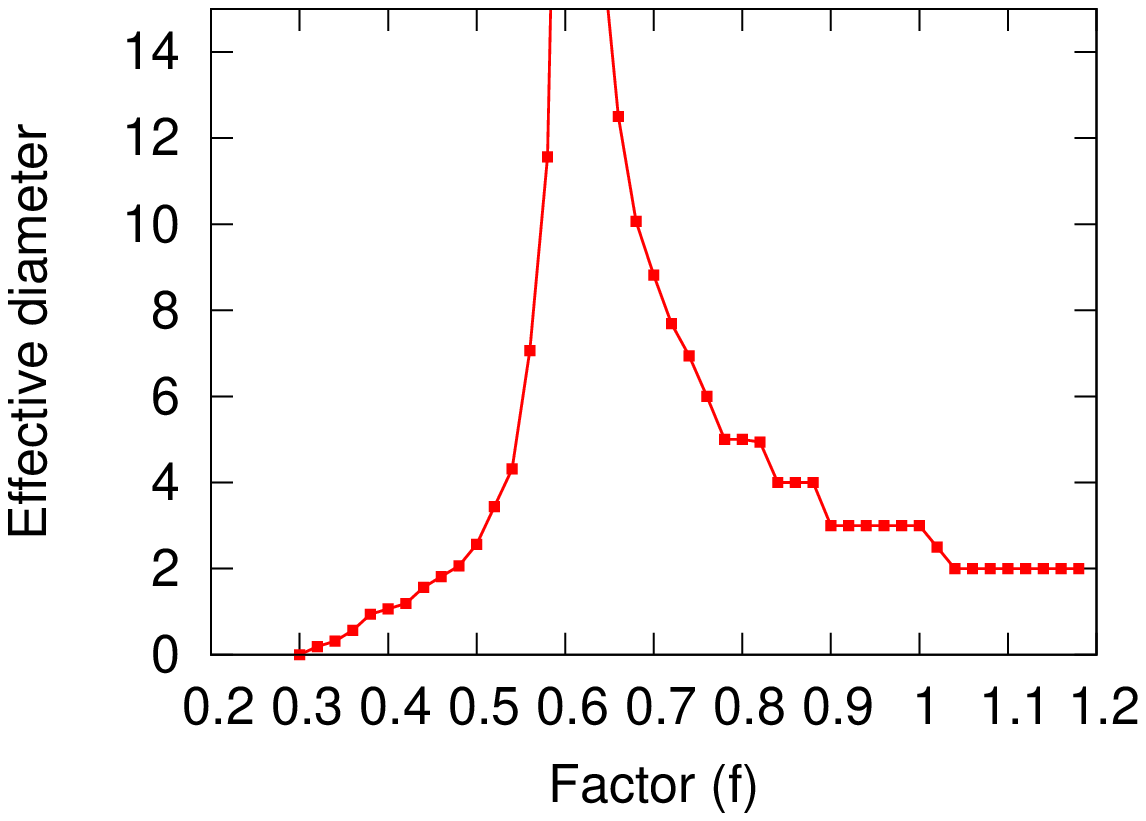} \\
\includegraphics[angle=90,width=0.04\textwidth]{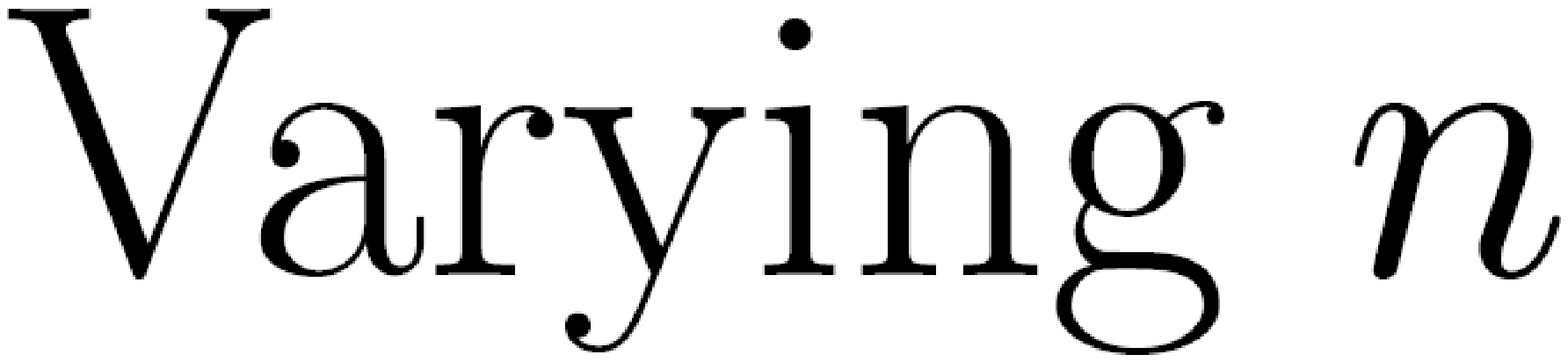}&
\includegraphics[width=0.28\textwidth]{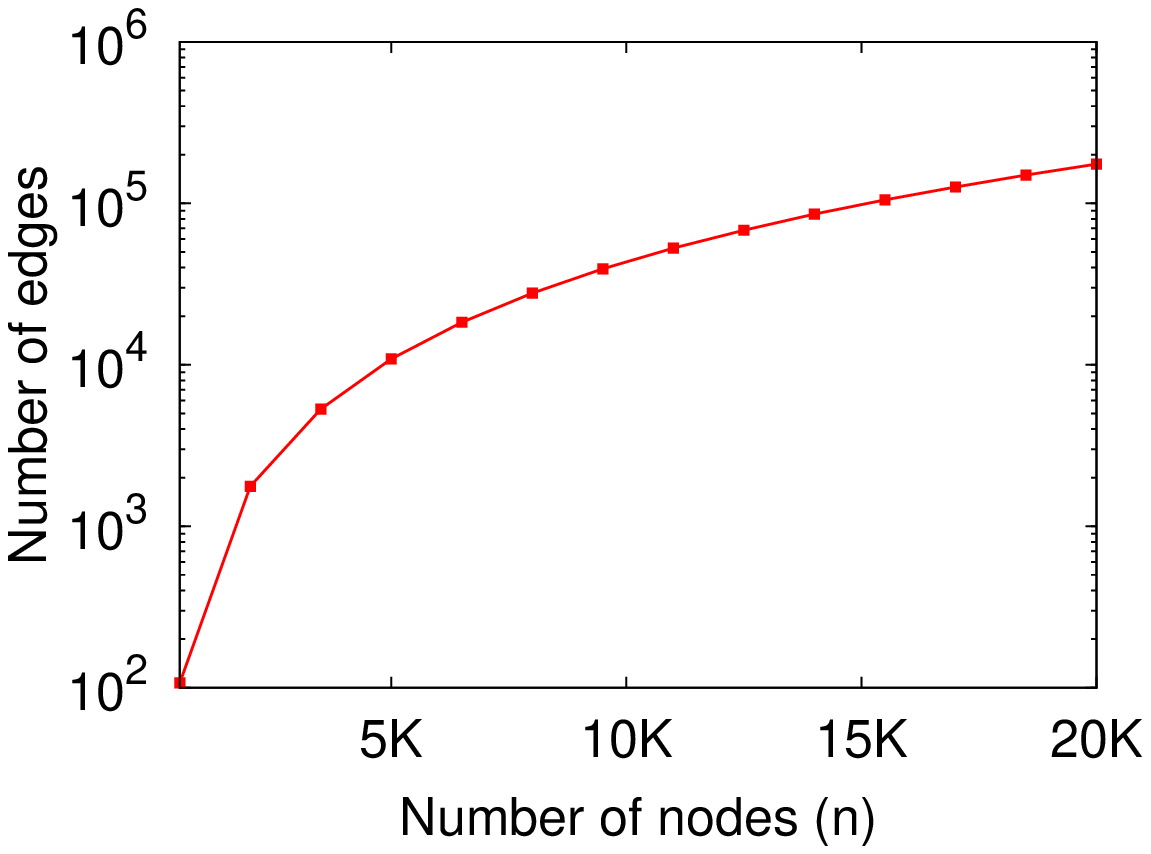} &
\includegraphics[width=0.28\textwidth]{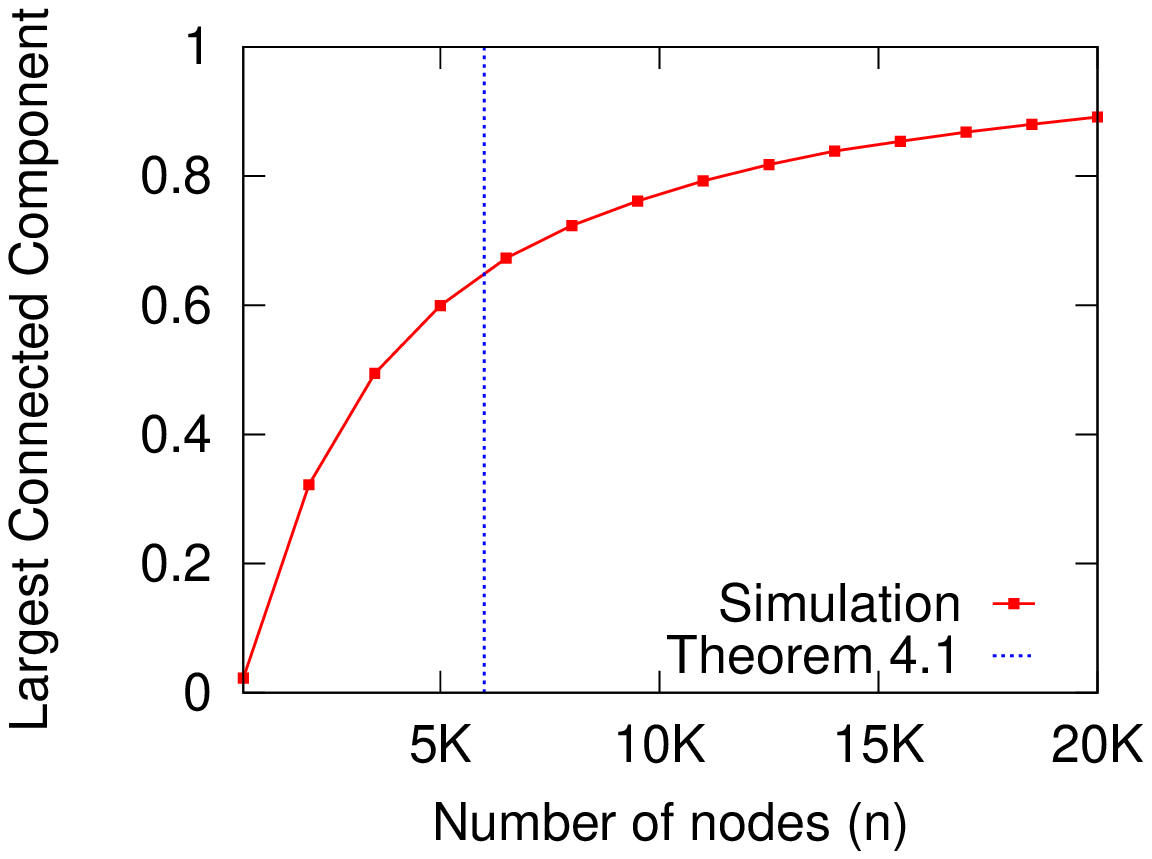} &
\includegraphics[width=0.28\textwidth]{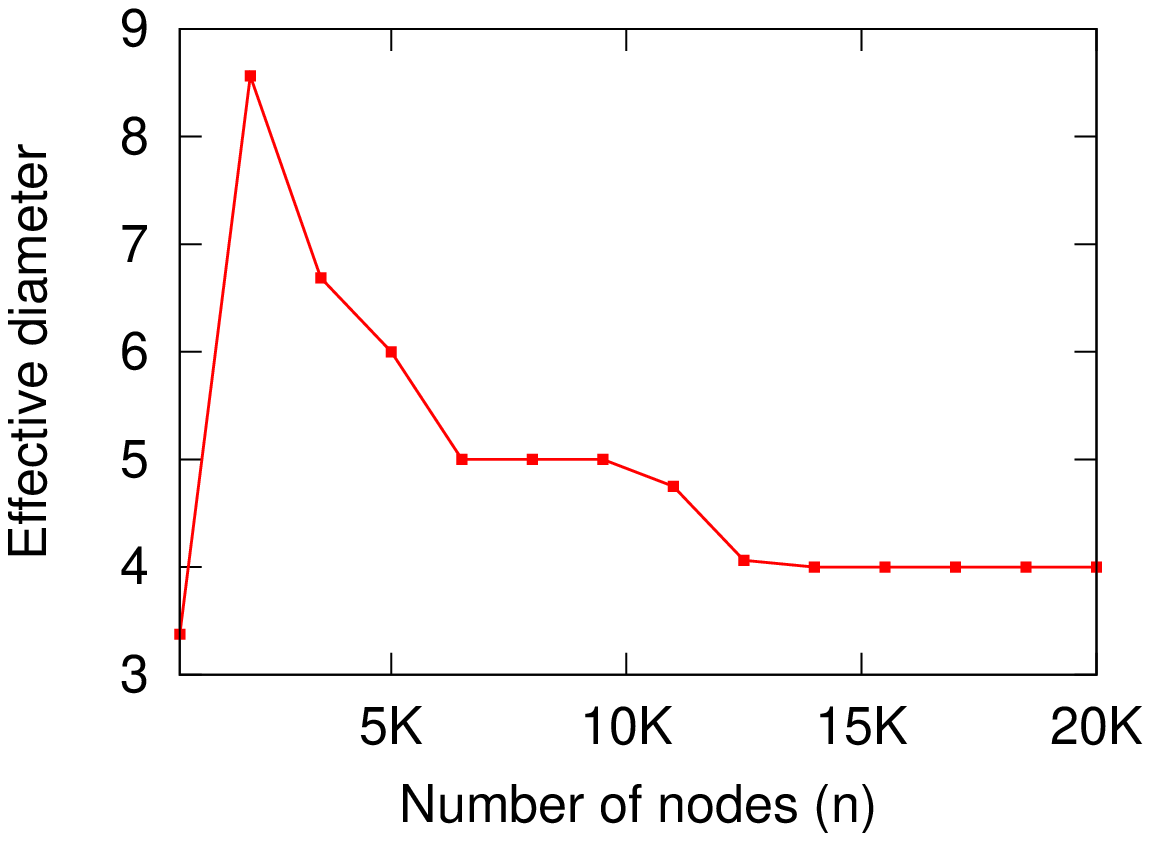} \\
%
& (a) Network size & (b) Largest connected component & (c) Effective diameter
\end{tabular}
\caption{Structural properties of a simplified \modelx~\modeldesc~when we fix $l$ and vary a single parameter one by one:~$\mu, \alpha, f$, or $n$.
As each parameter increases, in general, the synthetic network becomes denser so that a giant connected component emerges
and the diameter decreases to approach a constant.
}
\label{fig:params1d}
\end{figure}

Figure~\ref{fig:params1d} depicts the number of edges, the fraction of nodes in the largest connected component, and the effective diameter (the 90th-percentile distance~\cite{jure05dpl}) of the network as a function $\mu, \alpha, f$, and $n$ for fixed $l = 8$. First, we notice that the growth of network in the number of edges is slower than exponential since the curves on the plot grow sub-linearly in Figure~\ref{fig:params1d}(a) with log scaled $y$-axis.
Note that the network size is roughly proportional to 
$
n^{2} \left(\mu^{2}\alpha + 2 \mu(1-\mu)\beta + (1-\mu)^{2} \gamma\right)^{l}
$
from Theorem~\ref{thm:expedge}.
For example, by this formula, the network size is proportional to the $l$-th power of $f$, \ie~the eighth power of $f$ in this case.
As the expected number of edges is a polynomial function of each variable ($\mu, \alpha, f$ and $n$),
this sublinear growth on the log scale agrees with our analysis.
Furthermore, the larger the degree of the polynomial function for each variable is,
the closer to the straight line the network size curve becomes.
For instance, the network size grows by the polynomial function of degree 16 over $\mu$, whereas it grows by degree 2 over $n$.
In Figure~\ref{fig:params1d}(a), we thus observe that the network size growth over $\mu$ is even closer to the exponential curve than that over $n$.

%
%
Second, in Figure~\ref{fig:params1d}(b), the size of the largest component shows a sharp thresholding behavior, which indicates a rapid emergence of the giant component. This is very similar to thresholding behaviors observed in other network models such as the \GNP s model~\cite{erdos60random}.
The vertical line in the middle of each figure represents the theoretical theshold for the unique giant connected component.
As we analyzed, each network contains at least half size of giant connected component at its threshold.
%
%

Last, while the previous two network properties monotonically change,
in Figure~\ref{fig:params1d}(c) the effective diameter of the network increases quickly up to about the point where the giant connected component forms and then drops rapidly after that and approaches a constant value. This behavior is in accordance with empirical observations 
of the ``gelling'' point where the giant component forms and the diameter starts to decrease
in the evolution of real-world networks~\cite{jure05dpl,McGlohonAF08}.

\begin{figure}[ttp]
\centering
\begin{tabular}{cccc}
\centering 
\includegraphics[angle=90,width=0.04\textwidth]{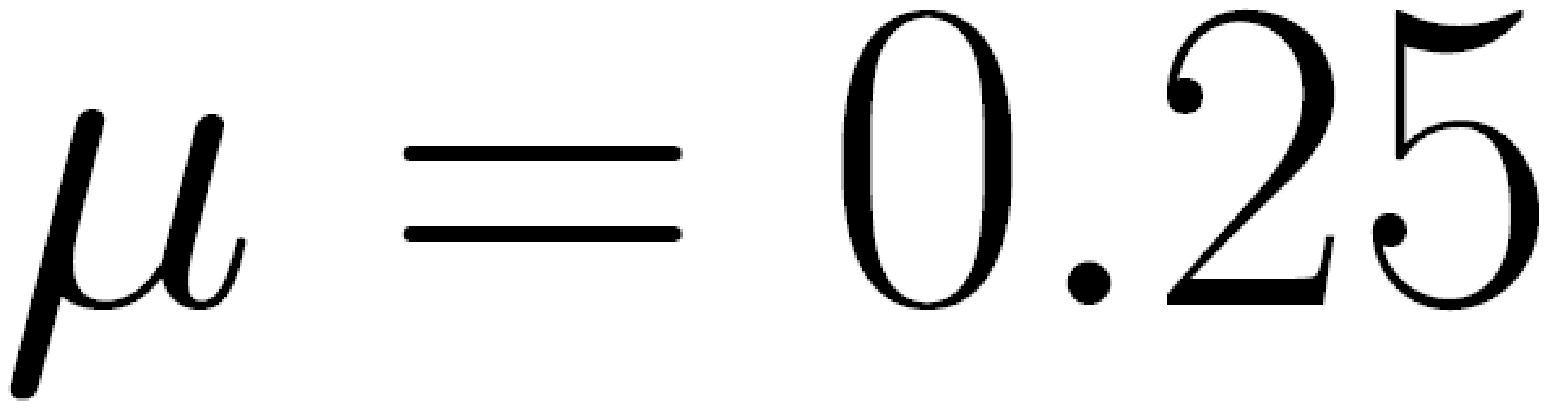}&
\includegraphics[width=0.28\textwidth]{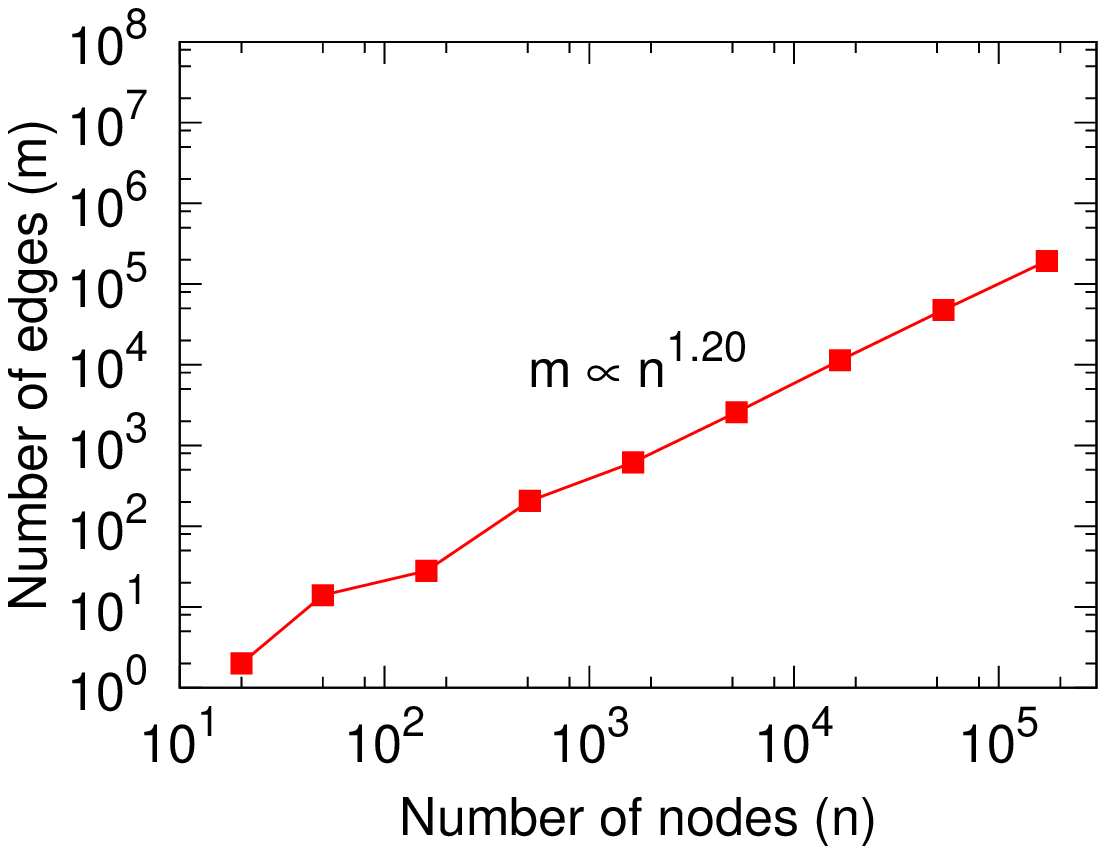}&
\includegraphics[width=0.28\textwidth]{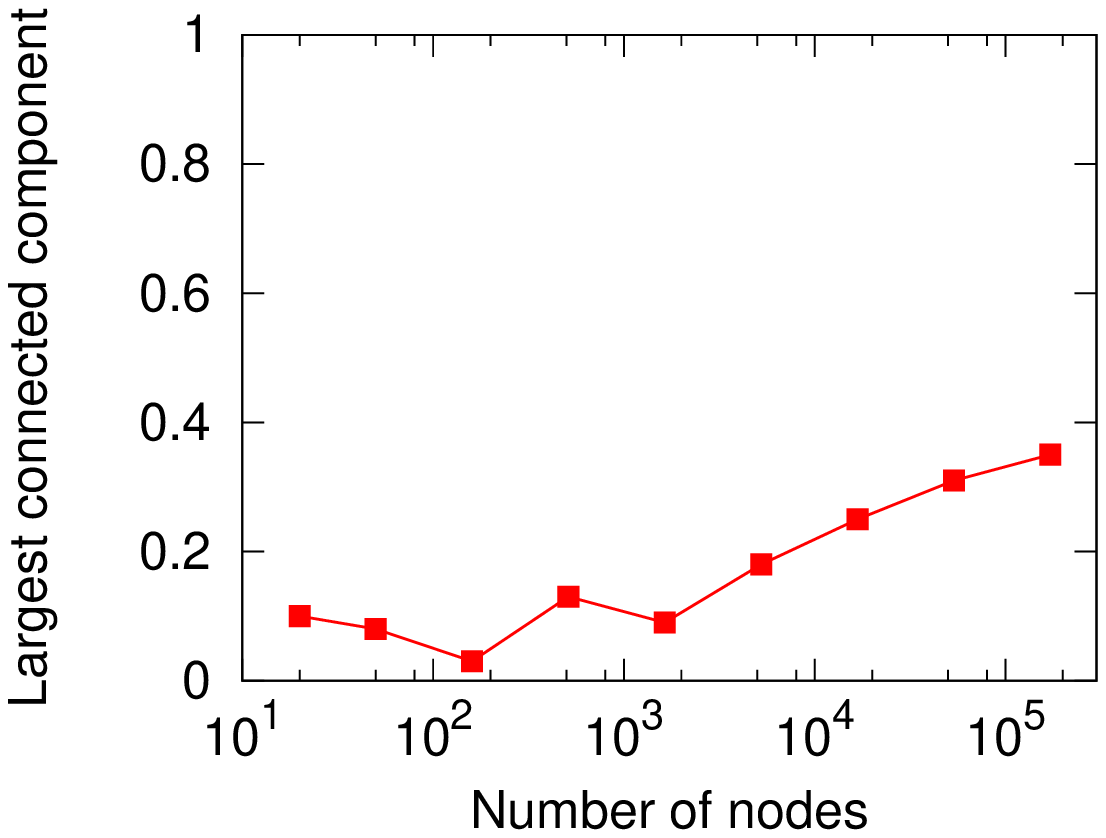} &
\includegraphics[width=0.28\textwidth]{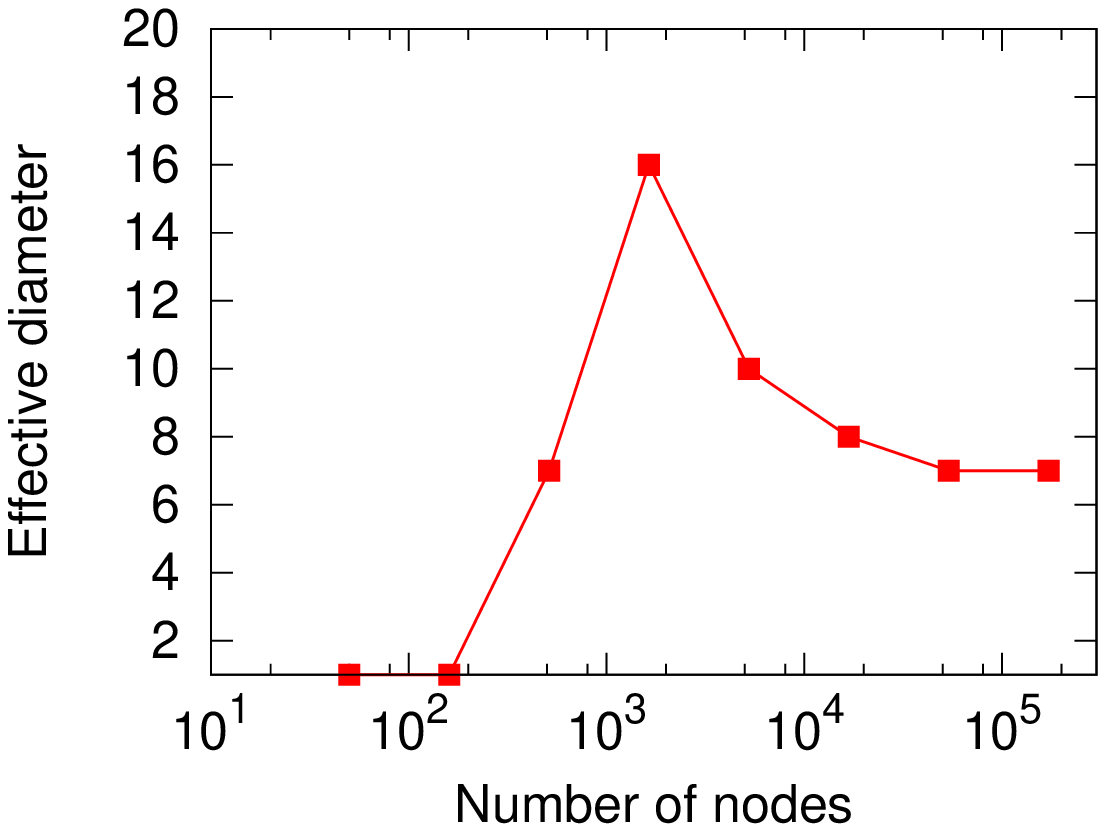} \\
\includegraphics[angle=90,width=0.04\textwidth]{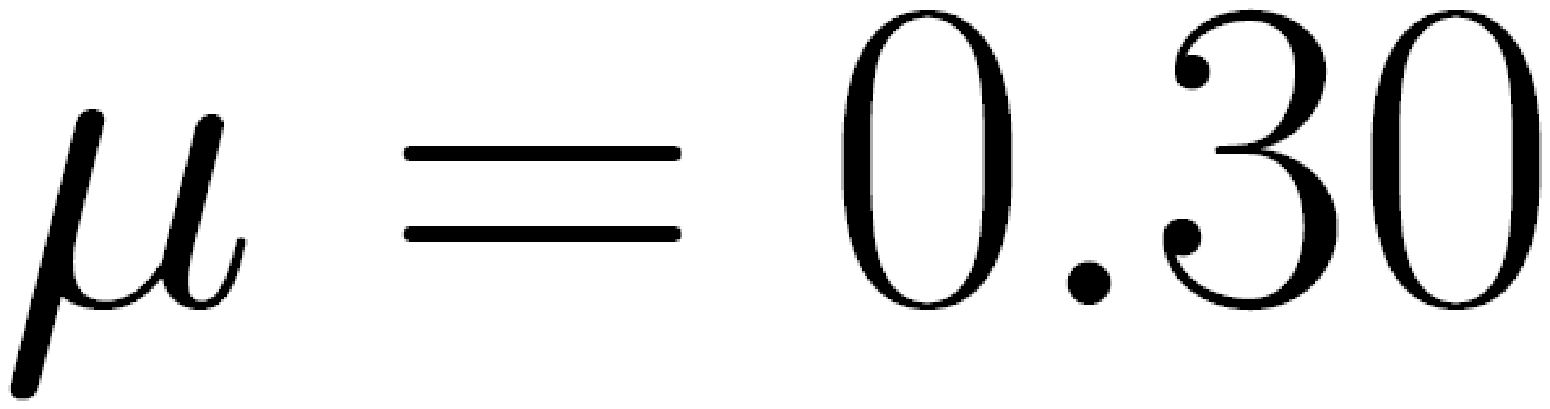}&
\includegraphics[width=0.28\textwidth]{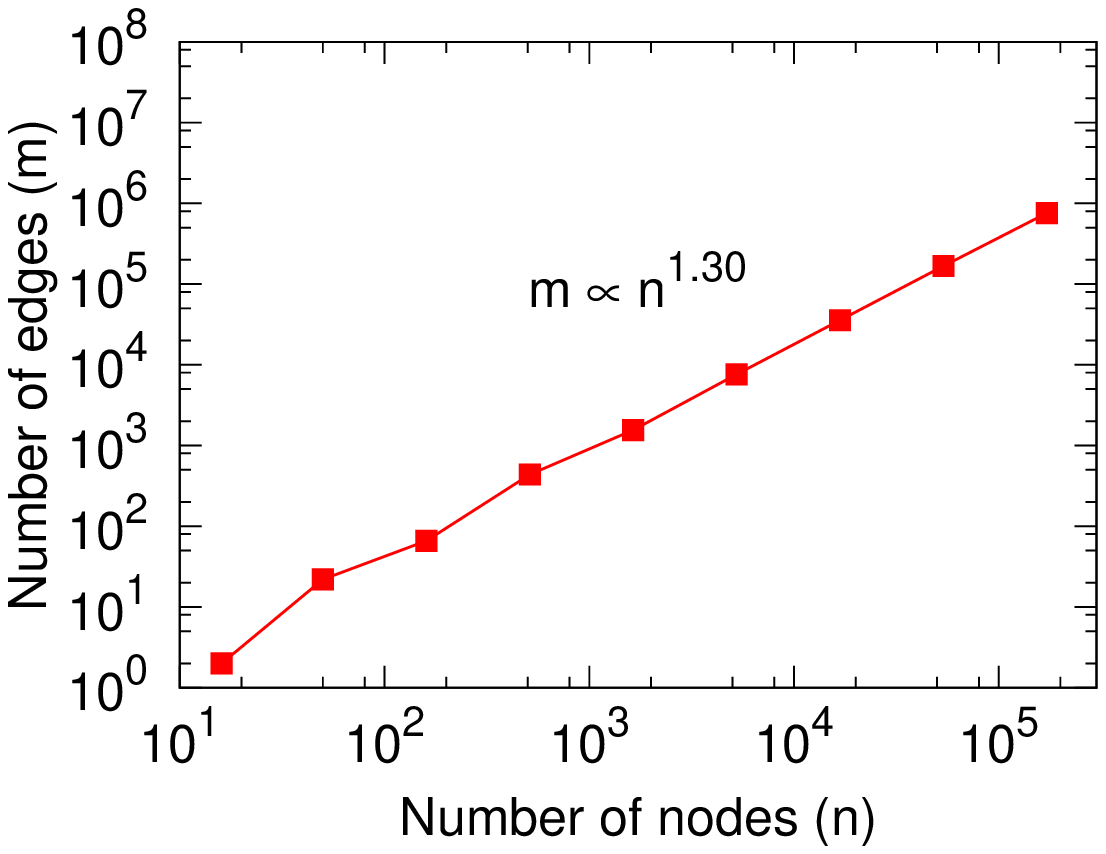} &
\includegraphics[width=0.28\textwidth]{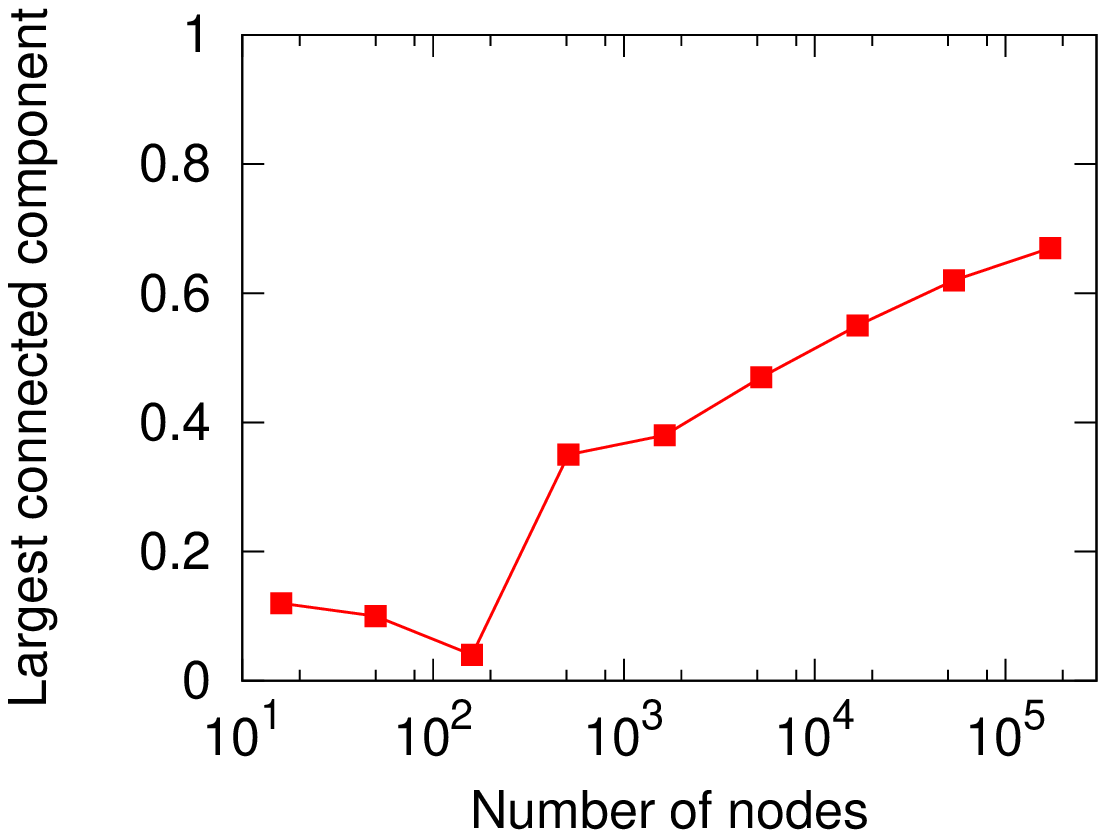} & 
\includegraphics[width=0.28\textwidth]{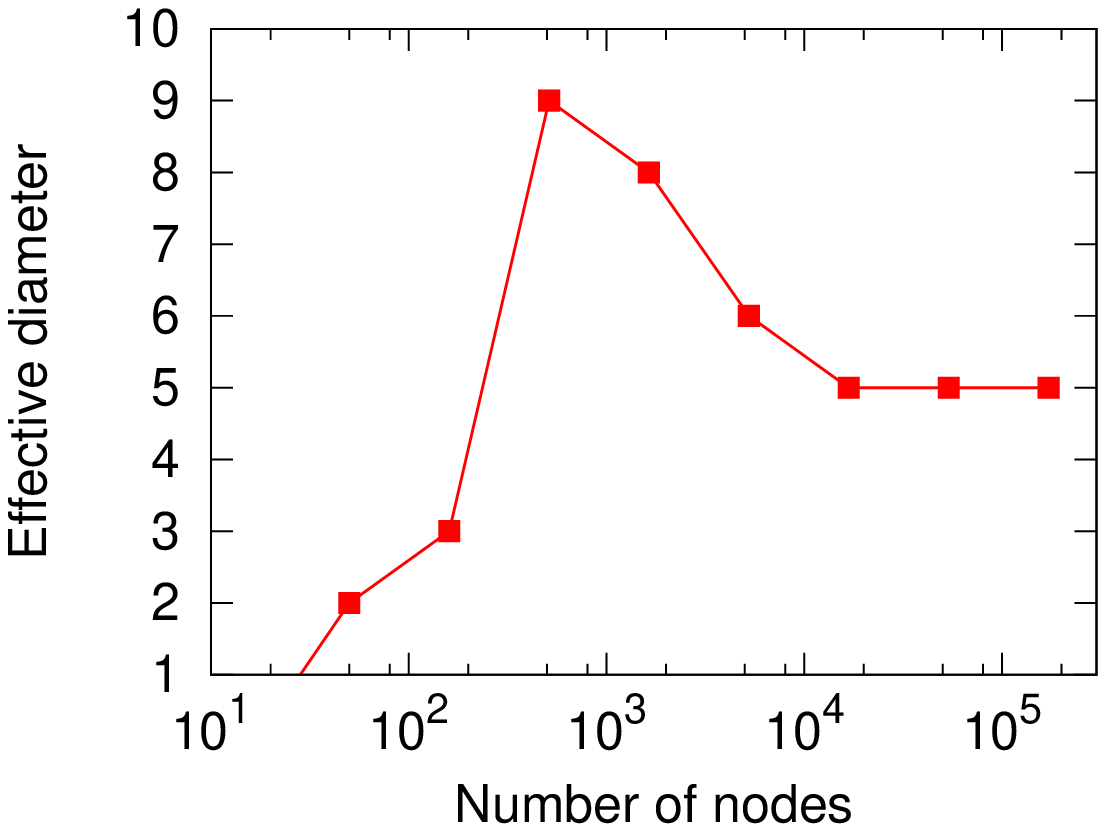} \\
\includegraphics[angle=90,width=0.04\textwidth]{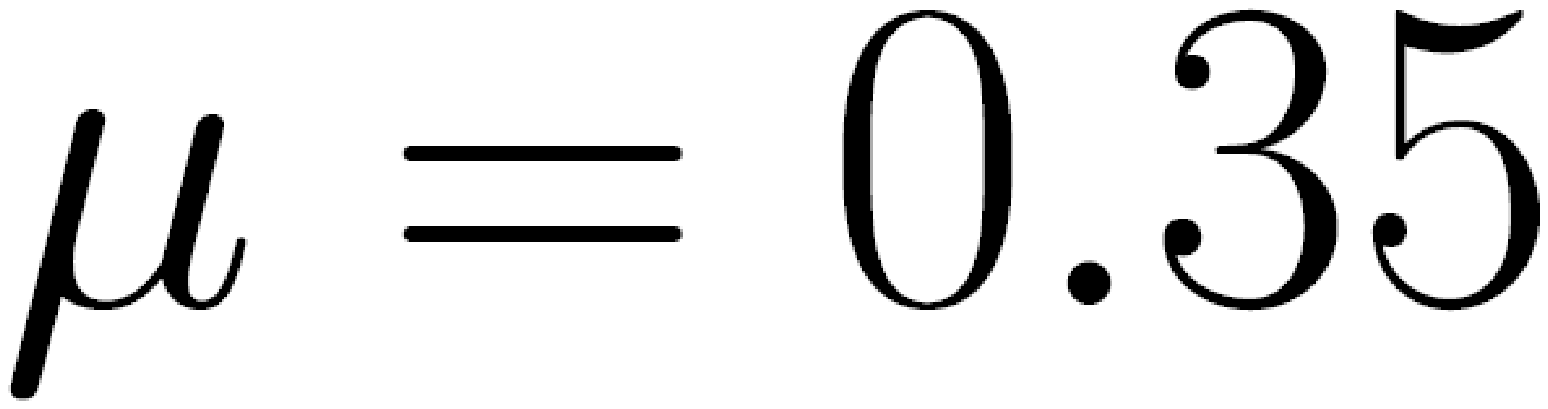}&
\includegraphics[width=0.28\textwidth]{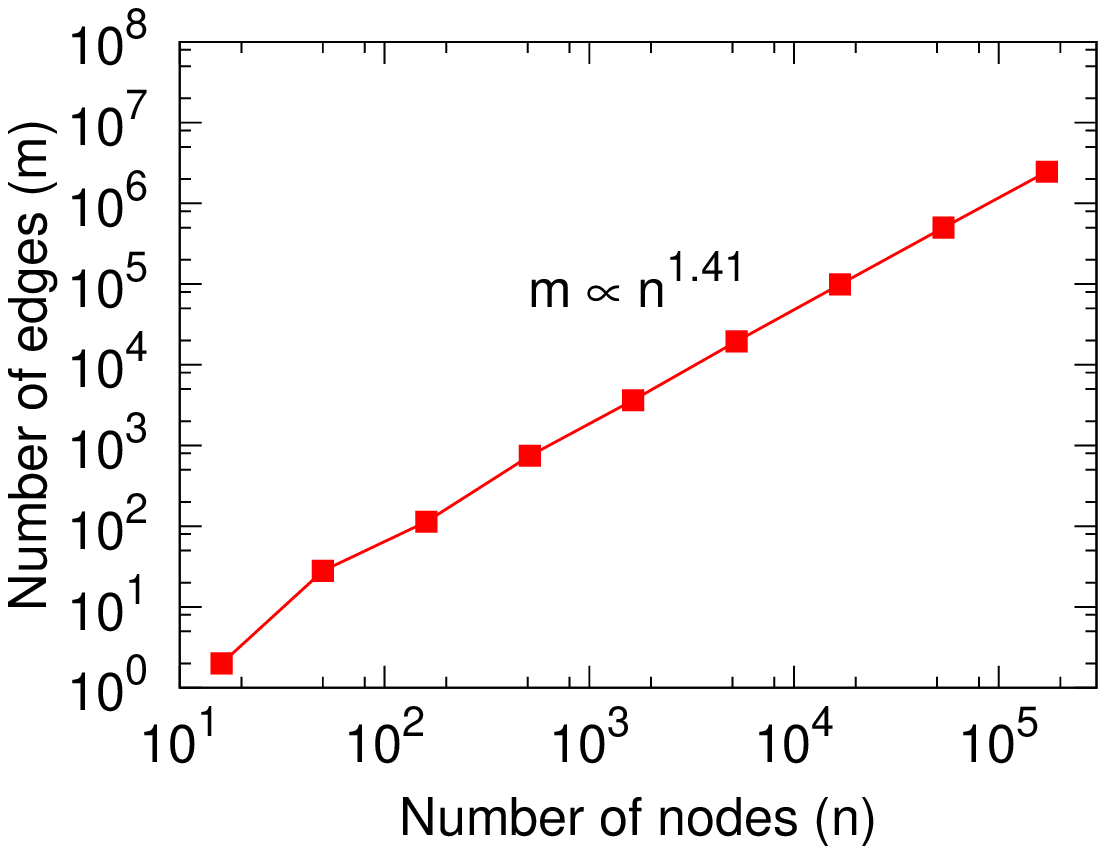} &
\includegraphics[width=0.28\textwidth]{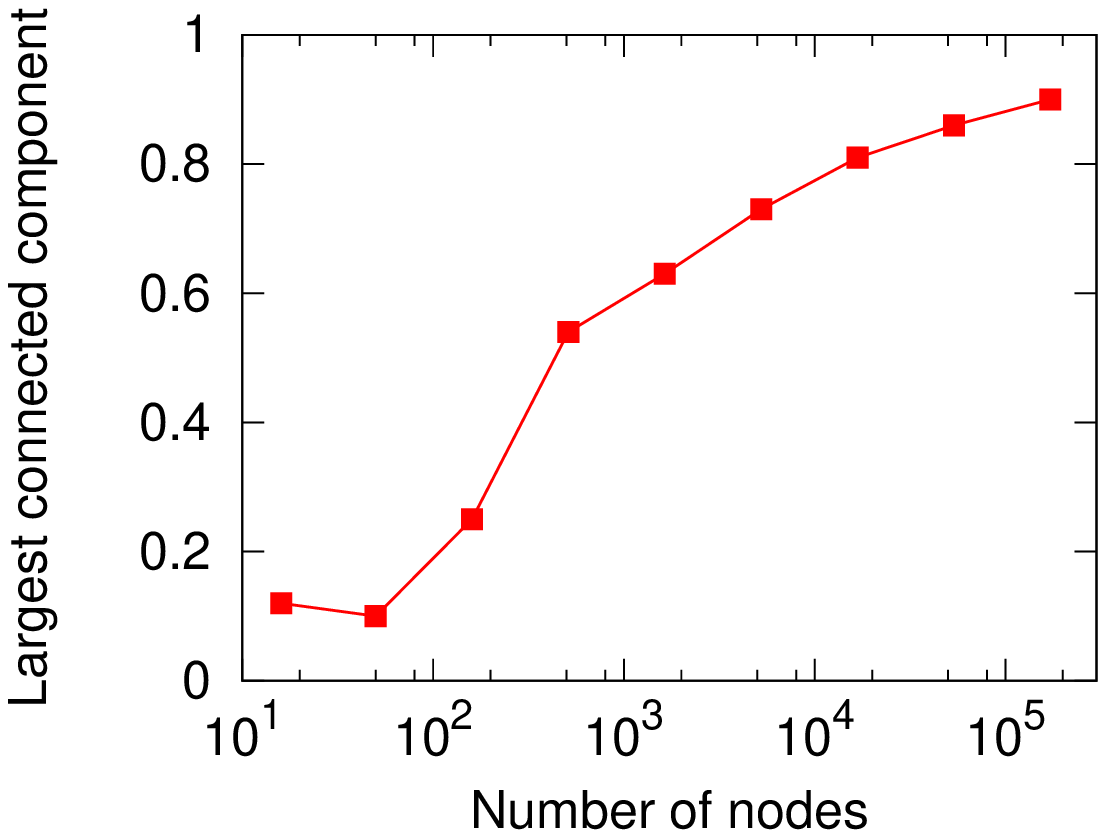} &
\includegraphics[width=0.28\textwidth]{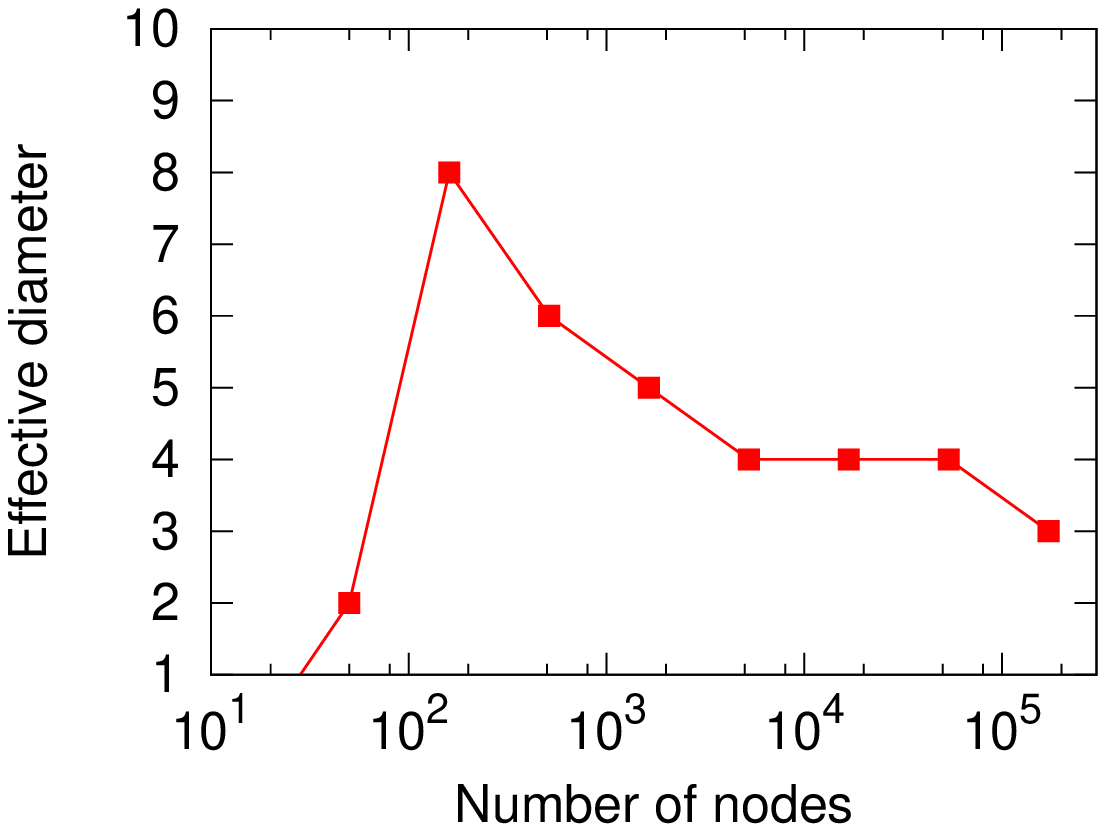} \\
\includegraphics[angle=90,width=0.04\textwidth]{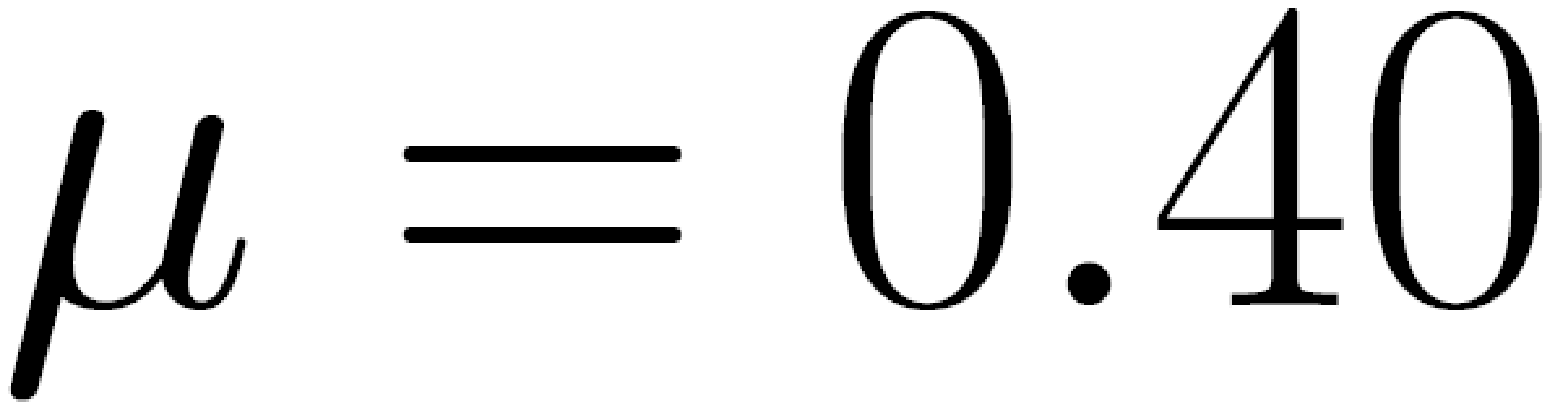}&
\includegraphics[width=0.28\textwidth]{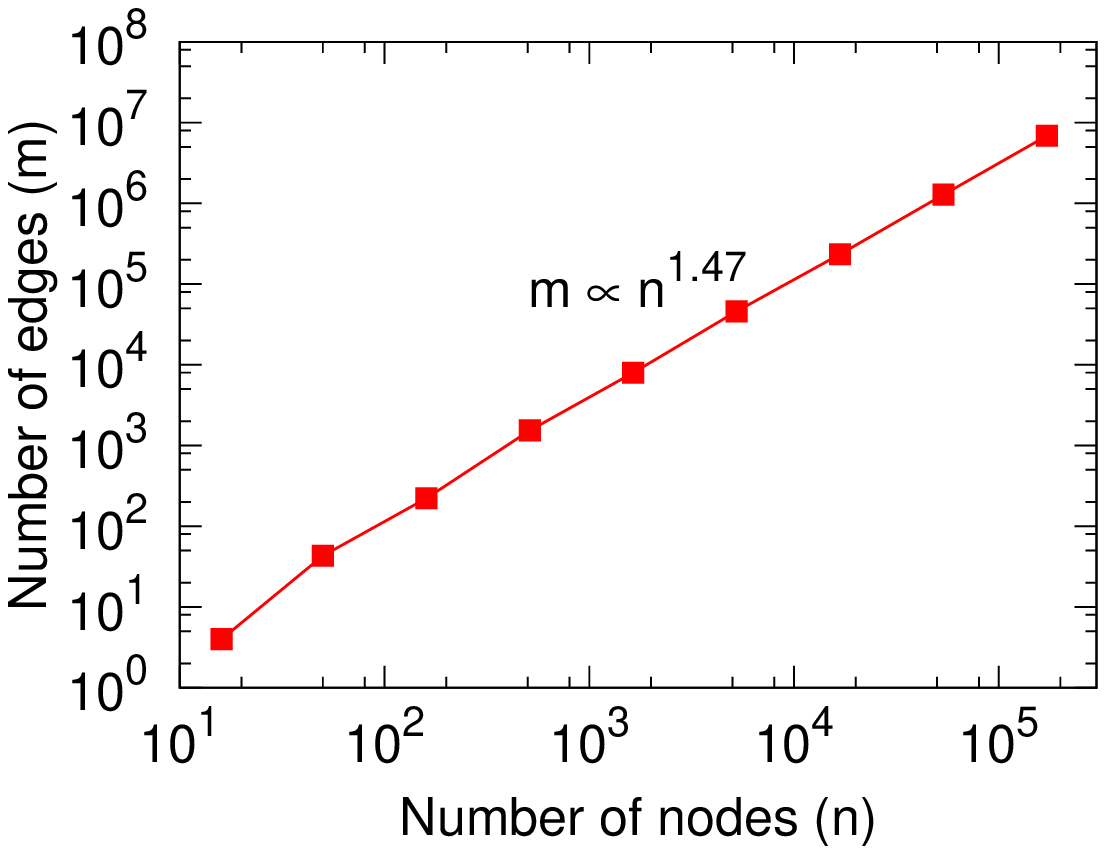} &
\includegraphics[width=0.28\textwidth]{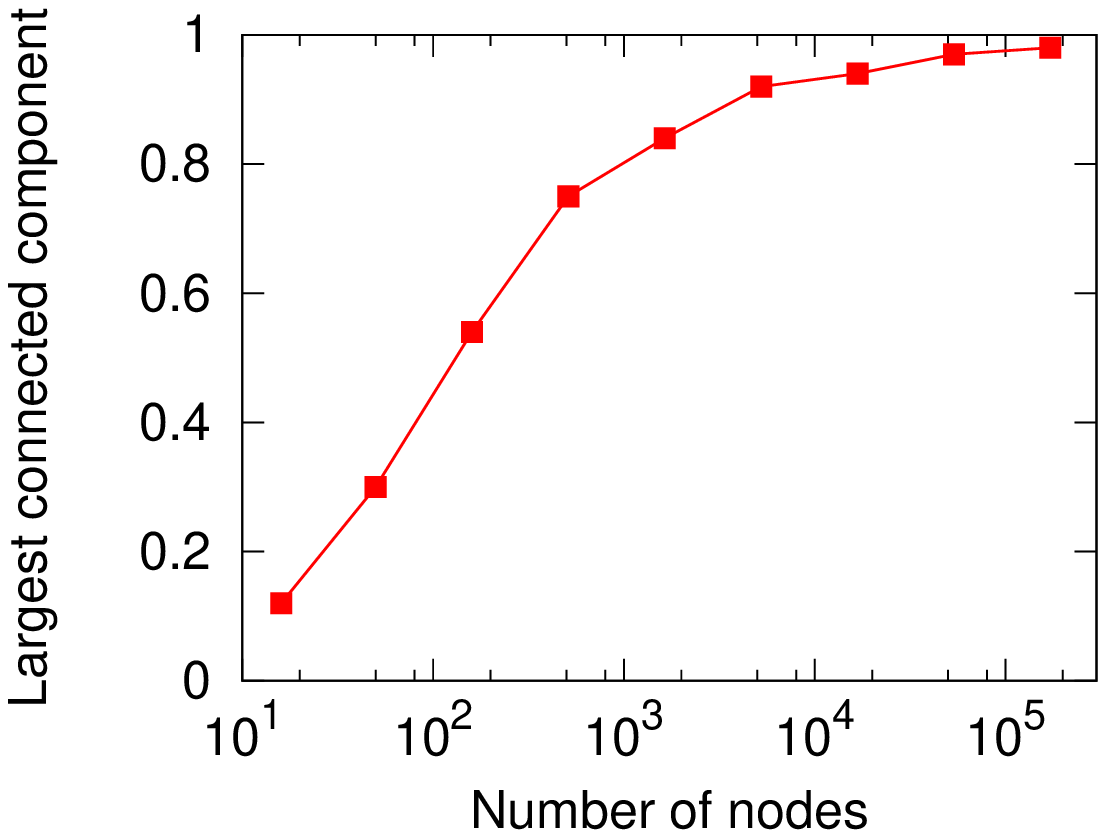} &
\includegraphics[width=0.28\textwidth]{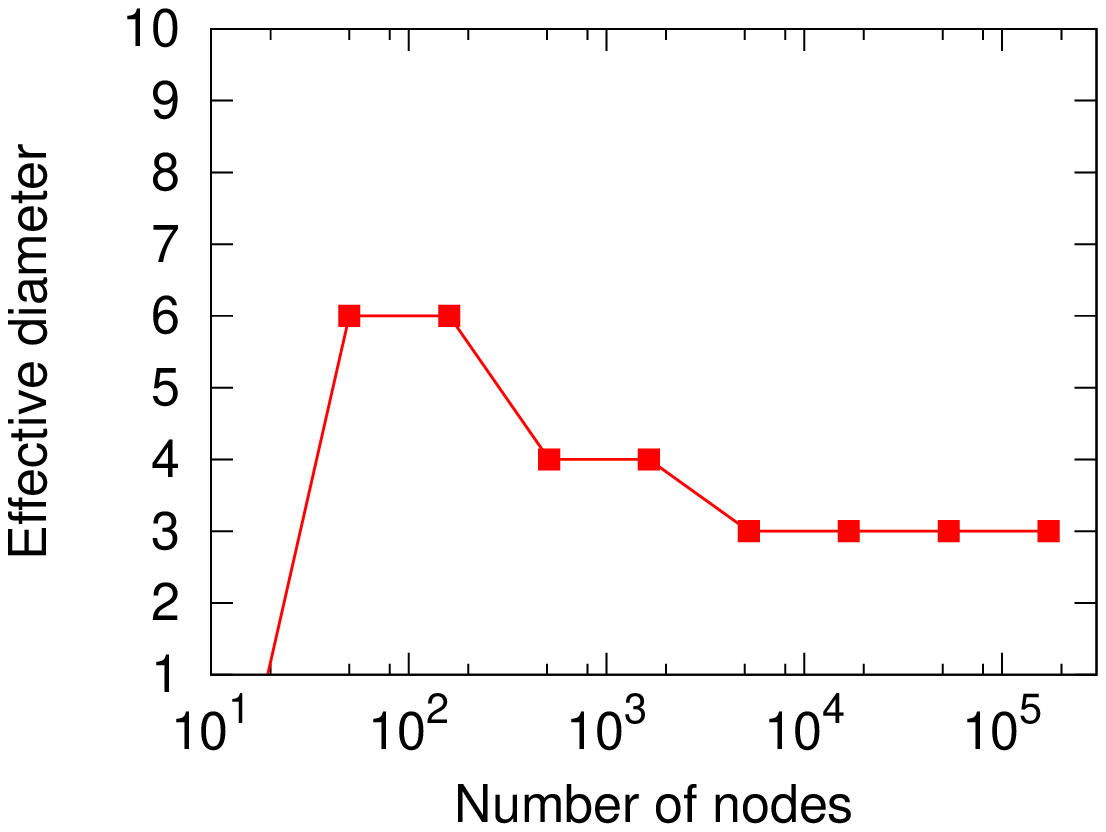} \\
& (a) Network size & (b) Largest connected component & (c) Effective diameter
\end{tabular}
\caption{Structural properties of a simplified \modelgraph~as a function of the number of nodes $n$ for different values of $\mu$
(we fix the affinity matrix $\Theta = [0.85~0.7;0.7~0.15]$ and
the ratio $\rho = l / \log n = 0.596$).
Observe not only that the relationship between the number of edges and nodes obeys Densification Power Law 
but also that the diameter begins shrinking after the giant component is formed~\cite{McGlohonAF08}.}
\label{fig:DPL}
\end{figure}

Furthermore, we also performed simulations where we fix $\Theta$ and $\mu$ 
but simultaneously increase both $n$ and $l$ by maintaining their ratio constant.
Figure~\ref{fig:DPL} plots the change in each network metric (network size, fraction of the largest connected component, and effective diameter) as a function of the number of nodes $n$ for different values of $\mu$.
Each plot effectively represents the evolution of the MAG network
as the number of nodes grows over time.
From the plots, we see that \modelx~follows densification power law (DPL) and the shrinking diameter properties of real-world networks~\cite{jure05dpl}.
Depending on the choice of $\mu$, one can control the rate of densification and the diameter.

%
%
%

%
\subsection{Degree Distributions}
In addition to the network size, connectivity, and diameter, we also examined the degree distributions of
\modelgraph s empricially.
We already proved that the \modelx~can give rise to networks that have either a log-normal or a power-law degree distribution depending on the model parameters.
Here we generate the two versions of networks and compare their degree distributions.

\begin{figure}[ttp]
\centering
\subfigure[Simplified MAG: \modeldesc. Log-normal degree distribution.]{
	\includegraphics[width=0.33\textwidth]{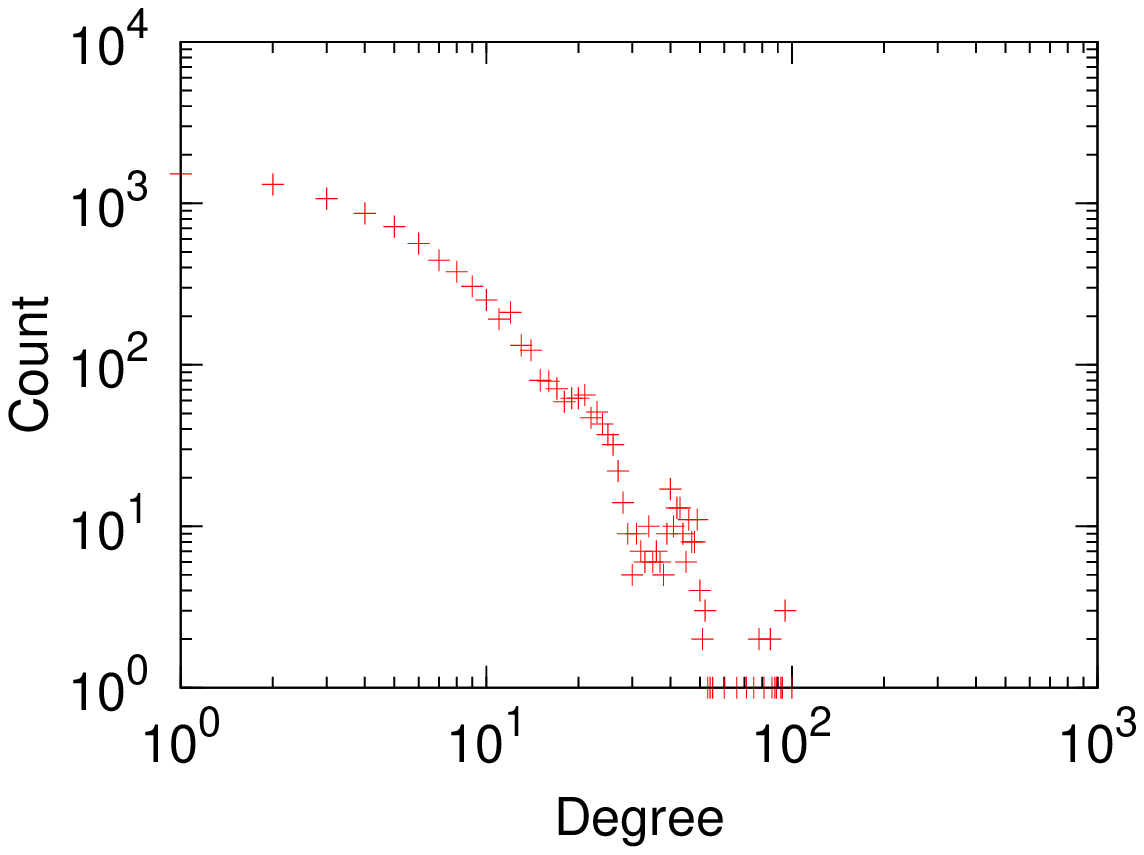}
	\includegraphics[width=0.33\textwidth]{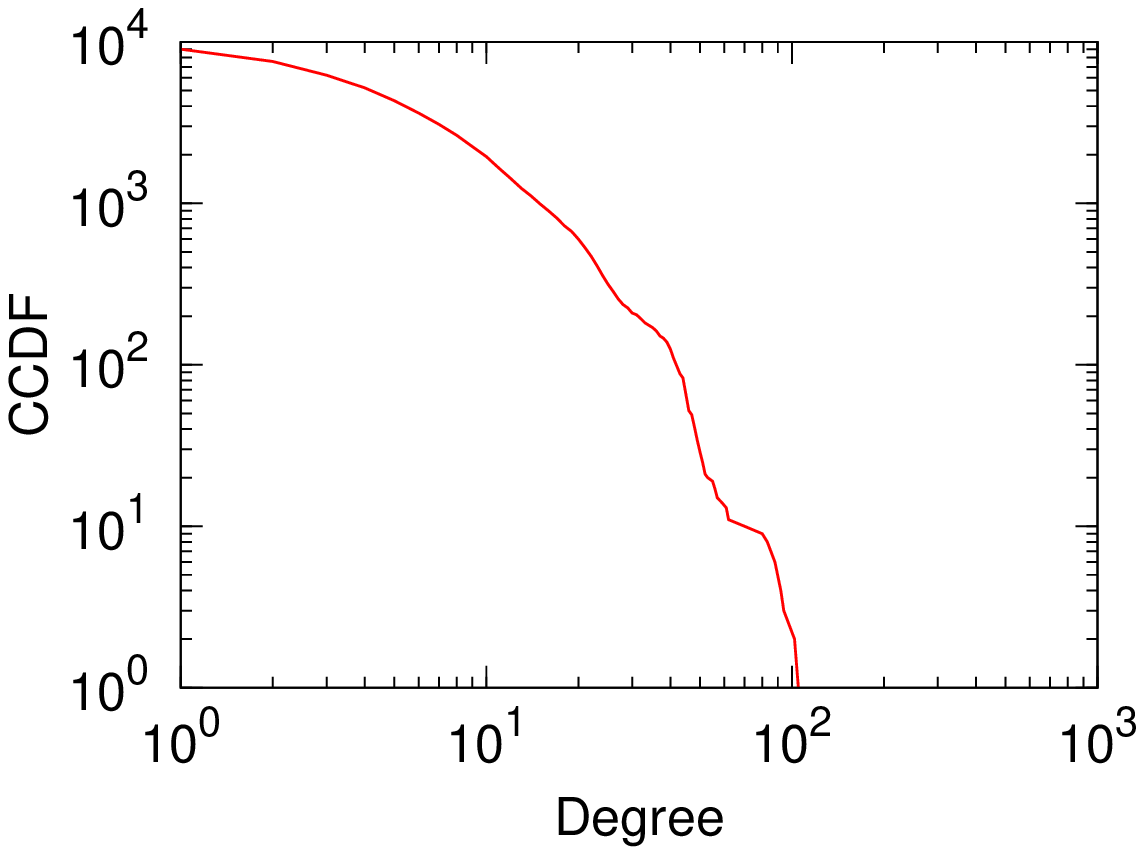}
}
\subfigure[Power-law MAG: \modelpower. Power-law degree distribution.]{
	\includegraphics[width=0.33\textwidth]{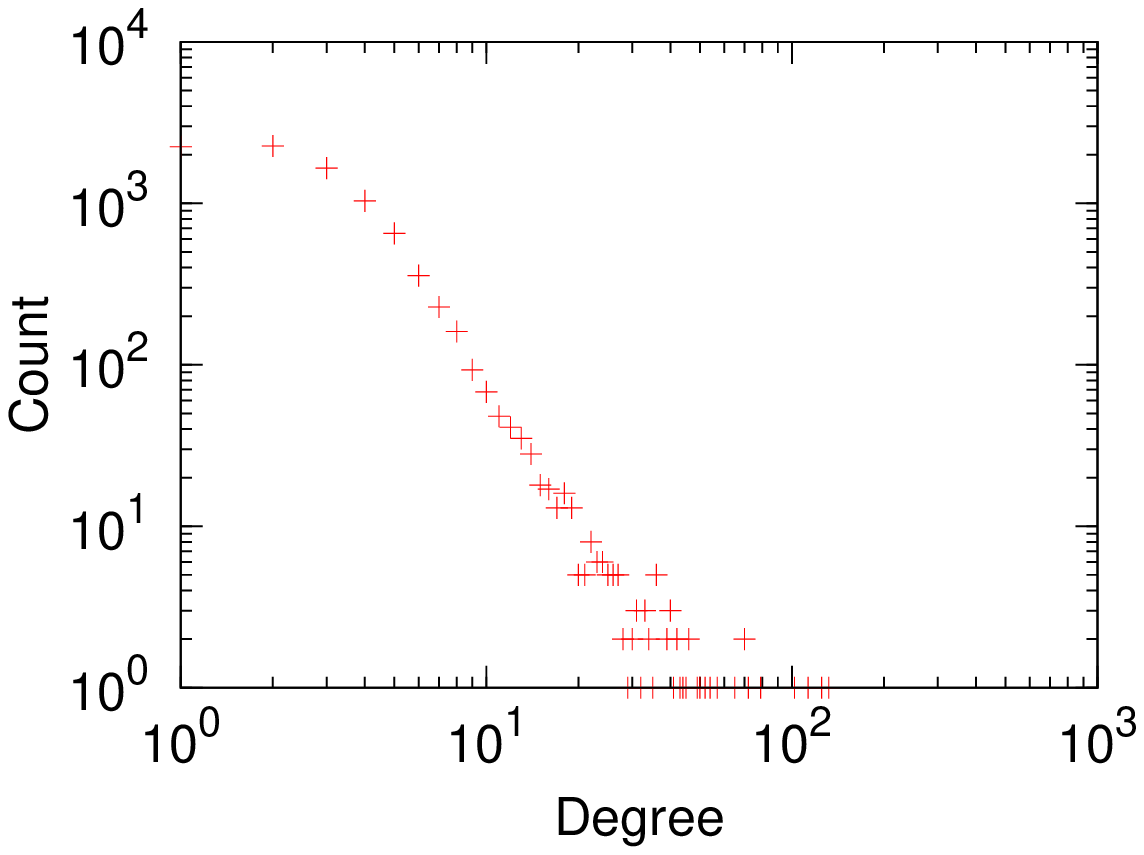}
	\includegraphics[width=0.33\textwidth]{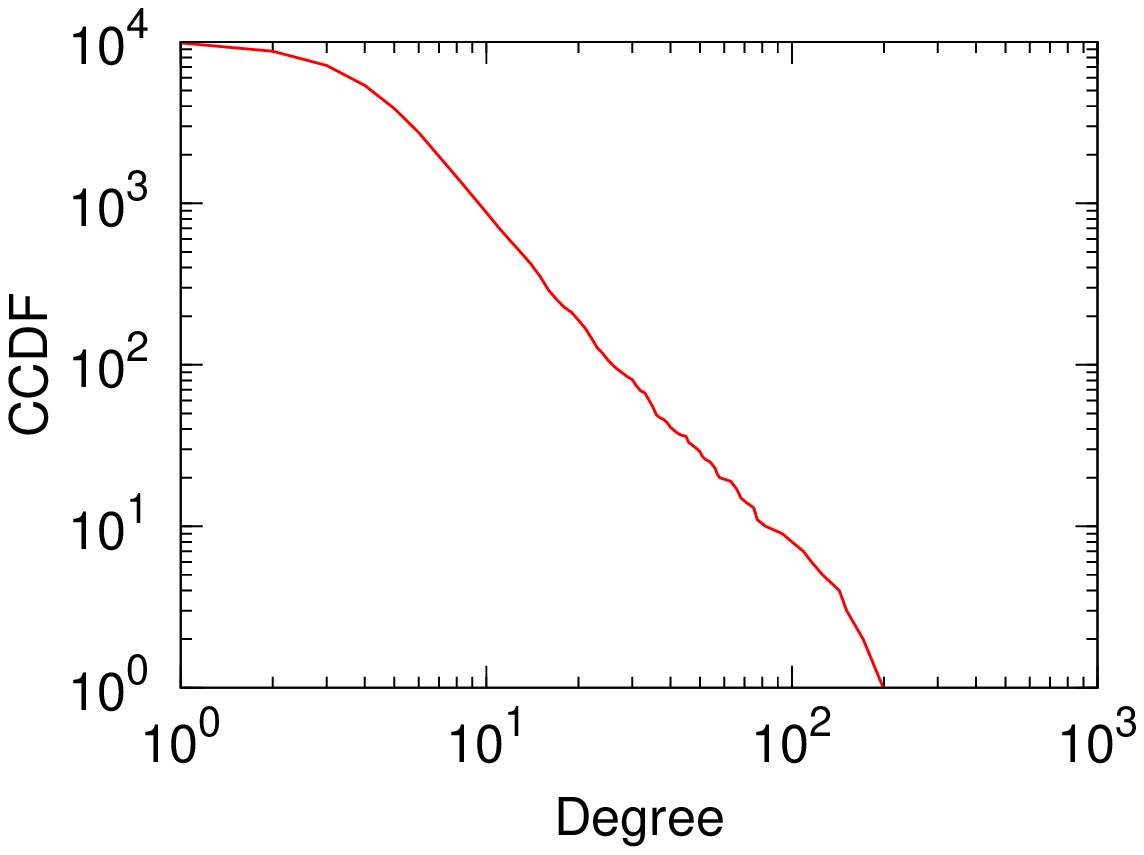}
}
\caption{Degree distributions of simplified and power-law version of \modelgraph s (see Section~\ref{sec:power}).
We plot both PDF and CCDF of the degree distribution.
The simplified version in Figure~(a) has parabolic shape on log-log scale, which is an indication
of a log-normal degree distribution.
In contrast, the power-law version in Figure~(b) shows a straight line on the same scale, which demonstrates a power-law degree distribution.}
\label{fig:power}
\end{figure}

Figure~\ref{fig:power} exhibits the degree distributions of the two types of \modelx.
While Figure~\ref{fig:power}(a) plots the degree distributions of the simplified \modelx~\modeldesc,
Figure~\ref{fig:power}(b) shows those of the power-law \modelx~\modelpower.
For each case, the left plot represents the raw form of degree histogram, whereas the right curve plots the \textit{complementary cumulative distribution}~(CCDF), which nicely removes the noisy factor.
In Figure~\ref{fig:power}(a), both raw and CCDF versions of distribution look parabolic on the log-log scale, which verifies that \modeldesc~has a log-normal degree distribution.
On the other hand, in Figure~\ref{fig:power}(b), both plots exhibit the straight line on the same scale, which indicates that the degree distribution of \modelpower~follows a power-law.
All these experimental results agree with our analyses in Section~\ref{sec:degree} and Section~\ref{sec:power}.

%


\subsection{Comparison to Real-world Networks}
%
Also, we qualitatively compare the structural properties of a specific real-world network and the corresponding MAG model.
This leads to interesting questions of how to find optimal \modelx~parameters so that synthetic network resembles the given real-world network. 
The full resolution of these questions lies beyond the scope of the present paper;
currently, we searched by brute force over (the relatively small number of) possible MAG parameter settings.
%
%
We manually selected some parameter settings (for $n, l, \mu, \Theta$) to synthesize the simplified \modelx~and obtained the properties of \modeldesc~to compare the \modelx~with a real-world network. Our goal is not to claim that these particular parameter values are in any way ``optimal'' for the given real-world network but rather to show that many properties of the MAG model exhibit qualitatively similar behavior as real-world networks.

\begin{figure}[ttp]
\centering
\begin{tabular}{ccc}
\includegraphics[width=0.25\textwidth]{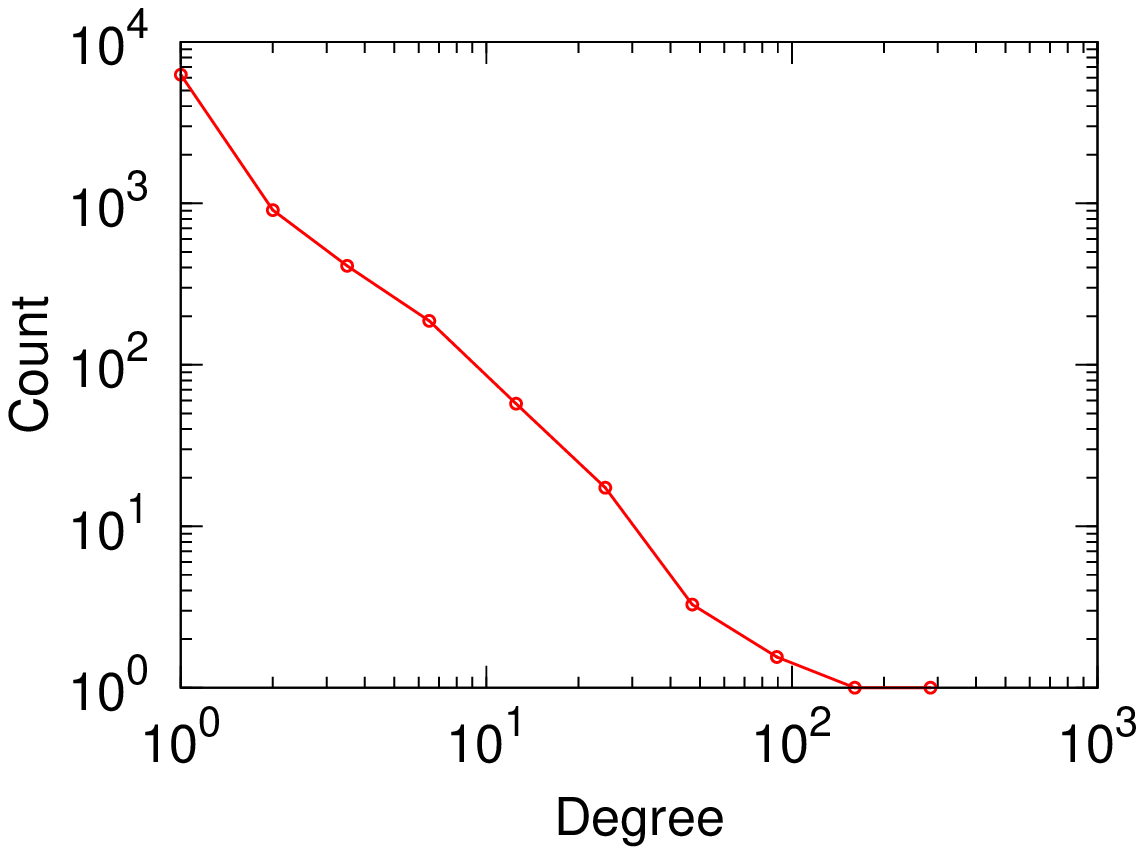} &
\includegraphics[width=0.25\textwidth]{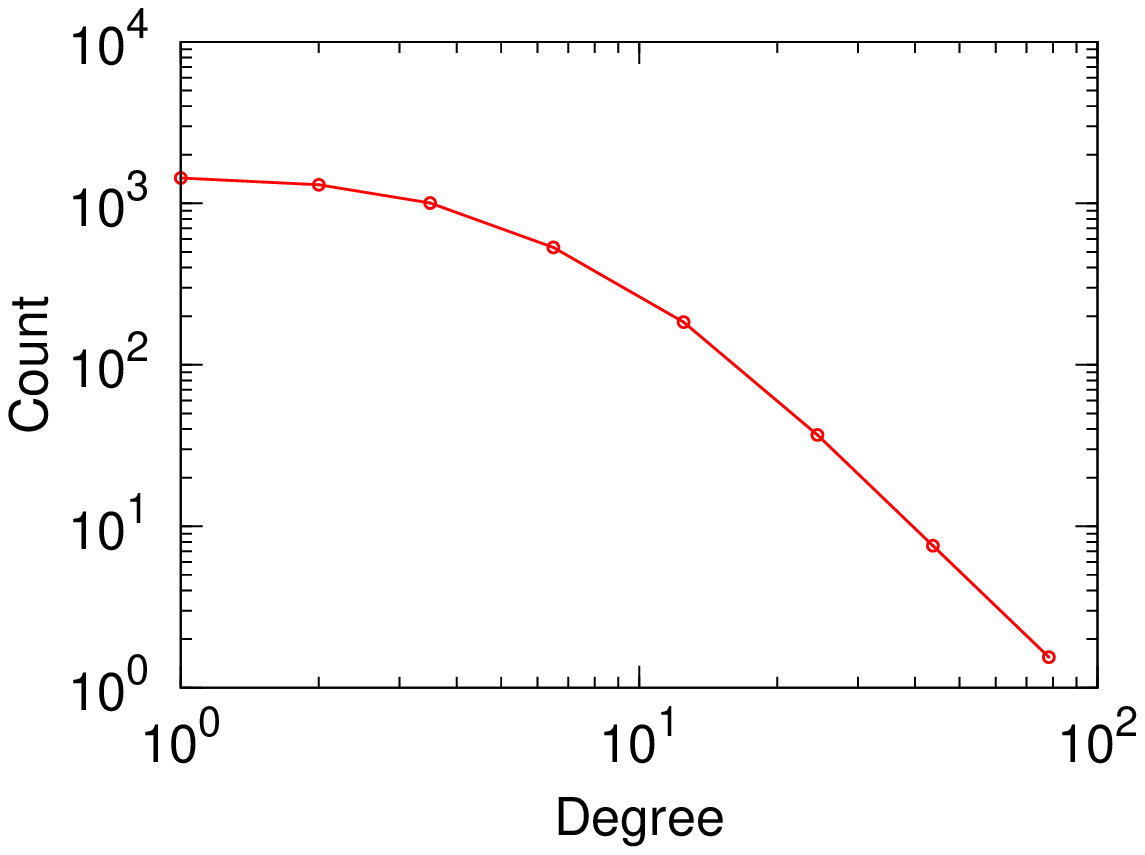} &
\includegraphics[width=0.25\textwidth]{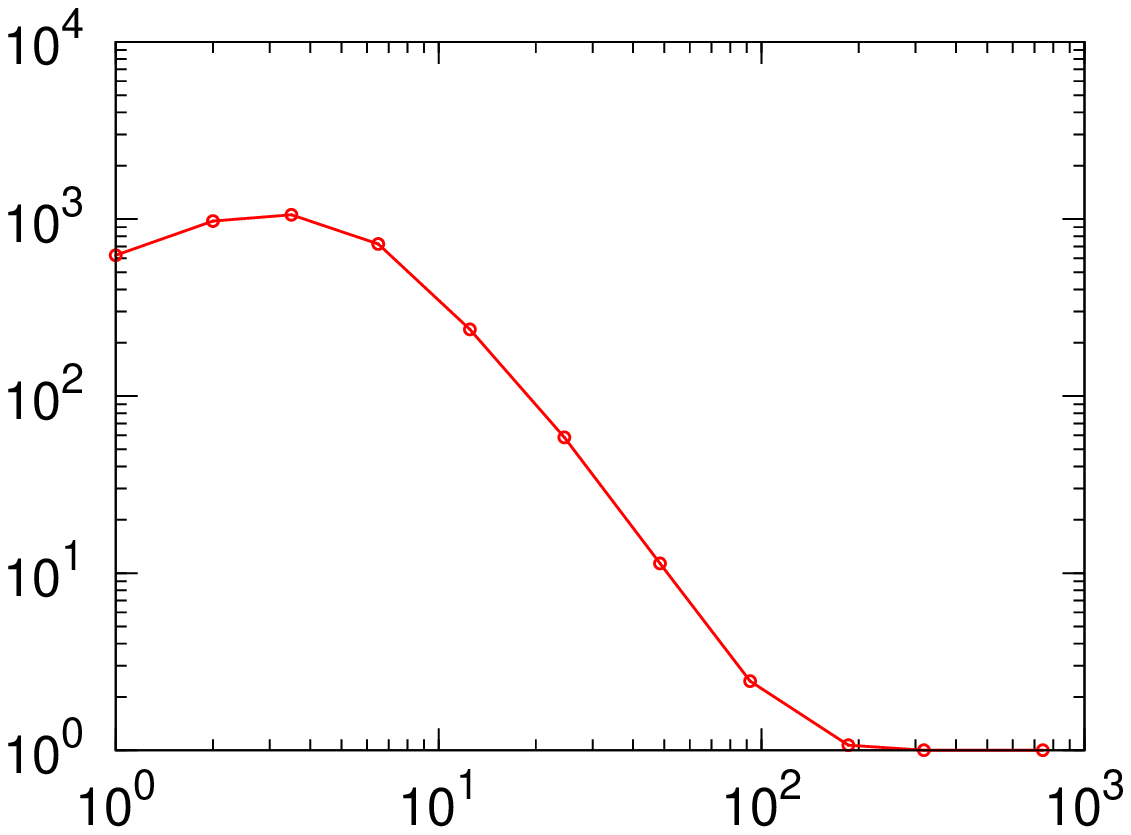} \\
\includegraphics[width=0.25\textwidth]{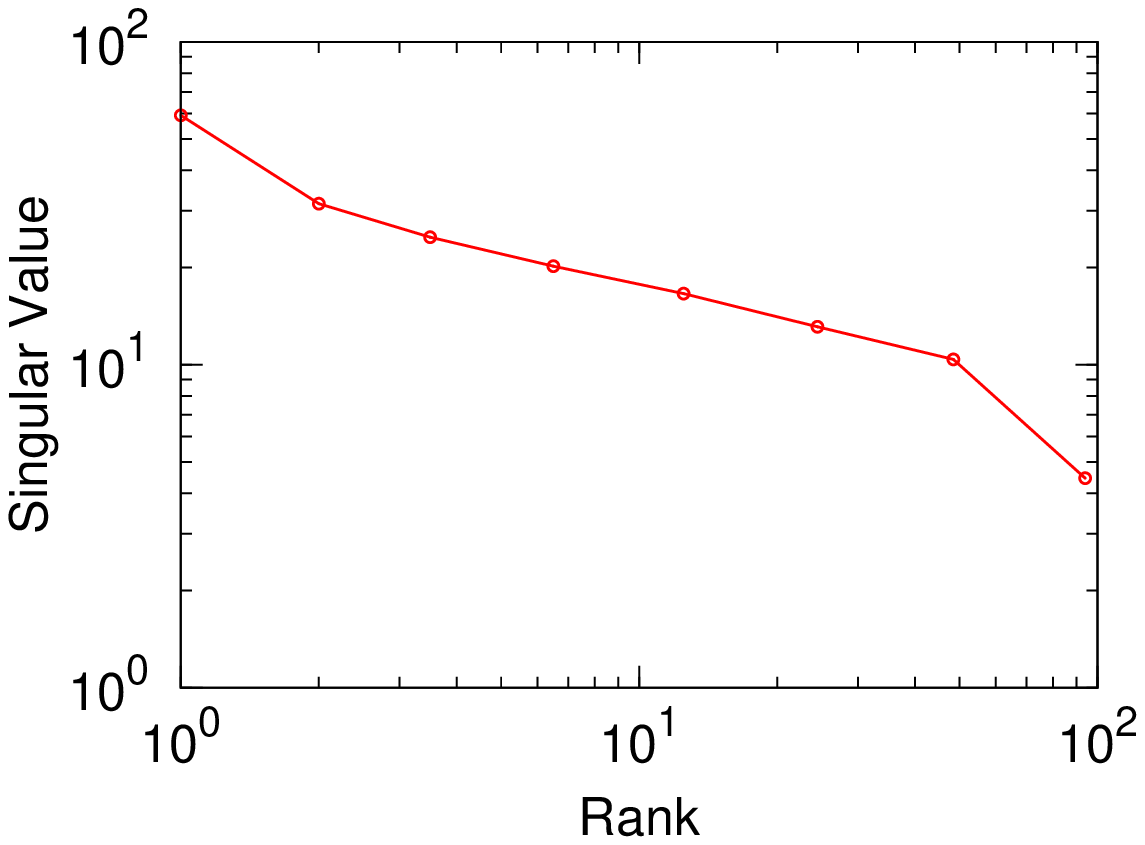} &
\includegraphics[width=0.25\textwidth]{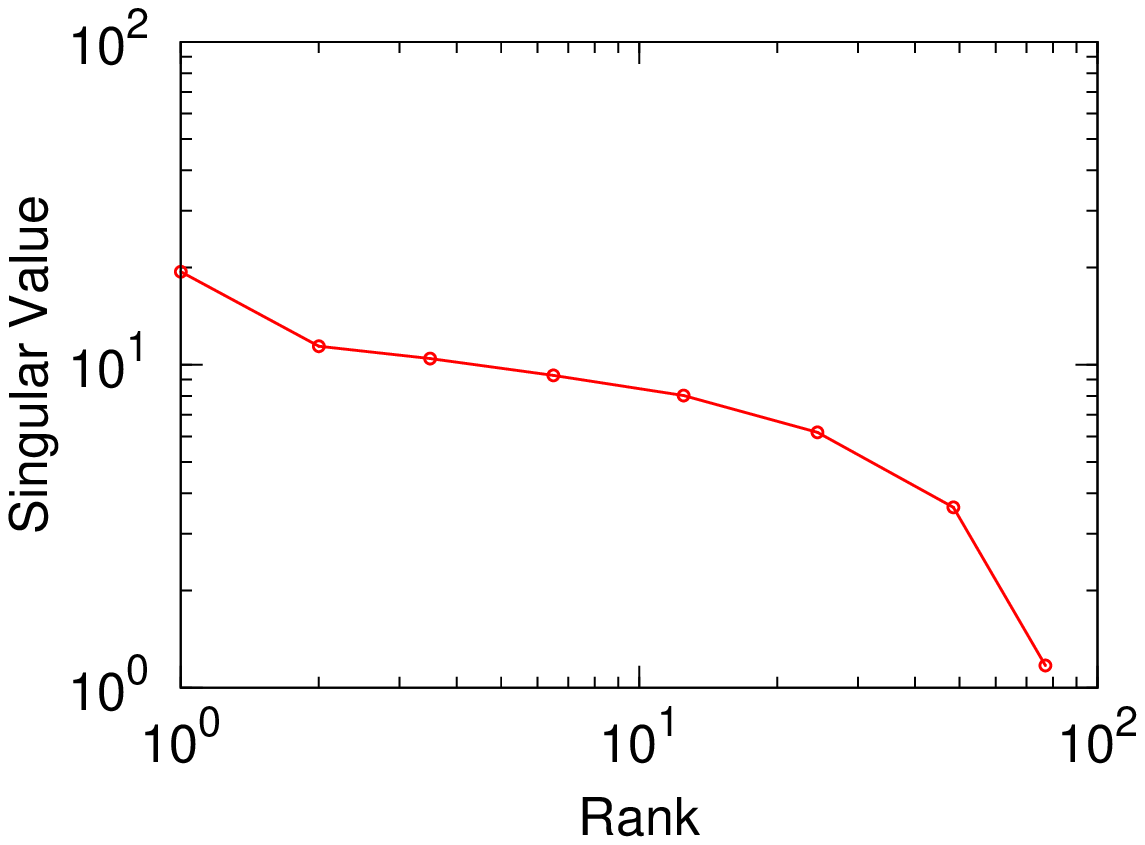} &
\includegraphics[width=0.25\textwidth]{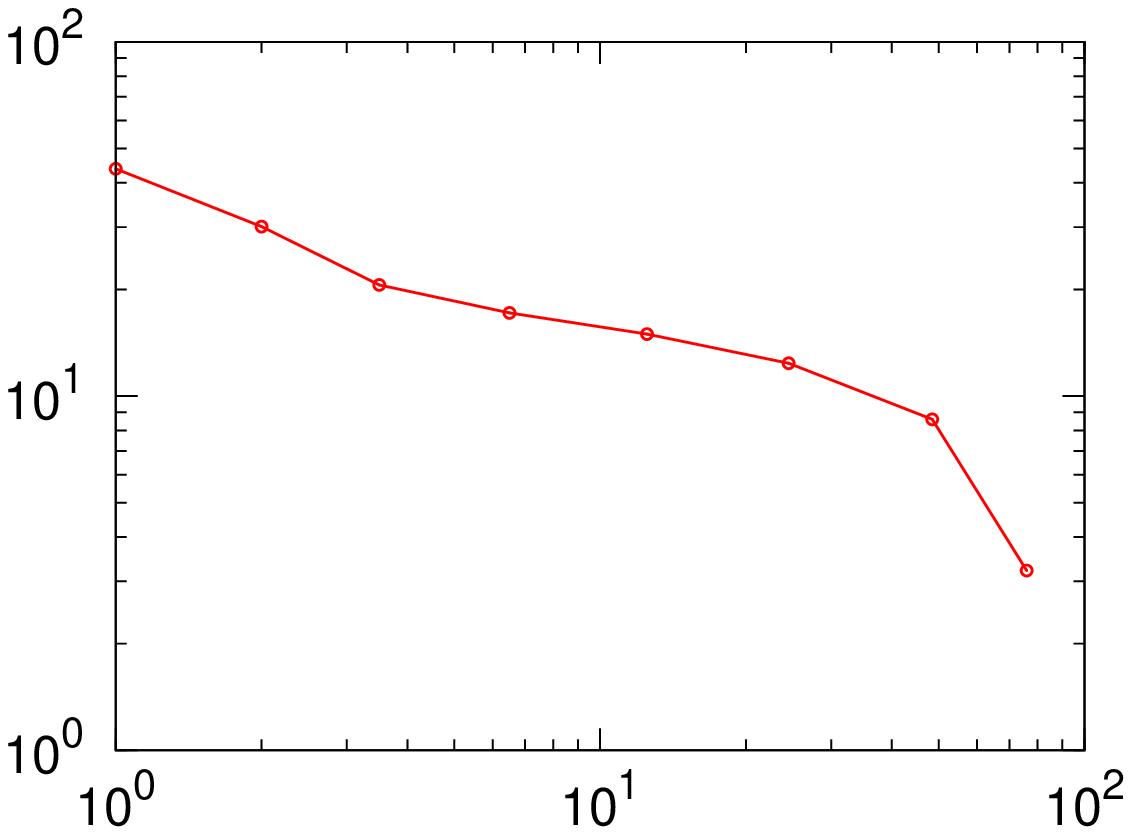} \\
\includegraphics[width=0.25\textwidth]{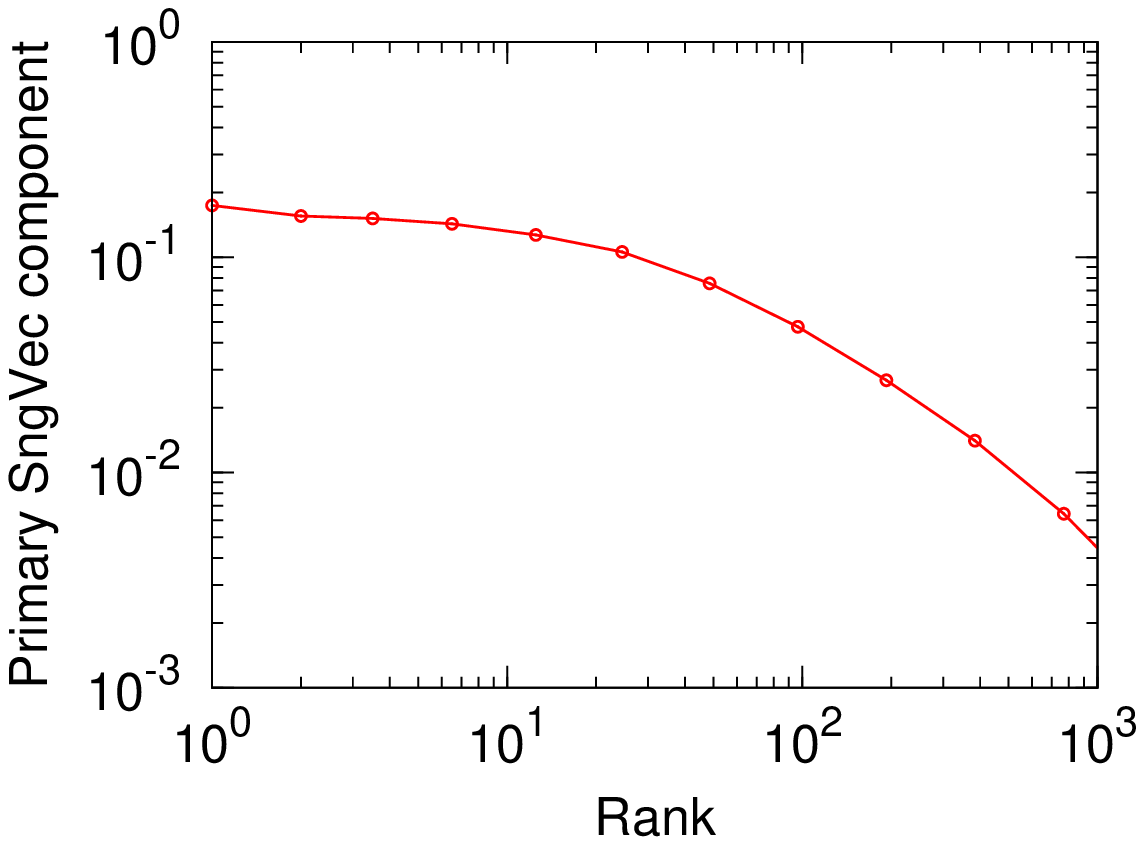} &
\includegraphics[width=0.25\textwidth]{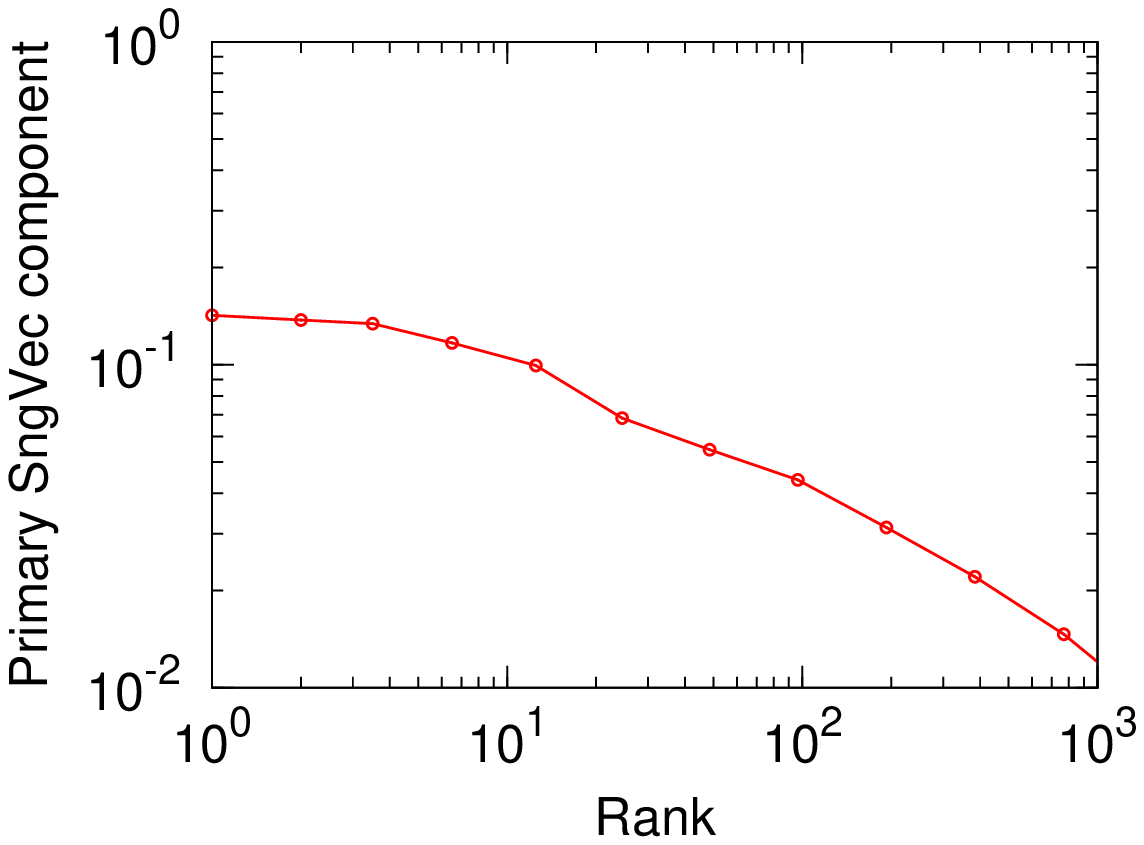} &
\includegraphics[width=0.25\textwidth]{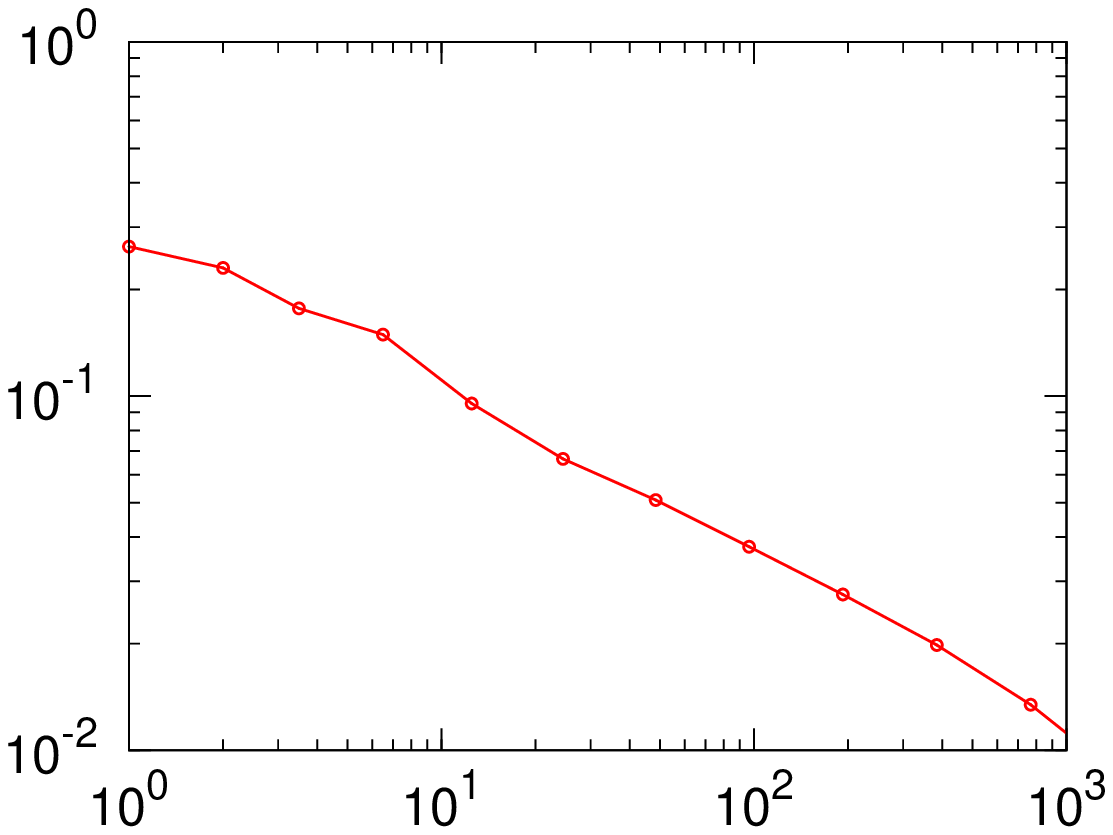} \\
\includegraphics[width=0.25\textwidth]{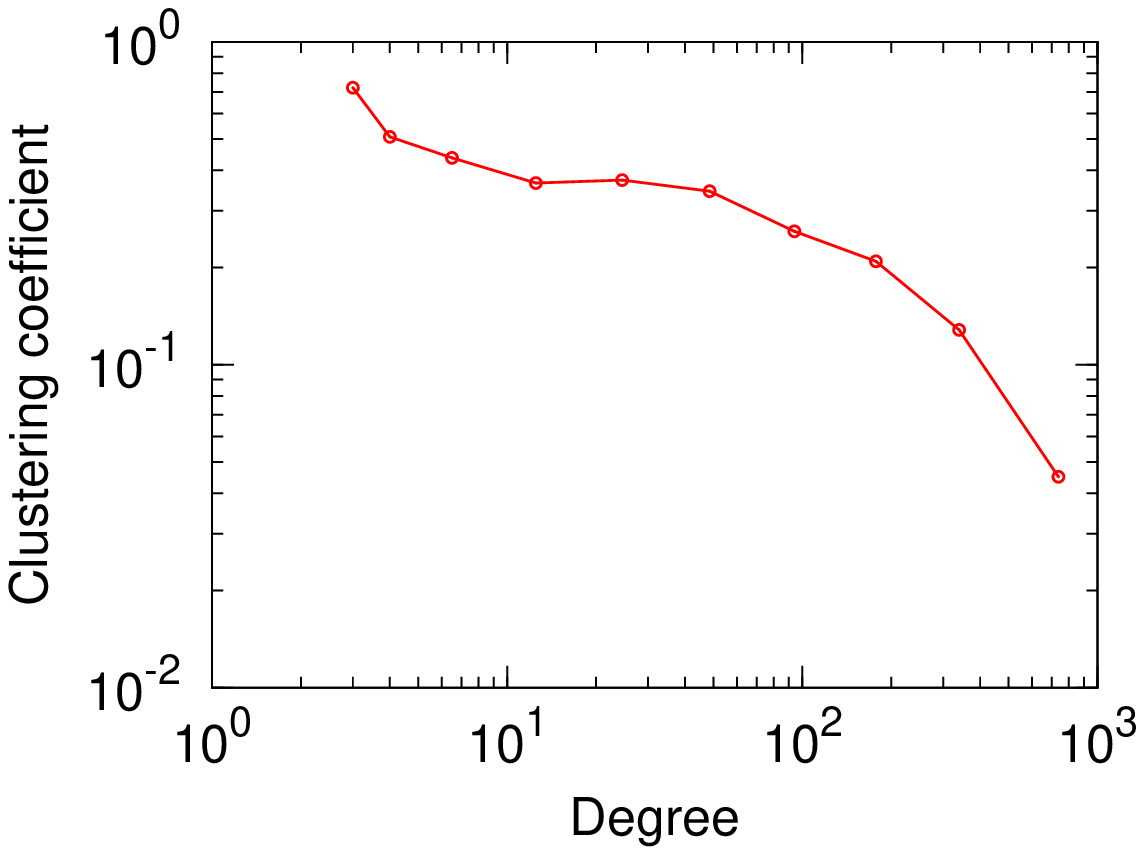} &
\includegraphics[width=0.25\textwidth]{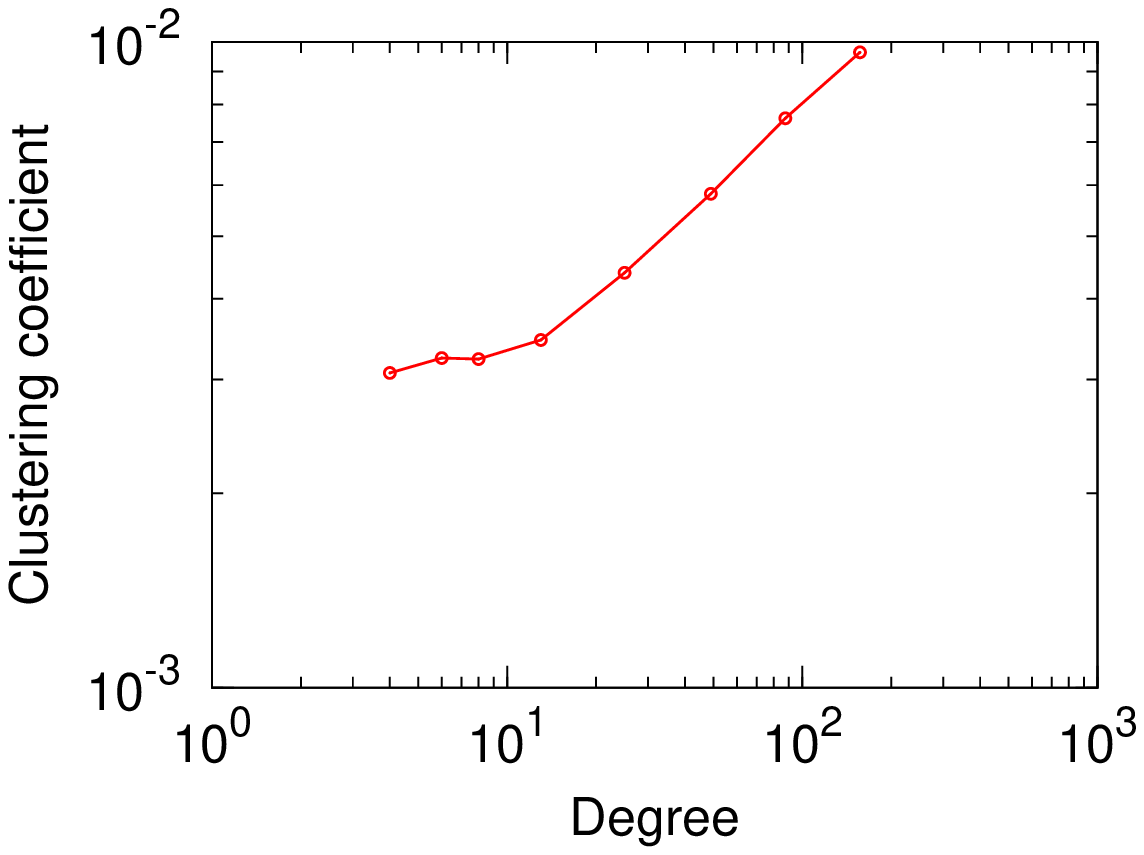} &
\includegraphics[width=0.25\textwidth]{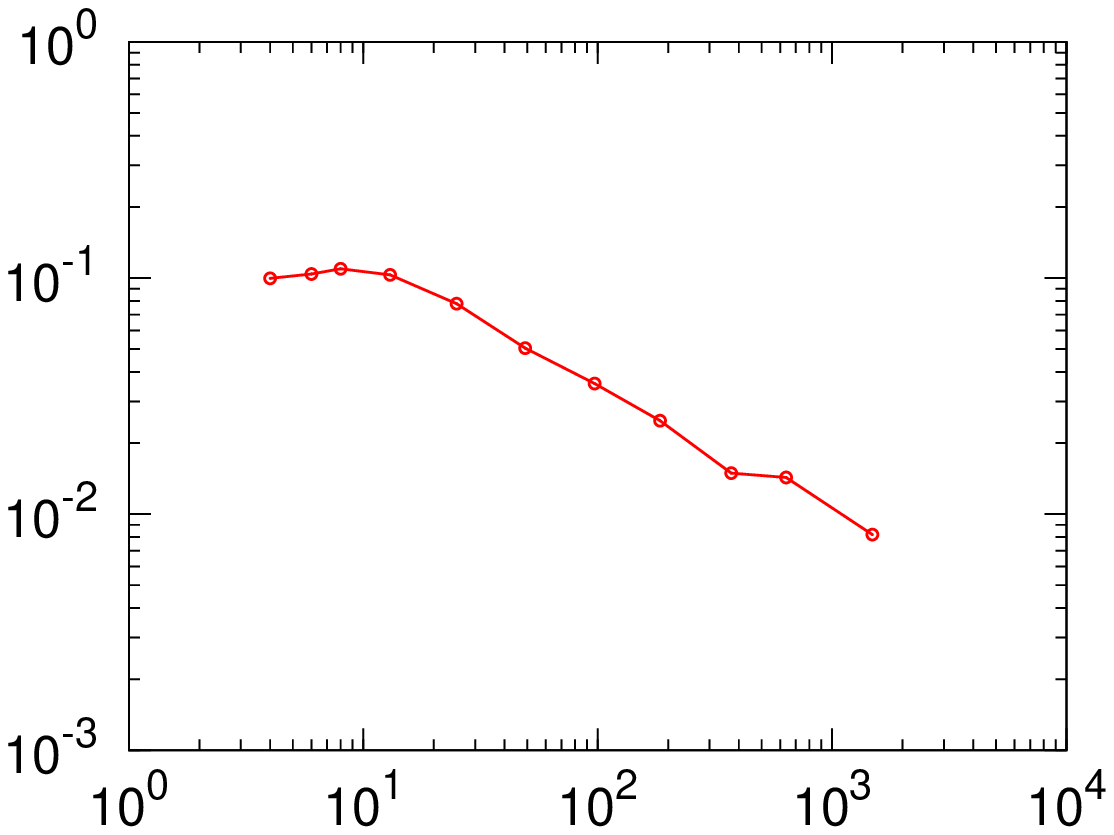} \\
\includegraphics[width=0.25\textwidth]{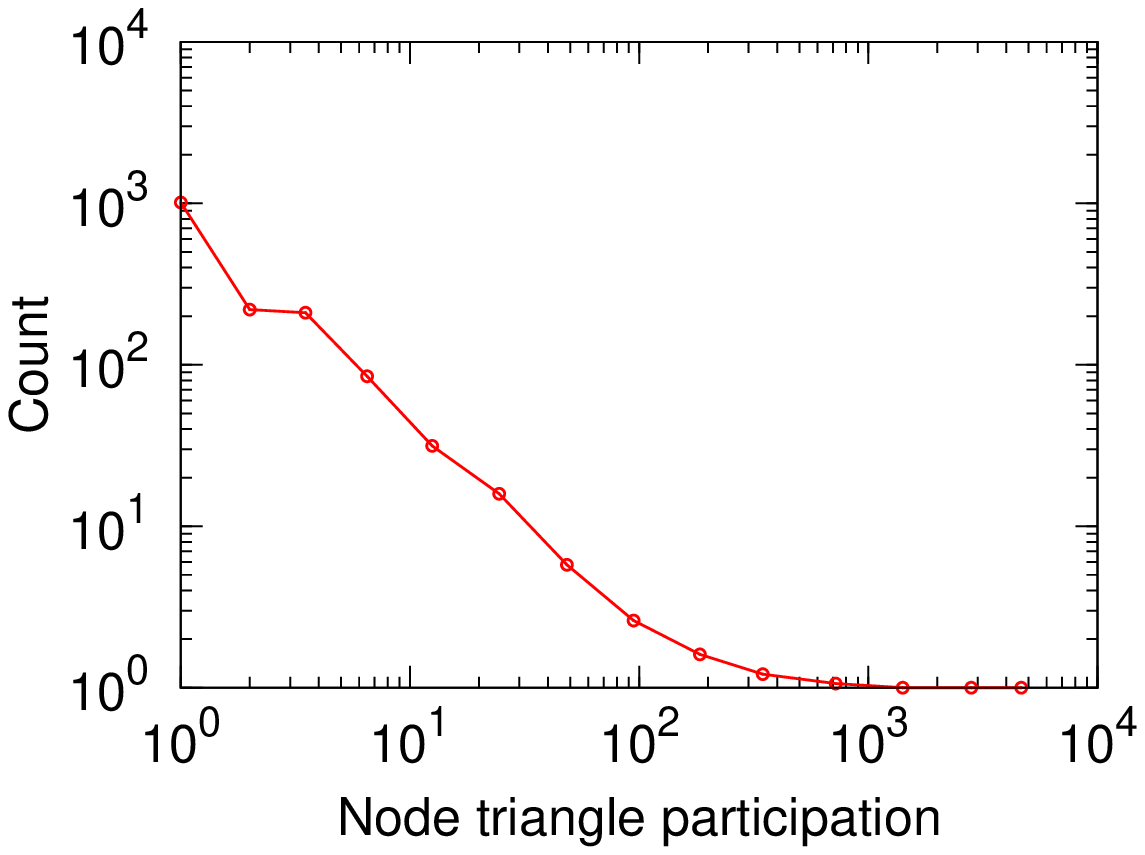} &
\includegraphics[width=0.25\textwidth]{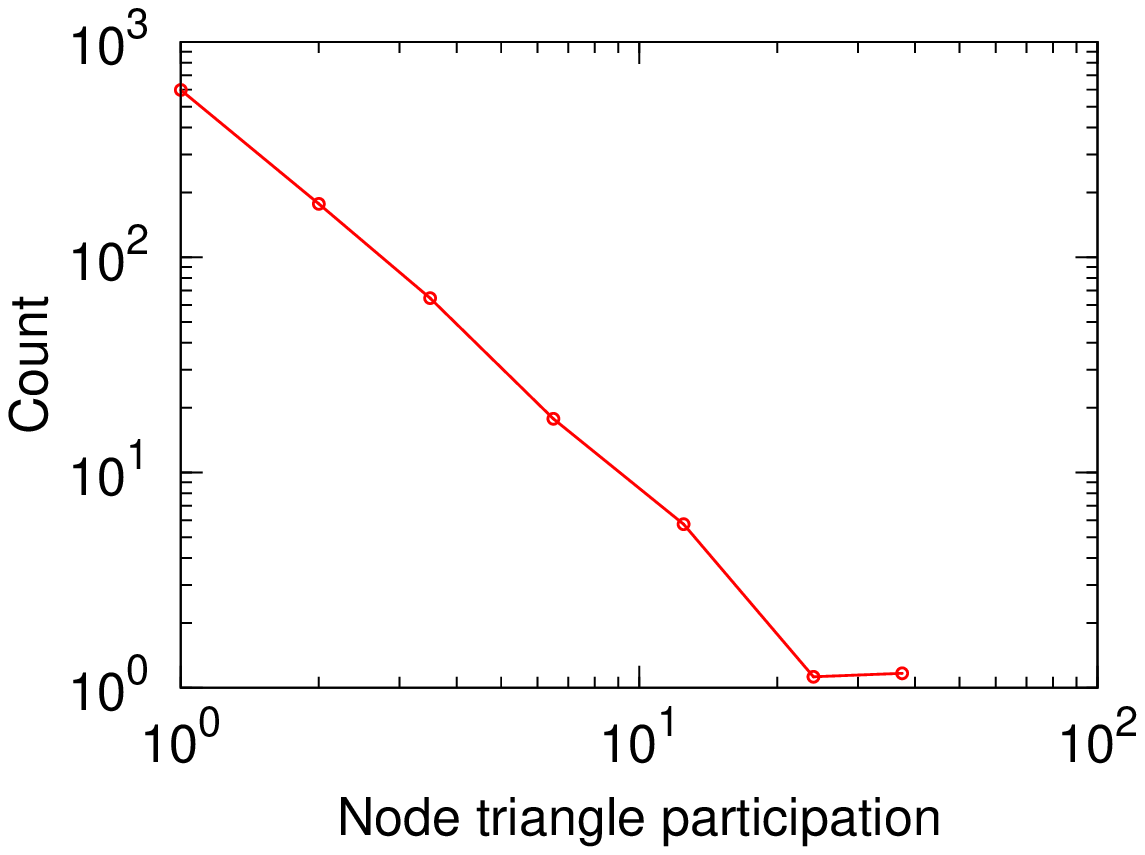} &
\includegraphics[width=0.25\textwidth]{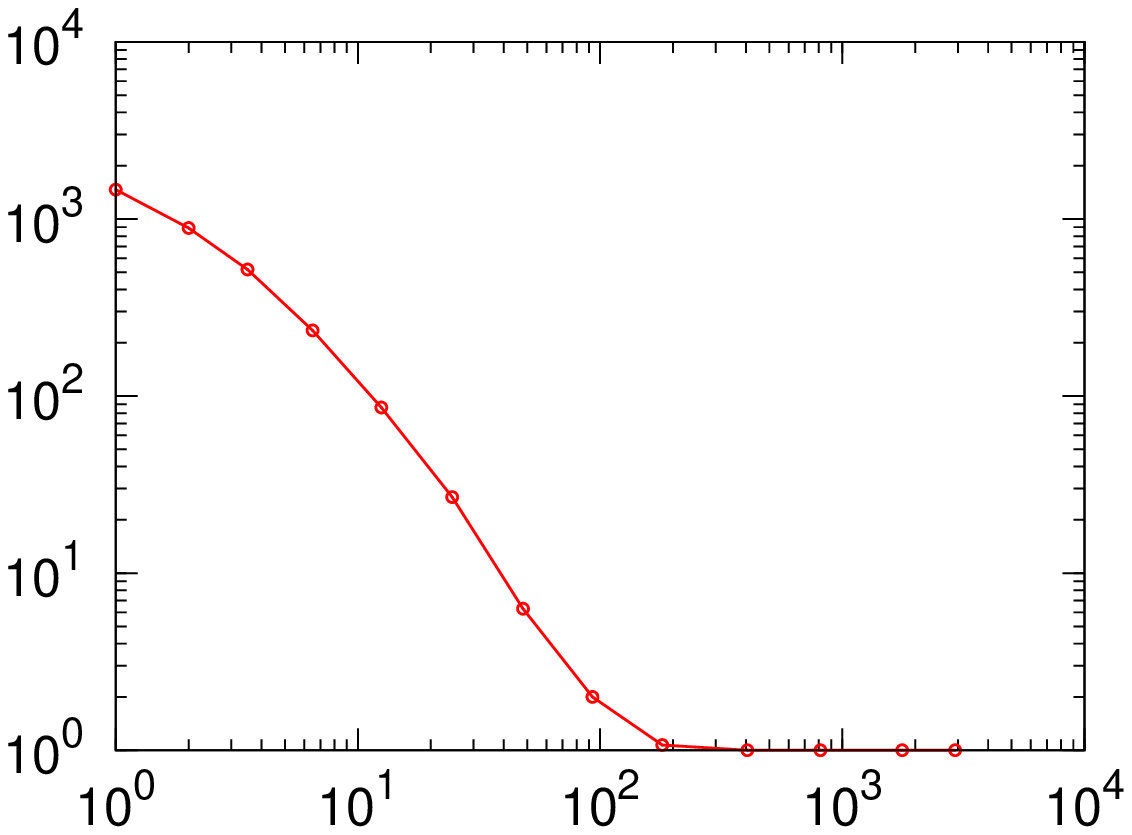} \\
\includegraphics[width=0.25\textwidth]{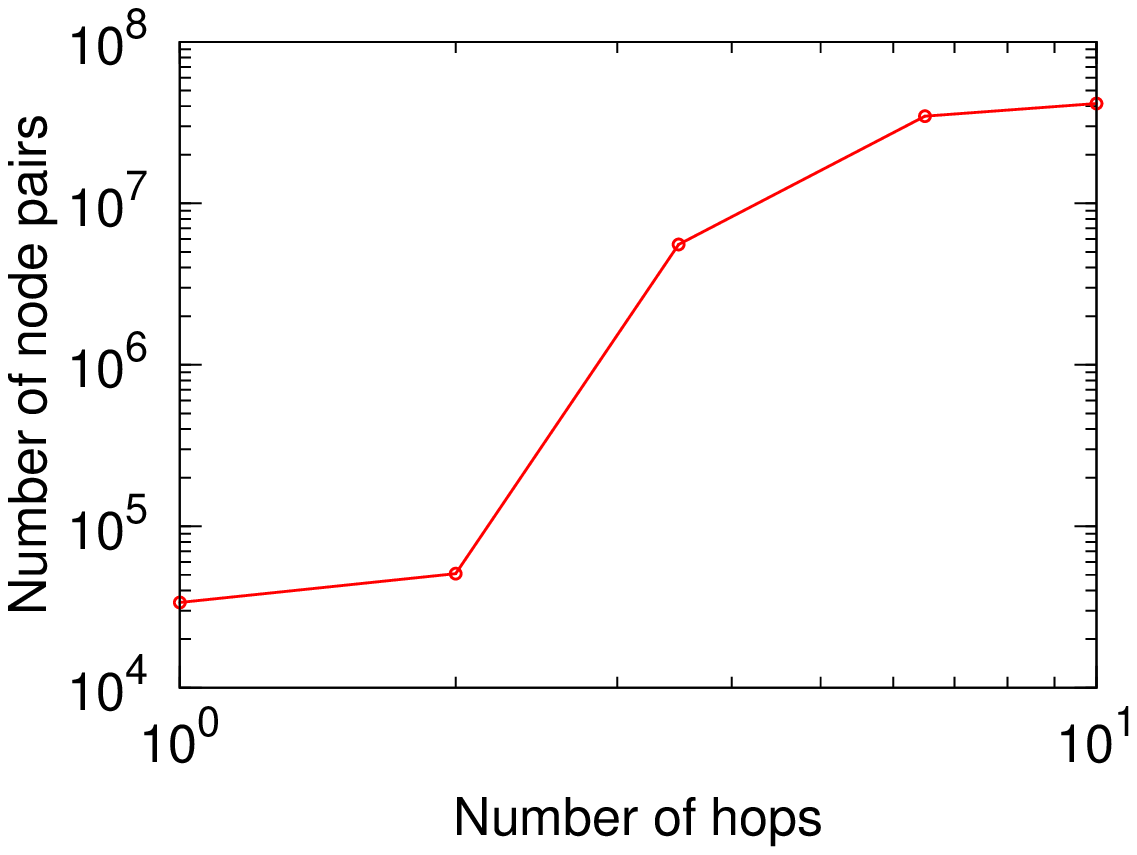} &
\includegraphics[width=0.25\textwidth]{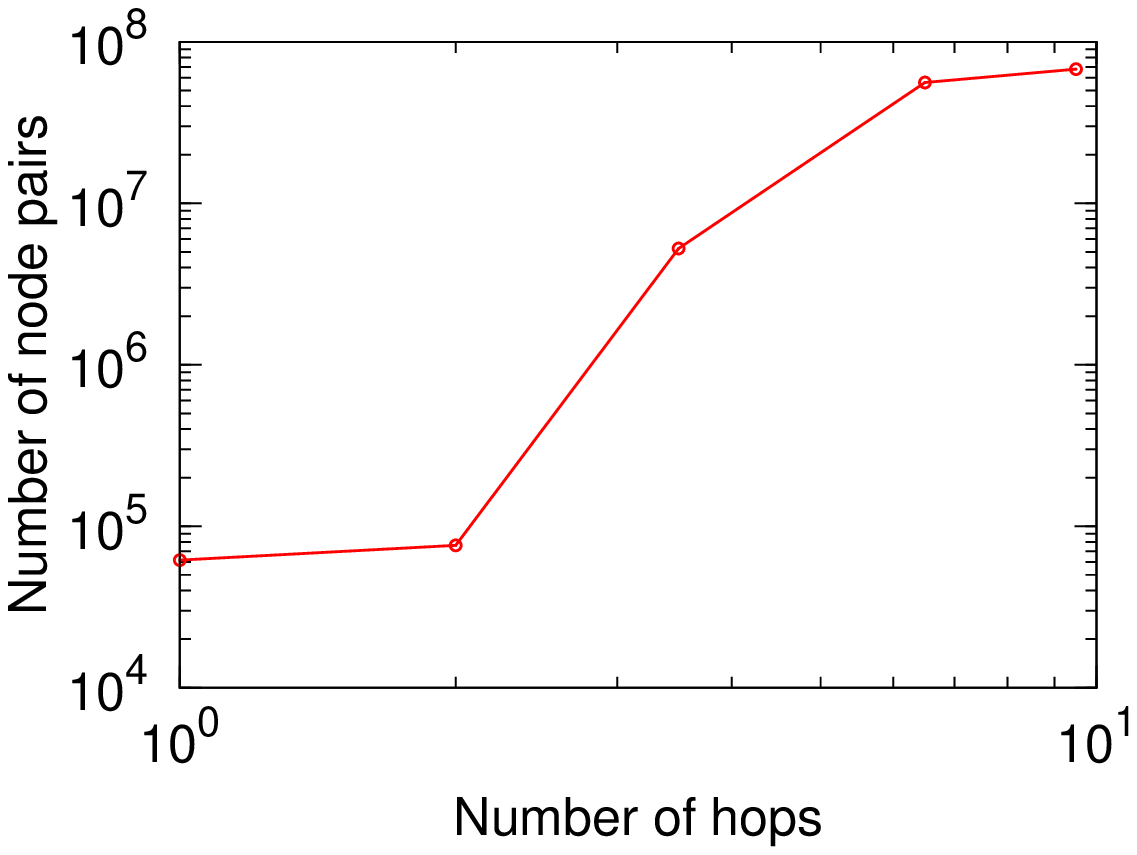} &
\includegraphics[width=0.25\textwidth]{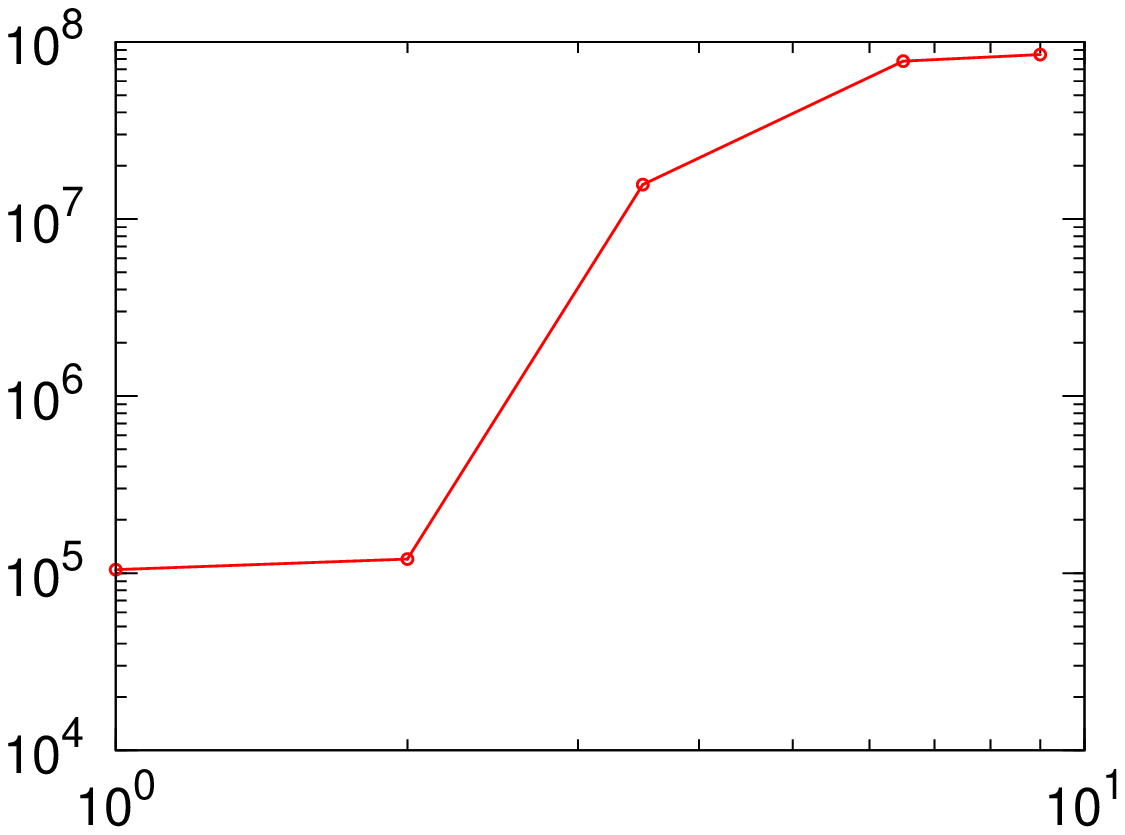} \\
(a) \social~network & (b) Simplified \modelx & (c) General \modelx \\
\end{tabular}
\caption{The comparison of network properties between real-world \social~online social network, a simplified \modelx~network, and a general version of \modelx. Except for clustering coefficient, the properties of MAG model qualitatively resemble those of the \social~network even when it is the simplified version in Figure~(b). Moreover, the general version of the \modelx~can represent all six network properties of similar shape to real-world networks in Figure~(c).}
\label{fig:netcomp}
\end{figure}

%
For the real-world network, we use the \social~online social network on 10,240 nodes and 44,800 edges.
For the simplified \modelx~\modeldesc, we used $l = 8, \mu = 0.45, \Theta = [0.85~0.30;0.30~0.25]$ with the same number of nodes $n = 10,240$.
Figure~\ref{fig:netcomp}(a) and (b) illustrate the following properties of the real-world and the corresponding synthetic network of the simplified \modelx~(in the same order of figures).

\begin{itemize}
  \item {\em Degree distribution} is a histogram of the
      number of edges of a node~\cite{faloutsos99powerlaw}.
  \item {\em Singular values} indicate the singular values of the
      adjacency matrix versus their rank~\cite{Spectra}.
  \item {\em Singular vector} represents the distribution of
      components in the left singular vector associated with the largest
      singular value~\cite{chakrabartiZF04}.
  \item {\em Clustering coefficient} represents the degree versus the
      average (local) clustering coefficient of nodes of a given
      degree~\cite{watts98smallworld}.
  \item {\em Triad participation} indicates the number of triangles
      that a node is adjacent to. It measures the transitivity in networks~\cite{Triad}.
  \item {\em Hop plot} shows the number of reachable pairs of nodes as the number of hops. It sketches how quickly the network expands~\cite{Palmer02anf,jure07kronfit}.
\end{itemize}

Figure~\ref{fig:netcomp} reveals that the plots of properities of \modelx~resemble those of \social~network. Notice qualitatively similar behavior of nearly all properties between Figure~\ref{fig:netcomp}(a) and (b). 
The only property where the simplified MAG model does not match the \social~network seems to be the clustering coefficient. 
As in real-world networks high degree nodes tend to have lower clustering, in the simplified MAG model the situation is reverse -- higher degree nodes also tend to have higher clustering. 
This is due to the fact that for all attributes we use the same affinity matrix $\Theta$ which represents only the core-perphery structure ($\alpha > \beta > \gamma$).  
Thus, the simpified \modelx~can only resemble the overall core-periphery shape of real-world networks.
However, in the \social~network,
we can also discover the local clustering effect of homophily and network community formation,
which views the network in the opposite way compared to the global core-periphery structure.

Hence, our hypothesis is that
the local clustering of nodes would naturally emerge
by mixing core-periphery affinity matrices ($\alpha>\beta>\gamma$) and homophily affinity matrices ($\alpha, \gamma > \beta$).
To investigate this, we also generated the synthetic network with more general version of \modelx, \modelpower.
Figure~\ref{fig:netcomp}(c) illustrates the network properties of this general version.
Note that this general version of the model nicely captures the heavy-tailed cluestering coefficient distribution that the real-world network shows 
while the simplified version cannot.
For the other properties, the general version still exhibits distributions 
which qualatatively seem similar to those of the real-world network.

By this experiment, we can find that \modelx~is capable of representing real-world networks.
Furthermore, we verify the flexibility of \modelx~in a sense that it can give rise to networks with different network properties depending on the \modelx~parameter configuration.

%% file: W060conclusion.tex
We presented the Multiplicative Attribute Graph model for real-world networks
which considers the categorical node attributes as well as the affinity of link formation depending on the values of node attributes.
We introduced the attribute-attribute affinity matrix to represent the affinity of link formation and 
provide the flexibility in the network structure.

On the other hand, the \modelx~is both analytically tractable and statistically interesting.
In this paper, we analytically showed several network properties observed in real-world networks.
We proved that the \modelx~obeys the Densification Power Law.
We also showed both the existence of unique giant connected component and a small diameter in the \modelx.
Furthermore, we mathematically analyzed that the \modelx~give rise to networks with either a log-normal or a power-law degree distribution.
Finally, we emprically verified our analytical results.

The \modelx~is statistically interesting in a sense that it can represent various types of network structure as well as 
lead a problem that aims to find such structures of the given real-world networks in terms of the \modelx~parameters.
However, we leave the parameter fitting problem as a venue of the future work.
Furthermore, future work includes other kinds of problems such as how to find underlying network structures and missing node attributes where node attributes are partially observed.

%% file: W070appendix.tex
\newcommand{\SN}[1]{N_{uv}^{#1}}
\newcommand{\Sn}[1]{n_{uv}^{#1}}

%
%
%
\section{Appendix: The Number of Edges}
~\\
\begin{appendproof}{lemma:expprob}
Let $\SN{0}$ be the number of attributes that take value $0$ in both $u$ and $v$. 
For instance, if $a(u) = [0~0~1~0]$ and $a(v) = [0~1~1~0]$, then $\SN{0} = 2$.
We similarly define $\SN{1}$ as the number of attributes that take value $1$ in both $u$ and $v$. 
Then, $\SN{0}, \SN{1} \geq 0$ and $\SN{0} + \SN{1} \leq l$ as $l$ indicates the number of attributes in each node. 

By definition of \modelx, the edge probability between $u$ and $v$ is 
\[
P[u, v] = \alpha^{\SN{0}} \beta^{l - \SN{0} - \SN{1}} \gamma^{\SN{1}} \,.
\]

Since both $\SN{0}$ and $\SN{1}$ are random variables, we need their conditional joint distribution to compute the expectation of the edge probability $P[u, v]$ given the weight of node $u$.
Note that $\SN{0}$ and $\SN{1}$ are independent of each other if the weight of $u$ is given.
Let the weight of $u$ be $i$, \ie~$u \in W_{i}$.
Since $u$ and $v$ can share value $0$ only for the attributes where $u$ already takes value $0$, 
$\SN{0}$ equivalently represents the number of heads in $i$ coin flips with probabiltiy $\mu$.
Therefore, $\SN{0}$ follows $Bin(i, \mu)$. 
Similarly, $\SN{1}$ follows $Bin(l-i, 1-\mu)$. 
Hence, their conditional joint probability is 
\begin{align*}
P(\SN{0}, \SN{1} | u \in W_{i}) & = {i \choose \SN{0}} \mu^{\SN{0}} (1-\mu)^{i-\SN{0}} {l-i \choose \SN{1}} \mu^{l-i-\SN{1}} (1 - \mu)^{\SN{1}} \, .
\end{align*}
Using this conditional probability, we can compute the expectation of $P[u, v]$ given the weight of $u$:
\begin{align*}
\ECP{u, v}{u \in W_{i}} & =  \EX{\alpha^{\SN{0}}\beta^{i-\SN{0}}\beta^{l-i-\SN{1}}\gamma^{\SN{1}} | u \in W_{i}} \\
& =  \sum_{\SN{0}=0}^{i} \sum_{\SN{1}=0}^{l-i} {i \choose \SN{0}}{l-i \choose \SN{1}} (\alpha \mu)^{\SN{0}} \left((1-\mu)\beta\right)^{i-\SN{0}} (\mu \beta)^{l-i-\SN{1}} \left((1-\mu)\gamma\right)^{\SN{1}} \\
& =  \left[ \sum_{\SN{0}=0}^{i} {i \choose \SN{0}}(\alpha \mu)^{\SN{0}} \left((1-\mu)\beta\right)^{i-\SN{0}}\right] \left[ \sum_{\SN{1} = 0}^{l-i} {l-i \choose \SN{1}} (\mu \beta)^{l-i-\SN{1}} \left((1-\mu)\gamma\right)^{\SN{1}} \right] \\
& =  \bkron{i} \, .
\end{align*}
\end{appendproof}

~\\
\begin{appendproof}{lemma:expdeg}
By Lemma~\ref{lemma:expprob} and the linearity of expectation, we sum this conditional probability over all nodes and result in the expectation of the degree given the weight of node $u$.
\end{appendproof}

~\\
\begin{appendthmproof}{thm:expedge}
We compute the number of edges, $\EX{m}$, by adding up the degrees of all nodes described in Lemma~\ref{lemma:expdeg},
\begin{align*}
\EX{m} & =  \EX{\frac{1}{2} \sum_{u \in V} deg(u)} \\
& = \frac{1}{2} n \sum_{j=0}^{l} P(W_{j}) \EX{deg(u) | u \in W_{j}} \\
& = \frac{1}{2} n \sum_{j=0}^{l} \EN{j} \EX{deg(u) | u \in W_{j}} \\
& = \frac{1}{2} n \sum_{j=0}^{l} {l \choose j} \left( (n-1)\bkron{j} + 2\alpha^{j}\mu^{j}\gamma^{l-j}(1-\mu)^{l-j}\right) \\
& = \frac{n(n-1)}{2} \left(\mu^{2}\alpha + 2 \mu(1-\mu)\beta + (1-\mu)^{2} \gamma\right)^{l} + n \left(\mu \alpha + (1-\mu) \gamma\right)^{l} \, .
\end{align*}
\end{appendthmproof}

~\\
\begin{appendcorproof}{cor:landn}
Suppose that $l = \left(\epsilon - \frac{1}{ \log \zeta}\right) \log n$~for $\zeta = \mu^{2} \alpha + 2\mu(1-\mu)\beta + (1-\mu)^{2}\gamma$~and $\epsilon > 0$.
By Theorem~\ref{thm:expedge}, the expected number of edges is $\Theta\left(n^{2} \zeta^{l}\right)$. 
Note that $\log \zeta < 0$~since $\zeta < 1$.
Therefore, the expected number of edges is 
\[
\Theta(n^{2} \zeta^{l}) = \Theta\left(\zeta^{l + \frac{2\log n}{\log \zeta}}\right) = \Theta(n^{1 + \epsilon \log \zeta }) = o(n) \, .
\]
\end{appendcorproof}

~\\
\begin{appendcorproof}{cor:landn2}
Under the situation that $l \in o(\log n)$, the expected number of edges is
\[
\Theta(n^{2} \zeta^{l}) = \Theta(n^{2 + (\frac{l}{\log n}) \log \zeta} ) = \Theta(n^{2 + o(1) \log \zeta}) = \Theta(n^{2 - o(1)}) \, .
\]
\end{appendcorproof}


\section{Appendix: Connectivity}

Since Theorem~\ref{thm:mono} is used to prove other theorems, we begin with the proof of it.

~\\
\begin{appendthmproof}{thm:mono}
If $j \geq i$, for any $v \in W_{i}$, we can generate a node $v^{(j)} \in W_{j}$ from $v$ by flipping $(j-i)$ attribute values that originally take $1$ in $v$. 
For example, if $a(v) = [0~1~1~0]$, then $a(v^{(3)}) = [0~0~1~0]$ or $[0~1~0~0]$.
Hence, $P[u, v^{(j)}] \geq P[u, v]$ for $v \in W_{i}$.

Here we note that $\ECP{u, v^{(j)}}{v \in W_{i}} = \ECP{u, v^{(j)}}{v^{(j)} \in W_{j}}$,
because each $v^{(j)}$ can be generated by $j \choose i$ different $a(v)$ sets with the same probability.
Therefore,
\[
\ECP{u, v}{v \in W_{j}} = \EX{\ECP{u, v^{(j)}}{v \in W_{i}}} 
\geq \EX{\ECP{u, v}{v \in W_{i}}} 
= \ECP{u, v}{v \in W_{i}} \, .
\]
\end{appendthmproof}


Next theorem plays a key role in proving Theorem~\ref{thm:giant} as well as Theorem~\ref{thm:conn}.

\begin{theorem}
Let $|S_{j}| \in \Theta(n)$ and $\ECP{u, V \xset u}{u \in W_{j}} \geq c \log n$ as $\LGN$~~for some $j$ and sufficiently large $c$.
Then, $S_{j}$ is connected with high probability as $\LGN$.
\label{thm:connected}
\end{theorem}
\begin{proof}
Let $S'$ be a subset of $S_{j}$ such that $S'$ is neither an empty set nor $S_{j}$ itself.
Then, the expected number of edges between $S'$ and $S_{j} \xset S'$ is 
\[
\ECP{S', S_{j} \xset S'}{~|S'| = k} = k \cdot (|S_{j}| - k) \cdot\ECP{u, v}{u, v \in S_{j}}
\]
for distinct $u$ and $v$.
By Theorem~\ref{thm:mono},
\begin{align*}
\ECP{u, v}{u, v \in S_{j}} & \geq \ECP{u, v}{u \in S_{j}, v \in V}\\
& \geq \ECP{u, v}{u \in W_{j}, v \in V \xset u}\\
& \geq \frac{c \log n}{n} \, .
\end{align*}
Given the size of $S'$ as $k$, the probability that there exists no edge between $S'$ and $S_{j} \xset S'$ is at most $\exp \left(-\frac{1}{2}\ECP{S', S_{j} \xset S'}{|S'| = k}\right)$ by Chernoff bound.
Therefore, the probability that $S_{j}$ is disconnected is bounded as follows:
\begin{align*}
P(S_{j}\mathrm{~is~disconnected}) & \leq \sum_{S' \subset S_{j}, S' \neq \emptyset, S_{j}} P(\mathrm{no~edge~between~}S', S_{j} \xset S')\\
& \leq \sum_{S' \subset S, S' \neq \emptyset, S_{j}} \exp \left(-\frac{1}{2}\ECP{S', S_{j} \xset S'}{|S'|}\right) \\
& \leq \sum_{S' \subset S, S' \neq \emptyset, S_{j}} \exp \left(-|S'| \left(|S_{j}| - |S'|\right)\frac{c \log n}{2n}\right)\\
& \leq 2 \sum_{1 \leq k \leq |S_{j}|/2} {|S_{j}| \choose i} \exp \left(-\frac{c |S_{j}| \log n}{4n}  k\right)\\
& \leq 2 \sum_{1 \leq k \leq |S_{j}|/2} |S_{j}|^{k}  \exp \left(-\frac{c |S_{j}| \log n}{4n}  k\right)\\
& \leq 2 \sum_{1 \leq k \leq |S_{j}|/2} \exp \left( \left( \log |S_{j}| - \frac{c |S_{j}| \log n}{4n} \right) k \right)\\
& = 2 \sum_{1 \leq k \leq |S_{j}|/2} \exp \left(- k \Theta(\log n) \right) \quad \quad (\because |S_{j}| \in \Theta(n)) \\
& = 2 \sum_{1 \leq k \leq |S_{j}|/2} \left(\frac{1}{n^{\Theta(1)}}\right)^{k} \\
& \approx \frac{1}{n^{\Theta(1)}}  \in o(1)
\end{align*}
as $\LGN$.
Therefore, $S_{j}$ is connected with high probability.
\end{proof}

~\\
Now we turn our attention to the giant connected component.
To show its existence, we investigate $S_{\mu l}$, $S_{\mu l + l^{1/6}}$, and $S_{\mu l + l^{2/3}}$ depending on the situation.
The following lemmas tell us the size of each subgraph.
~\\
\begin{lemma}
\label{lemma:lccsize1}
$|S_{\mu l}| \geq \frac{n}{2} - o(n)$ \WHP~as $\LGN$.
\end{lemma}
\begin{proof}
By Central Limit Theorem, $\frac{|u| - \mu l}{\sqrt{l \mu (1-\mu)}} \sim N(0, 1)$ as $\LGN$, \ie~$l \rightarrow \infty$. Therefore, $P(|u| \geq \mu l)$ is at least $\frac{1}{2} - o(1)$ so $|S_{\mu l}| \geq \frac{n}{2} - o(n)$ with high probability as $\LGN$.
\end{proof}
~\\
\begin{lemma}
\label{lemma:lccsize3}
$|S_{\mu l + l^{1/6}}| \in \Theta(n)$ \WHP~as $\LGN$.
\end{lemma}
\begin{proof}
By Central Limit Theorem mentioned in Lemma~\ref{lemma:lccsize1},
\[
P( \mu l \leq |u| < \mu l + l^{1/6} ) \approx \Phi ( \frac{l^{1/6}}{\sqrt{l \mu (1-\mu)}} ) - \Phi(0) \in o(1)
\]
as $l \rightarrow \infty$ where $\Phi(z)$ represents the cdf of the standard normal distribution.
\\
Since $P(|u| \geq \mu l + l^{1/6})$ is still at least $\frac{1}{2} - o(1)$, the size of $S_{\mu l + l^{1/6}}$ is $\Theta(n)$ with high probability as $l \rightarrow \infty$, \ie~$\LGN$.
\end{proof}
~\\
\begin{lemma}
\label{lemma:lccsize2}
$|S_{\mu l + l^{2/3}}| \in o(n)$ \WHP~as $\LGN$.
\end{lemma}
\begin{proof}
By Chernoff bound, $P(|u| \geq \mu l + l^{2/3})$ is $o(1)$ as $l \rightarrow \infty$, thus $|S_{\mu l + l^{2/3}}|$ is $o(n)$ with high probability as $\LGN$.
\end{proof}

~\\
Using the above lemmas, we show the existence and the uniqueness of the giant connected component under the given condition.

~\\
\begin{appendthmproof}{thm:giant}
\xhdr{(Existence)}
First, if~$\crilcc > \HALF$, then by Lemma~\ref{lemma:expdeg},
\[
\ECP{u, V \xset u}{u \in W_{\mu l}} \approx \left[ 2 \crilcc \right]^{\log n} = (1+\epsilon)^{\log n} > c \log n
\]
for some constant $\epsilon > 0$ and $c > 0$.
Since $|S_{\mu l}| \in \Theta(n)$ by Lemma~\ref{lemma:lccsize1}, 
$S_{\mu l}$ is connected with high probability as $\LGN$~by Theorem~\ref{thm:connected}.
In other words, we are able to extract out a connected component of size at least $\frac{n}{2} - o(n)$.

Second, when~$\crilcc = \HALF$, we can apply the same argument for $S_{\mu l + l^{1/6}}$.
Because $|S_{\mu l + l^{1/6}}| \in \Theta(n)$ by Lemma~\ref{lemma:lccsize3},
\begin{align*}
& ~ \ECP{u, V \xset u}{u \in W_{\mu l + l^{1/6}}} \\
& \approx  \left[ 2 \crilcc \right]^{\log n} \left( \frac{\lwokron}{\swokron}\right)^{(\rho \log n)^{1/6}} \\
& = \left( \frac{\lwokron}{\swokron}\right)^{(\rho \log n)^{1/6}} \\
& =  (1 + \epsilon')^{{\rho \log n}^{1/6}}
\end{align*}
which is also greater than $c \log n$ as $\LGN$~for some constant $\epsilon' > 0$.
Thus, $S_{\mu l + l^{1/6}}$ is connected with high probability by Theorem~\ref{thm:connected}.

Last, on the contrary, when~$\crilcc < \HALF$, 
for $u \in W_{\mu l + l^{2/3}}$,
\begin{align*}
& ~ \ECP{u, V \xset u}{u \in W_{\mu l + l^{2/3}}} \\
& \approx  \left[ 2 \crilcc \right]^{\log n} \left( \frac{\lwokron}{\swokron}\right)^{(\rho \log n)^{2/3}} \\
& =  \left[ \left( 1 - \epsilon''\right)^{\rho^{-2/3} (\log n)^{1/3}}\left(\frac{\lwokron}{\swokron}\right)\right]^{(\rho \log n)^{2/3}} 
\end{align*}
is $o(1)$ as $\LGN$ for some constant $\epsilon'' > 0$.
Therefore, by Theorem~\ref{thm:mono}, the expected degree of a node with weight less than $\mu l + l^{2/3}$ is $o(1)$. 
However, since $S_{\mu l + l^{2/3}}$ is $o(n)$ by Lemma~\ref{lemma:lccsize2}, $n - o(n)$ nodes have less than $\mu l + l^{2/3}$ weights. 
Hence, most of $n - o(n)$ nodes are isolated so that the size of the largest component cannot be $\Theta(n)$.

\xhdr{(Uniqueness)}
We already pointed out that either $S_{\mu l}$ or $S_{\mu l + l^{1/6}}$ is the subset of $\Theta(n)$ component when the giant connected component exists.
Let this component be $H$.
Without loss of generality, suppose that $S_{\mu l} \subset H$. Then, for any fixed node $u$, 
\begin{align*}
P[u, H] & \geq P[u, S_{\mu l}] \quad \quad (\because S_{\mu l} \subset H) \\
& = |S_{\mu l}| \cdot \ECP{u, v}{v \in S_{\mu l}} \\
& \geq |S_{\mu l}| \cdot \ECP{u, v}{v \in V \xset S_{\mu l}} \quad \quad (\mbox{By Theorem}~\ref{thm:mono}) \\
& = \frac{|S_{\mu l}|}{n - |S_{\mu l}|} P[u, V \xset S_{\mu l}]
\end{align*}

Since $V \xset H \subset V \xset S_{\mu l}$,
\[
\EP{u, V \xset H} \leq \EP{u, V \xset S_{\mu l}} \leq \left(\frac{n - |S_{\mu l}|}{|S_{\mu l}|}\right) \EP{u, H}
\]
holds for every $u \in V$.

Suppose that another connected component $H'$ also contains $\Theta(n)$ nodes.
We will show the condtradiction if $H$ and $H'$ are not connected with high probability as $\LGN$.
To see $\EP{H, H'}$,
\begin{align*}
\EP{H, H'} & = |H'| \cdot \ECP{u, H}{u \in H'}\\
& \geq \frac{|H'| \cdot |S_{\mu l}|}{n-|S_{\mu l}|} \ECP{u, V \xset S_{\mu l}}{u \in H'}\\
& \geq \frac{|H'| \cdot |S_{\mu l}|}{n-|S_{\mu l}|} \ECP{u, H'}{u \in H'} \quad \quad (\because H' \subset V \xset H \subset V \xset S_{\mu l}) \, .
\end{align*}
However, $\ECP{u, H'}{u \in H'} \in \Omega (1)$.
Otherwise, since the probability that $u \in H'$ is connected to $H'$ is not greater than $\ECP{u, H'}{u \in H'}$ by Markov Inequality,
$u$ is disconnected from $H'$ with high probability as $\LGN$.
$H'$ thus includes at least one isolated node with high probability as $\LGN$.
This is contradiction to the connectedness of $H'$.

On the other hand, if~$\ECP{u, H'}{u \in H'} \in \Omega(1)$, then $\EP{H, H'} \in \Omega (n)$.
In this case, by Chernoff bound, $H$ and $H'$ are connected with high probability as $\LGN$.
This is also contradiction.
Therefore, there is no $\Theta(n)$ connected component other than $H$ with high probability as $\LGN$.
\end{appendthmproof}

~\\
Next, the proofs for the connectedness follow.
Before the main proof, we present some necessary lemmas and prove them.
~\\
\begin{lemma}
\label{lemma:entropy}
$\RATIO{x}$ is a monotonically increasing function of $x$ over $(0, \mu)$.
\end{lemma}
\begin{proof}
Let $f(x)$ be the log-value of the given function, \ie
\[
f(x) = x \left(\log \mu - \log x\right) + \left(1-x\right) \left(\log \left(1-\mu\right) - \log \left( 1- x \right)\right) \, .
\]
To take the derivative of $f(x)$,
\[
f'(x) = \left( \log \mu - \log x\right) + \left( \log \left(1-x\right) - \log \left(1-\mu\right) \right) \, .
\]
Since $x < \mu$ and $1-\mu < 1-x$, ~$f'(x) > 0$ where $0 < x < \mu$. This implies that $f(x)$ is strictly increasing, so the given function is also strictly increasing over $(0, \mu)$.
\end{proof}

~\\
\begin{lemma}
\label{lemma:minweight}
If $(1 - \mu)^{\rho} \geq \HALF$,
then $\frac{V_{\min}}{l} \rightarrow 0$ \WHP~as $\LGN$.
Otherwise, 
if $(1 - \mu)^{\rho} < \HALF$,
$\frac{V_{\min}}{l} \rightarrow \nu$ \WHP~as $\LGN$
where
$\nu$ is a solution of the equation
$
\left[ \RATIO{\nu} \right]^{\rho} = \HALF
$ in $(0, \mu)$.
\end{lemma}
\begin{proof}
First, we assume that $(1 - \mu)^{\rho} \geq \HALF$, which indicates $n (1-\mu)^{\rho} \geq 1$ by defition.
Then, the probability that $|W_{i}| = 0$ is at most $\exp (-\frac{1}{2} \EX{|W_{i}|})$ by Chernoff bound.
However, for fixed $\mu$, 
\[
\EX{|W_{1}|} = n \EN{1} \geq \frac{\mu}{1 - \mu} l \in O(l) \,.
\]
Therefore, by Chernoff bound, $P(|W_{1}| = 0) \rightarrow 0$ as $l \rightarrow \infty$. This implies that $V_{\min}$ is $o(l)$ with high probability as $\LGN$.

Second, we look at the case that $(1 - \mu)^{\rho} < \HALF$.
For any $\epsilon \in (0, \mu - \nu)$, to use Stirling's approximation,
\begin{align*}
\EX{|W_{(\nu + \epsilon) l}|} & \approx n \EL{(\nu + \epsilon)} \\
& \approx \ACHL{(\nu + \epsilon)} \\
& \qquad \times n \mu^{(\nu + \epsilon)l} (1 - \mu)^{(1 - (\nu + \epsilon))l}\\
& = \frac{n}{\ACHSIML{(\nu + \epsilon)}} \left[ \RATIC{\nu + \epsilon} \right]^{l} \, .
\end{align*}
Since $\RATIO{x}$ is a increasing function of $x$ over $(0, \mu)$ by Lemma~\ref{lemma:entropy}, 
\[
\RATIC{\nu + \epsilon} = (1+\epsilon') \left(\HALF\right)^{1 / \rho} = (1+\epsilon') n^{-1/l}
\]
for some constant $\epsilon' > 0$. Therefore, 
\[
\EX{|W_{(\nu + \epsilon) l}|} = \frac{(1+\epsilon')^{l}}{\ACHSIML{(\nu + \epsilon)}}
\]
exponentially increases as $l$ increases. By Chernoff bound, $|W_{(\nu + \epsilon)l}|$ is not zero with high probability as $l \rightarrow \infty$, \ie~$\LGN$.

In a similar way, $\EX{|W_{(\nu - \epsilon) l}|} = \frac{(1-\epsilon')^{l}}{\ACHSIML{(\nu - \epsilon)}}$ exponentially decreases as $l$ increases. 
Since $\EX{|W_{i}|} \geq \EX{|W_{j}|}$ if $\mu l \geq i \geq j$, the expected number of nodes with at most weight $(\nu - \epsilon)l$ is less than $(\nu - \epsilon)l ~\EX{|W_{(\nu - \epsilon)l}|}$ 
and its value goes to zero as $l \rightarrow \infty$. 
Hence, by Chernoff bound, there exists no node of the weight less than $(\nu - \epsilon)l$ with high probability as $\LGN$.

To sum up, $\frac{V_{\min}}{l}$ goes to $\nu$ with high probability as $l \rightarrow \infty$, \ie~$\LGN$.
\end{proof}

~\\
Using the above lemmas, we show the condition that the network is connected.

~\\
\begin{appendthmproof}{thm:conn}
\hide{
Assume that $|S_{j}| \in \Theta(n)$ for some $j$.
Then, we hope to claim that $S_{j}$ is connected with high probability if $\ECP{u, V \xset u}{u \in W_{j}} \geq c \log n$ for sufficiently large $c$ as $\LGN$.
Let's think of a subset $S' \subset S_{j}$ such that $S'$ is neither an empty set nor $S_{j}$ itself.
Then, the expected number of edges between $S'$ and $S_{j} \xset S'$ is $\EP{S', S_{j} \xset S'} = |S'|\cdot|S_{j} - S'|\cdot\ECP{u, v}{u, v \in S_{j}}$ for distinct $u$ and $v$.
By Theorem~\ref{thm:mono},
\begin{align*}
\ECP{u, v}{u, v \in S_{j}} & \geq \ECP{u, v}{u \in S_{j}, v \in V}\\
& \geq \ECP{u, v}{u \in W_{j}, v \in V \xset u}\\
& \geq \frac{c \log n}{n} \, .
\end{align*}
Since the probability that there exists no edge between $S'$ and $S_{j} \xset S'$ is at most $\exp \left(-\frac{1}{2}\EP{S', S_{j} \xset S'}\right)$ by Chernoff bound, it is bounded as follows:
\begin{align*}
P(S\mathrm{~is~disconnected}) & \leq \sum_{S' \subset S_{j}, S' \neq \emptyset, S_{j}} P(\mathrm{no~edge~between~}S', S_{j} \xset S')\\
& \leq \sum_{S' \subset S_{j}, S' \neq \emptyset, S_{j}} \exp \left(-\frac{c \log n}{2n} |S'|\cdot|S_{j} \xset S'|\right)\\
& \leq 2 \sum_{1 \leq i \leq |S_{j}|/2} {|S_{j}| \choose i} \exp \left(-\frac{c |S_{j}| \log n}{2n}  i\right)\\
& \leq 2 \sum_{1 \leq i \leq |S_{j}|/2} \exp \left( \left( \log |S_{j}| - \frac{c |S_{j}| \log n}{2n} \right) i \right)\\
& = 2 \sum_{1 \leq i \leq |S_{j}|/2} \exp \left(- O(i \log n) \right) \in o(1)
\end{align*}
as $\LGN$.
Therefore, $S_{j}$ is connected with high probability.
}
Let $\frac{V_{\min}}{l} \rightarrow t$ for a constant $t \in [0, \mu)$ as $\LGN$.

If $\cricpxconn{t} > \HALF$, by Lemma~\ref{lemma:expdeg},
\begin{align*}
\ECP{u, V \xset u}{u \in W_{V_{\min}}} 
& \approx \ECP{u, V \xset u}{u \in W_{tl}} \\
& \approx \left[2 \cricpxconn{t}\right]^{\log n} \\
& = (1 + \epsilon)^{\log n} \\
& \geq c \log n
\end{align*}
for some $\epsilon > 0$ and sufficiently large $c$.
Note that $S_{V_{\min}}$ indicates the entire network by definition of $V_{\min}$.
Since $|S_{V_{\min}}|$ is $\Theta(n)$, ~$S_{V_{\min}}$ is connected with high probability as $\LGN$ by Theorem~\ref{thm:connected}.
Equivalently, the entire network is also connected with high probability $\LGN$.

On the other hand, when $\criconn{V_{\min}} < \HALF$, the expected degree of a node with $|V_{\min}|$ weight is $o(1)$ 
because from the above relationship $\ECP{u, V\xset u}{u \in W_{V_{\min}}} \approx (1-\epsilon')^{\log n}$ for some $\epsilon' > 0$.
Thus, in this case, some node in $W_{V_{\min}}$ is isolated with high probability so the network is disconnected.
\end{appendthmproof}

\section{Appendix: Diameter}

\begin{theorem}
\label{thm:randomdiam}
\cite{bollobas90diameter,klee81diameter}
For an \GNP~$G(n, p)$, if $(pn)^{d-1}/n \rightarrow 0$ and $(pn)^{d}/n \rightarrow \infty$ for a fixed integer $d$, then $G(n, p)$ has diameter $d$ with probability approaching 1 as $\LGN$.
\end{theorem}

~\\
\begin{appendproof}{lemma:directedge}
Let $A^{G}$ and $A^{H}$ be the probabilistic adjacency matrix of random graphs $G$ and $H$, respectively.
If $A^{G}_{ij} \geq A^{H}_{ij}$ for every $i, j$ and $H$ has a constant diameter with high probability, then so does $G$.
It can be understood in the following way.
To generate a network with $A^{G}$, we first generate edges with $A^{H}$ and further create edges with $(A^{G} - A^{H})$.
However, as the edges created in the first step already result in the constant diameter with high probability, $G$ has a constant diameter.

Note that $\min_{u, v \in S_{\lambda l}} P[u, v] \geq \beta^{\lambda l} \gamma^{(1-\lambda)l}$.
Thus, it is sufficient to prove that the \GNP~$G(|S_{\lambda l}|, \beta^{\lambda l} \gamma^{(1-\lambda)l})$ has a constant diameter with high probability as $\LGN$.
However,
\begin{align*}
\EX{|W_{\lambda l}|} \beta^{\lambda l} \gamma^{(1-\lambda)l} & = n \EL{\lambda} \beta^{\lambda l} \gamma^{(1-\lambda)l} \\
& \approx \frac{n}{\ACHSIML{\lambda}} \left( \frac{\mu \beta}{\lambda} \right)^{\lambda l} \left( \frac{(1-\mu)\gamma}{1-\lambda}\right)^{(1-\lambda)l} \quad \quad (\mbox{By Stirling approximation})\\
& = \frac{n}{\ACHSIML{\lambda}} \skron^{l} \quad \quad (\because \lambda = \frac{\mu \beta}{\mu \beta + (1-\mu) \gamma})\\
& = \frac{1}{\ACHSIML{\lambda}} \left( 2 \crisimconn \right)^{\log n}\\
& = \frac{1}{\ACHSIML{\lambda}} \left( 1 + \epsilon \right)^{\log n}
\end{align*}
for some $\epsilon > 0$.

Since this value goes to infinity as $\LGN$, so does $\EX{W_{\lambda l}}$.
Therefore, by Chernoff bound,
$|W_{\lambda l}| \geq c \EX{W_{\lambda l}}$ with high probability as $\LGN$ for some constant $c$.
Then,
\begin{align*}
|S_{\lambda l}| \beta^{\lambda l} \gamma^{(1-\lambda)l}
& \geq |W_{\lambda l}|\beta^{\lambda l} \gamma^{(1-\lambda)l} \\
& \geq c \EX{|W_{\lambda l}|}\beta^{\lambda l} \gamma^{(1-\lambda)l} \\
& \approx \frac{c}{\ACHSIML{\lambda}} \left( 1 + \epsilon \right)^{\log n} \, .
\end{align*}

By Theorem~\ref{thm:randomdiam}, an \GNP~$G(|S_{\lambda l}|, \frac{c(1+\epsilon)^{\log n}}{|S_{\lambda l}| \ACHSIML{\lambda} })$ has a diameter of at most $\left( 1 + \frac{\ln 2}{\epsilon} \right)$ with high probability as $\LGN$. Thus, the diamters of $G(|S_{\lambda l}|, \beta^{\lambda l} \gamma^{(1-\lambda)l})$ as well as $S_{\lambda l}$ are also bounded by a constant with high probability as $\LGN$.
\end{appendproof}

~\\
\begin{appendproof}{lemma:directedge}
For any $u \in V$,
\begin{align*}
P[u, S_{\lambda l}] & \geq \sum_{j = \lambda l}^{l} n \EN{j} \beta^{j} \gamma^{l - j}\\
& = \sum_{j = \lambda l}^{l} n {l \choose j} \lambda^{j} (1-\lambda)^{l-j} \left(\frac{\mu \beta}{\lambda}\right)^{j} \left(\frac{(1-\mu) \gamma}{1-\lambda}\right)^{l - j}\\
& = \sum_{j = \lambda l}^{l} n {l \choose j} \lambda^{j} (1-\lambda)^{l-j} \left(\mu \beta + (1-\mu) \gamma\right)^{l}\\
& = \left( 2 \crisimconn \right)^{\log n} \left( \sum_{j = \lambda l}^{l} {l \choose j} \lambda^{j} (1-\lambda)^{l-j} \right) \, .
\end{align*}
By Centeral Limit Theorem, $\sum_{j = \lambda l}^{l} {l \choose j}\lambda^{j} (1 - \lambda)^{l-j}$ converges to $\frac{1}{2}$ as $l \rightarrow \infty$. 
Therefore, $P[u, S_{\lambda l}]$ is greater than $c \log n$ for a constant $c$, and then, by Chernoff bound, $u$ is directly connected to $S_{\lambda l}$ with high probability as $\LGN$.
\end{appendproof}


\section{Appendix: Degree Distribution}

\begin{theorem}
\label{thm:young}
\cite{young07dotprod}
$
P\left( deg(u) = k \right) = \int_{u \in V} {{n-1} \choose k} \left( \EP{u, v} \right)^{k} \left(1-\EP{u, v}\right)^{n-1-k} du
$ \, .
\end{theorem}

\begin{corollary}
\label{cor:degdist}
For $E_{j} = \lkron^{j}\skron^{l-j}$,\\
the probability of degree $k$ in \modeldesc~is~
$
p_{k} = \sum_{j = 0}^{l} \EN{j}  {{n-1} \choose k} E_{j}^{k} \left( 1 - E_{j} \right)^{n - 1 - k}
$ \, .
\end{corollary}
\begin{proof}
To reformulate Theorem~\ref{thm:young},
\[
P\left( deg(u) = k \right) = \sum_{j = 0}^{l} P(u \in W_{j}) {{n-1} \choose k} \left( \ECP{u, v}{u \in W_{j}} \right)^{k} \left(1-\ECP{u, v}{u \in W_{j}} \right)^{n-1-k}  \,.
\]
Therefore, by applying Lemma~\ref{lemma:expprob}, we obtain the desired formula.
\end{proof}

~\\
\begin{appendthmproof}{thm:degdist}
To reduce the space, we begin by defining some notations as follow:
\begin{align*}
x & = \lwokron \\
y & = \swokron \\
f_{j}(k) & = {{n-1} \choose k} \left( \XYA{j} \right)^{k} \left( 1 - \XYA{j} \right)^{n - 1 - k} \\
g_{j}(k) & = \EN{j} f_{j}(k) \, .
\end{align*}
By Corollary~\ref{cor:degdist}, we can restate $p_{k}$ as $\sum_{j = 0}^{l} g_{j}(k)$.

If most of those terms turn out to be insignificant under our assumptions,
the probability $p_k$ can be approximately proportional to one or few dominant terms.
In this case, what we need to do is thus 
to seek for $j$ that maximizes $g_{j}(k) = \EN{j} f_{j}(k)$ and 
find its approximate formula.
 
We start with the approximation of $f_{j}(k)$. 
For large $n$ and $k$, by Stirling approximation,
\begin{align*}
f_{j}(k) & \approx \frac{\sqrt{2 \pi n} (n/e)^{n} \left(\XYA{j}\right)^{k} \left( 1- \XYA{j} \right)^{n-k} }
{\sqrt{2\pi k}(k/e)^{k} \sqrt{2\pi (n-k)}\left((n-k)/e\right)^{n-k}}\\
& = \frac{1}{\sqrt{2 \pi k \left(1 - \frac{k}{n}\right)}} \left(\frac{n \XYA{j}}{k} \right)^{k} \left(\frac{1-\XYA{j}}{1-k/n}\right)^{n-k} \, .
\end{align*}

However, the expected degree of maximum weight node is $O(n \lkron^{l})$, so is the expected maximum degree.
$k$ is thus $o(n)$ with high probability as $\LGN$, \ie~$l \rightarrow \infty$.
\[
\therefore~~
\left(\frac{1-\XYA{j}}{1-k/n}\right)^{n-k} \approx \exp \left( -(n-k)\XYA{j} + (n-k)k/n \right) \approx \exp ( -n \XYA{j} + k ) \, .
\]

For sufficiently large $l$, we can further simplify $g_{j}(k)$ by normal approximation of the binomial distribution:
\begin{align*}
\ln g_{j}(k) & = \ln \EN{j} + \ln f_{j} (k) \\
& \approx -\frac{1}{2} \ln \left( 2 \pi l \mu (1-\mu) \right)- \frac{1}{2 l \mu (1-\mu)} (j - \mu l)^{2} + \ln f_{j} (k) \\
& \approx C - \frac{1}{2 l \mu (1-\mu)} (j - \mu l )^{2} - \frac{1}{2} \ln k - k \ln \frac{k}{n \XYA{j}}  + k \left( 1 - \frac{n \XYA{j}}{k}\right)
\end{align*}
for some constant $C$.
When $k = n \XYA{\tau}$ for $\tau \geq \mu l$ and $R = \frac{x}{y}$,
\begin{align*}
\ln g_{j}(k) & \approx C - \frac{1}{2 l \mu (1-\mu)} (j - \mu l )^{2} - \frac{1}{2} \ln k + k (j - \tau) \ln R + k \left ( 1 - R^{j - \tau}\right) \,.
\end{align*}
Using $(j - \mu l)^{2} = (j - \tau)^{2} + (\tau - \mu l)^{2} + 2(j- \tau) (\tau - \mu l)$,
\[
\ln g_{j}(k) \approx C_{\tau}
- \frac{(j - \tau)^{2}}{2l \mu (1-\mu)} + (j-\tau) \left(k \ln R - \frac{\tau - \mu l}{l \mu (1-\mu)} \right) + k\left(1 - R^{j - \tau}\right) - \frac{1}{2} \ln k
\]
for $C_{\tau} = C - \frac{(\tau - \mu l)^{2}}{2l \mu (1-\mu)}$.

Considering $g_{j}(k)$ as a function of $j$, not $k$, 
now we find $j$ that maximizes $g_{j}(k)$
for $k = n \XYA{\tau}$.
However, the median weight is approximately equal to $\mu l$ by Central Limit Theorem.
If we focus on the higher half degrees.
we can thus let $\tau \geq \mu l$.
In this case, since $\crilcc > \HALF$,
\[
\therefore~~
k \geq \cricpxconn{\mu} \in \Omega(l) \, .
\]

If we differentiate $\ln g_{j}(k)$ over $j$,
\[
\left( \ln g_{j}(k) \right)' \approx - \frac{j - \tau}{l \mu (1-\mu)} + \left(k \ln R - \frac{\tau - \mu l}{l \mu (1-\mu)} \right) - k R^{j - \tau} \ln R = 0 \, .
\]

Because $k \in \Omega(l)$ and $j, \tau \in O(l)$,
we can conclude that $R^{j-\tau} \approx 1$ as $\LGN$; 
otherwise, $| \left(\ln g_{j}(k) \right)'|$ grows as large as $\Omega(k)$.
Therefore, when $j \approx \tau$, $g_{j}(k)$ is maximized.

Furthermore, since $|\frac{j-\tau}{2 l \mu (1-\mu)}| \ll k \ln R$ as $\LGN$,
the first quadratic term $\frac{(j-\tau)^2}{2l \mu (1-\mu)}$ in $\ln g_{j}(k)$ is negligible.
As a result, when $R$ is practical (close to $1.6 \sim 3$), 
$\ln g_{\tau + \Delta}$ would be at most $\left(\Theta(-k|\Delta|) - \ln g_{\tau}\right)$ for $\Delta \geq 1$. 
After all, $g_{\tau}$ effectively dominates the probability $p_{k}$, \ie~$\ln p_{k}$ is roughly proportional to $\ln g_{\tau}$. 
By assigning $\tau = \frac{\ln k - \ln ny^{l}}{\ln R}$, we obtain
\begin{align*}
\ln p_{k} 
& \approx C - \frac{1}{2 l \mu (1-\mu)} \left(\frac{\ln k - \ln ny^{l}}{\ln R} - \mu l \right)^{2} - \frac{1}{2} \ln k \\
& = C' - \frac{1}{2 l \mu (1-\mu) (\ln R)^{2}} \left(\ln k - \ln ny^{l} - l \mu \ln R - \HALF l \mu (1-\mu) (\ln R)^{2} \right)^{2} - \ln k \, .
\end{align*}
for some constant $C'$.
Therefore, the degree distribution $p_{k}$ approximately follows the log-normal
as described in Theorem~\ref{thm:degdist}.
\end{appendthmproof}


\section{Appendix: Power-law Distribution}

~\\
\begin{appendproof}{lemma:generalprob}
Since $a_{i}$'s are independently distributed Bernoulli random variables,
Lemma~\ref{lemma:generalprob} holds.
\end{appendproof}

~\\
\begin{appendproof}{lemma:generaldeg}
Let's define $P_{j}(u, v)$ as the edge probability between $u$ and $v$ when considering only up to the $j$-th attribute, \ie
\[
P_{j}(u, v) = \prod_{i = 1}^{j} \Theta_{i}\left[ a_{i}(u), a_{i}(v) \right] \, .
\]
Thus, what we aim to show is that for a node $v$, 
\[
\EX{P_{l}(u, v)} = \prod_{i=1}^{l} \left(\mu_{i} \alpha_{i} + (1-\mu_{i})\beta_{i}\right)^{\mathbf{1}\left\{a_{i}(u) = 0\right\}}
\left(\mu_{i} \beta_{i} + (1-\mu_{i}) \gamma_{i} \right)^{\mathbf{1}\left\{a_{i}(u) = 1\right\}} \, .
\]

When $l = 1$, it is trivially true by Lemma~\ref{lemma:expprob}.
When $l > 1$, suppose that the above formula holds for $l = 1, 2, \cdots, k$.
Since $P_{k+1}(u, v) = P_{k}(u, v) \Theta_{k+1}\left[a_{k+1}(u), a_{k+1}(v)\right]$,
\begin{align*}
& \EX{P_{k+1}(u, v)} \\
& = \EX{P_{k}(u, v)} \EX{\Theta_{k+1}[ a_{k+1}(u), a_{k+1}(v)]} \\
& = \EX{P_{k}(u, v)} (\mu_{k+1} \alpha_{k+1} + (1-\mu_{k+1})\beta_{k+1})^{\mathbf{1}\left\{a_{k+1}(u) = 0\right\}} (\mu_{k+1} \beta_{k+1} + (1-\mu_{k+1})\gamma_{k+1})^{\mathbf{1}\left\{a_{k+1}(u) = 1\right\}} \\
& = \prod_{i=1}^{k+1} \left(\mu_{i} \alpha_{i} + (1-\mu_{i})\beta_{i}\right)^{\mathbf{1}\left\{a_{i}(u) = 0\right\}}
\left(\mu_{i} \beta_{i} + (1-\mu_{i}) \gamma_{i} \right)^{\mathbf{1}\left\{a_{i}(u) = 1\right\}} \, .
\end{align*}

Therefore, the expected degree formula described in Lemma~\ref{lemma:generaldeg} holds for every $l \geq 1$.
\end{appendproof}


~\\
\begin{appendthmproof}{thm:power}
Before the main argument, we need to define the ordered probability mass of attribute vectors as $\OP{j}$ for $j = 1, 2, \cdots, 2^{l}$.
For example, if the probability of each attribute vector ($00, 01, 10, 11$) is respectively $0.2, 0.3, 0.4$, and $0.1$ when $l = 2$, 
the ordered probability mass is $\OP{1} = 0.1$, $\OP{2} = 0.2$, $\OP{3} = 0.3$, and $\OP{4} = 0.4$.

Then, by Theorem~\ref{thm:young}, we can express the probability of degree $k$, $p_k$, as follows:
\begin{equation}
\label{eq:power}
p_{k} = {n-1 \choose k} \sum_{j = 1}^{2^{l}} \OP{j} (E_{j})^{k} ( 1 - E_{j})^{n-1-k}
\end{equation}
where $E_{j}$ denotes the average edge probability of the node which has the attribute vector corresponding to $\OP{j}$.
If $p_{(j)}$'s and $E_{j}$'s are configured so that few terms dominate the probability, we may approximate $p_{k}$ as
${n-1 \choose k} \OP{\tau} (E_{\tau})^{k}(1 - E_{\tau})^{n-1-k}$
for $\tau = \arg \max_{j} p_{(j)} \left( E_{j} \right)^{k} \left( 1 - E_{j} \right)^{n-1-k}$.
Assuming that this approximation holds, we will propose a sufficient condition for the power-law degree distribution and suggest an example for this condition.


To simplify computations,
we propose a condtion that $p_{(j)} \propto E_{j}^{-\delta}$ for a constant $\delta$.
Then, the $j$-th term is
\[
{n-1 \choose k} p_{(j)} \left( E_{j} \right)^{k} \left( 1 - E_{j} \right)^{n-1-k}
\propto
\left(\left(E_{j}\right)^{k - \delta} \left(1 - E_{j}\right)^{n-1-k}\right)\, ,
\]
which is maximized when $E_{j} \approx \frac{k - \delta}{n - 1 - \delta}$. 
Moreover, under this condition, 
%
if $E_{j+1} / E_{j}$ is at least $(1 + z)$ for a constant $z > 0$, then
\[
\frac{p_{(\tau + \Delta)} \left(E_{\tau + \Delta} \right)^{k} \left(1 - E_{\tau + \Delta} \right)^{n-1-k}}
{p_{(\tau)} \left(E_{\tau} \right)^{k} \left(1 - E_{\tau} \right)^{n-1-k}}
\]
is $o(1)$ for $\Delta \geq 1$ as $\LGN$. Therefore, the $\tau$-th term dominates the Equation~(\ref{eq:power}).

Next, by the Stirling approximation with above conditions,
\begin{align*}
p_{k} & \approx {n-1 \choose k} \left( \frac{k - \delta}{n-1-\delta} \right)^{k-\delta} \left(\frac{n-1-k}{n-1-\delta}\right)^{n-1-k} \\
& \propto
\frac{1}{\sqrt{k (n-1-k)}}
\left(k - \delta \right)^{-\delta}
\left(\frac{(n-1)(k-\delta)}{k(n-1-\delta)} \right)^{k} 
\left(\frac{n-1}{n-1-\delta}\right)^{n-1-k} \\
& \propto k^{-1/2}
\left(k - \delta \right)^{-\delta}
\left(1 - \frac{\delta}{k} \right)^{k}  \\
& \approx k^{-\delta - 1/2} \exp(-\delta)
\end{align*}
for sufficiently large $k$ and $n$.
Thus, $p_{k}$ is approximately proportional to $k^{-\HALF-\delta}$ for large $k$ as $\LGN$.

Last, we prove that the two conditions for the power-law degree distribution are simultaneously feasible by providing an example configuration.

If every $\OP{j}$ is distinct and
$
\frac{\mu_{i}}{1-\mu_{i}} = \left( \frac{\mu_{i} \alpha_{i} + (1-\mu_{i}) \beta_{i}}{\mu_{i} \beta_{i} + (1-\mu_{i}) \gamma_{i}} \right)^{-\delta}
$,
then we satisfy the condition that $p_{(j)} \propto (E_{j})^{-\delta}$ by Lemma~\ref{lemma:generalprob} and Lemma~\ref{lemma:generaldeg}.
On the other hand, if we set $\frac{\mu_{i}}{1-\mu{i}} = \left(1+z\right)^{-2^{i}\delta}$, then the other condition, $E_{j+1} / E_{j} \geq (1+z)$ is also satisfied. 
Since we are free to configure $\mu_{i}$'s and $\Theta_{i}$'s independently, the sufficient condition for the power law degree distribution is feasible.
\end{appendthmproof}